%% file: Thesis.tex
\documentclass[a4paper, 8pt, twoside]{Thesis} % Use the "Thesis" style, based on the ECS Thesis style by Steve Gunn
\graphicspath{Figures/}  % Location of the graphics files (set up for graphics to be in PDF format)
\usepackage[round]{natbib}   % omit 'round' option if you prefer square brackets
\usepackage{graphicx}
\usepackage{caption}
\usepackage{booktabs}
\usepackage{amsmath}
\usepackage{colortbl}
\usepackage{rotating}
\usepackage{amssymb}
\usepackage{latexsym}
\usepackage{ifthen}
\usepackage{enumitem}
\def\be{\begin{equation}}
\def\bea{\begin{eqnarray}}
\def\eea{\end{eqnarray}}
\def\ee{\end{equation}}

\def\md{\dot{\cal M}}
\def\mdtj{{\dot {\cal M}}}
\def\mdtjc{{\dot {\cal M}}_c}

\def\bc{\begin{center}}
\def\ec{\end{center}}

\def\etal{{\em et al.}}
\def\ie{{\em i.e.}}
\def\ep{{e^--p^+}}
\def\el{{e^--e^+}}
\def\s{{\rm C}}
\def\sk{{\rm D}}
\def\kd{{\rm K}}
\def\mbh{M_{\rm B}}
\def\rg{r_{\rm g}}
\def\rs{r_{\rm s}}

\def\tg{t_{\rm g}}
\def\xsh{x_{\rm sh}}
\def\bsh{B_{\rm sh}}

\def\rsh{r_{\rm sh}}
\def\ep{{{\rm e}^--{\rm p}^+}}

\def\gamt{\gamma_{\rm \small T}}
\def\vt{v_{\rm \small T}}
\def\veq{v_{\rm eq}}

\def\od{\rm \small D}
\def\msol{M_\odot}
\def\frad{{\cal F}_{\rm rd}}

\def\gamt{\gamma_{\rm \small T}}
\def\gsim{\lower.5ex\hbox{$\; \buildrel > \over \sim \;$}}

\def\gmt{\gamma_{\infty}}
\def\Hsh{H_{\rm sh}}
\def\lsim{\lower.5ex\hbox{$\; \buildrel < \over \sim \;$}}

\begin{document}
%\oneside
%\usepackage{setspace}
\frontmatter      % Begin Roman style (i, ii, iii, iv...) page numbering

% Set up the Title Page
\title  {Astrophysical Jets in Relativistic Regime : Thermal and Radiative Driving}
\authors  {\texorpdfstring
            {\href{mukesh.vyas@aries.res.in}{Mukesh Kumar Vyas}}
            {Mukesh Kumar Vyas}
            }
\addresses  {\groupname\\\deptname\\\univname}  % Do not change this here, instead these must be set in the "Thesis.cls" file, please look through it instead
\date       {August 2019}
\subject    {}
\keywords   {}

\maketitle
%% ----------------------------------------------------------------

%\setstretch{1.3}  % It is better to have smaller font and larger line spacing than the other way round

% Define the page headers using the FancyHdr package and set up for one-sided printing
\fancyhead{}  % Clears all page headers and footers
\rhead{\thepage}  % Sets the right side header to show the page number
\lhead{}  % Clears the left side page header

\addtocontents{toc}{\vspace{0em}}
%\vfill\vfill\vfill\vfill\vfill\vfill\null
\clearpage  % Funny Quote page ended, start a new page
%% ----------------------------------------------------------------

% The Abstract Page
%\frontmatter

%\frontmatter
%\abstract{g}
%\vspace{2em}

\input{abs.tex}
\addtotoc{Abstract}  % Add the "Abstract" page entry to the Contents
%\abstract{
\addtocontents{toc}{\vspace{0em}}  % Add a gap in the Contents, for aesthetics
%\ldots
%}
%\clearpage  % Abstract ended, start a new page
%% ----------------------------------------------------------------

\setstretch{1.3}  % Reset the line-spacing to 1.3 for body text (if it has changed)

% The Acknowledgements page, for thanking everyone
%\acknowledgements{s

%\addtocontents{toc}{\vspace{0em}}  % Add a gap in the Contents, for aesthetics

%The acknowledgments and the people to thank go here, don't forget to include your project advisor\ldots
%\input{ackn.tex}
%\addtotoc{Acknowledgement}  % Add the "Abstract" page entry to the Contents
%\abstract{
%\addtocontents{toc}{\vspace{0em}}  % Add a gap in the Contents, for aesthetics

%}
\clearpage  % End of the Acknowledgements
%% ----------------------------------------------------------------

\pagestyle{fancy}  %The page style headers have been "empty" all this time, now use the "fancy" headers as defined before to bring them back

%% ----------------------------------------------------------------
\lhead{\emph{Contents}}  % Set the left side page header to "Contents"
\small
\tableofcontents  % Write out the Table of Contents

%% ----------------------------------------------------------------
%@\lhead{\emph{List of Figures}}  % Set the left side page header to "List if Figures"
%@\listoffigures  % Write out the List of Figures

%% ----------------------------------------------------------------
%@\lhead{\emph{List of Tables}}  % Set the left side page header to "List of Tables"
%@\listoftables  % Write out the List of Tables

%% ----------------------------------------------------------------
\setstretch{1.5}  % Set the line spacing to 1.5, this makes the following tables easier to read
\clearpage  % Start a new page
\lhead{\emph{Abbreviations}}  % Set the left side page header to "Abbreviations"
\addtotoc{Acknowledgement}
\listofsymbols{ll}  % Include a list of Abbreviations (a table of two columns)
{
% \textbf{Acronym} & \textbf{W}hat (it) \textbf{S}tands \textbf{F}or \\
\textbf{ADAF}  &~~~~~~~~~~~~ \textbf{A}dvection \textbf{D}ominated \textbf{A}ccretion \textbf{F}low\\
\textbf{AGN}  &~~~~~~~~~~~~ \textbf{A}ctive \textbf{G}alactic \textbf{N}ucleus\\
\textbf{BH} &~~~~~~~~~~~~ \textbf{B}lack \textbf{H}ole \\% \textbf{H}ere \\
\textbf{EoM} &~~~~~~~~~~~~ \textbf{E}quations \textbf{o}f \textbf{M}otion \\
\textbf{Eq} &~~~~~~~~~~~~ \textbf{E}quation\\
\textbf{EoS} &~~~~~~~~~~~~ \textbf{E}quation \textbf{o}f \textbf{S}tate \\
\textbf{GR} &~~~~~~~~~~~~ \textbf{G}eneral \textbf{R}elativity\\
\textbf{GRB} &~~~~~~~~~~~~ \textbf{G}amma-\textbf{R}ay \textbf{B}urst\\
\textbf{GRMHD} &~~~~~~~~~~~~ \textbf{G}eneral-\textbf{R}elativistic \textbf{M}agneto-\textbf{H}ydro-\textbf{D}ynamics\\
\textbf{KD}  &~~~~~~~~~~~~ \textbf{K}eplerian  \textbf{D}isc\\
\textbf{L.H.S} &~~~~~~~~~~~~ \textbf{L}eft \textbf{H}and \textbf{S}ide\\
\textbf{MHD} &~~~~~~~~~~~~ \textbf{M}agneto-\textbf{H}ydro-\textbf{D}ynamics\\
\textbf{pNp} &~~~~~~~~~~~~ \textbf{p}seudo \textbf{N}ewtonian \textbf{p}otential \\
\textbf{PSD} &~~~~~~~~~~~~ \textbf{P}ost-\textbf{S}hock \textbf{D}isk\\
\textbf{PW} &~~~~~~~~~~~~ \textbf{P}aczy\'nski-\textbf{W}iita\\
\textbf{RHD}    &~~~~~~~~~~~~ \textbf{D}adiation \textbf{H}ydro-\textbf{D}ynamics\\
\textbf{R.H.S} &~~~~~~~~~~~~ \textbf{R}eft \textbf{H}and \textbf{S}ide \\
\textbf{SKD} &~~~~~~~~~~~~ \textbf{S}ub-\textbf{K}eplerian \textbf{D}isc\\
\textbf{SR} &~~~~~~~~~~~~ \textbf{S}pecial \textbf{R}elativity\\
\textbf{XRB} &~~~~~~~~~~~~ \textbf{Y}-\textbf{R}ay \textbf{B}inary\\
\textbf{YSO} &~~~~~~~~~~~~ \textbf{Y}oung-\textbf{S}tellar \textbf{O}bject\\

}

%% ----------------------------------------------------------------
\clearpage  % Start a new page
%\lhead{\emph{Physical Constants}}  % Set the left side page header to "Physical Constants"
%\listofconstants{lrcl}  % Include a list of Physical Constants (a four column table)
%{
%% Constant Name & Symbol & = & Constant Value (with units) \\
%Speed of Light & $c$ & $=$ & $2.997\ 924\ 58\times10^{8}\ \mbox{ms}^{-\mbox{s}}$ (exact)\\
%
%}

%% ----------------------------------------------------------------
%\clearpage  %Start a new page
%\lhead{\emph{Relativistic Jets}}  % Set the left side page header to "Symbols"
%\listofnomenclature{lll}  % Include a list of Symbols (a three column table)
%{
%% symbol & name & unit \\
%$a$ & distance & m \\
%$P$ & power & W (Js$^{-1}$) \\
%& & \\ % Gap to separate the Roman symbols from the Greek
%$\omega$ & angular frequency & rads$^{-1}$ \\
%}
%% ----------------------------------------------------------------
% End of the pre-able, contents and lists of things
% Begin the Dedication page

 % Add a gap in the Contents, for aesthetics

%% ----------------------------------------------------------------
\lhead{\emph{Relativistic Jets}}
\mainmatter	  % Begin normal, numeric (1,2,3...) page numbering
\pagestyle{fancy}  % Return the page headers back to the "fancy" style
\fontfamily{qbk}\selectfont
%\fontfamily{qpl}\selectfont
%\small
\chapter{Introduction}
\label{Ch:intro}
\doublespacing
\setlength{\parindent}{4ex}
\section{Overview}
\citet{c18}, while analysing an optical image of M$87$, noticed something interesting and unusual phenomena. Following which he made a note --,
\begin{center}
 \textit {``A curious straight ray... connected with the nucleus".}\\

\end{center}
 This was later recognized and termed as `relativistic jet' \citep{bm54}.
Since then, the observational study of jets has been revolutionized in later decades and
these objects are now well established as ubiquitous astrophysical phenomena associated with various classes of astrophysical objects
% (Fig. \ref{lab:image})
like active galactic
nuclei (AGN, e.g., Messier $87$, PKS B$1545-321$,  J$1247+6723$), young stellar 
objects (YSO, e.g., HH $30$, HH $34$, HH $80-81$, NGC $2071-$IRS$3$), X-ray binaries (XRBs, e.g., SS$433$, Cyg
X$-3$, GRS $1915+105$, GRO $1655-40$) and Gamma ray bursts (e. g., GRB $980519$, GRB $020813$, GRB $080916$C), etc. Out of these classes, jets associated with AGNs, XRBs and GRBs are found to be relativistic and highly collimated.

Relativistic jets are beamed outflows generally bipolar in nature. They seem to be defying gravity as they emerge against strong gravitational pull. These are associated with systems that have a compact accreting massive object at the centre. In such systems, jets can only emerge from the matter being accreted as black holes (BHs) neither have hard surfaces nor they are capable of emission. This fact is further supported by strong correlation observed between spectral states of the accretion disc and jet \citep{gfp03,fgr10,rsfp10}. %The emergence and intensity of the jet is obtained to be in stark contrast with various spectral states of the accretion disc.

%\begin {figure}
%\begin{center}
%% \includegraphics[trim={0cm 0 0cm 0}, width=5.cm]{Curtis.eps}
%\includegraphics[trim={0cm 0 0cm 0}, width=6.4cm]{800px-Messier_87_Hubble_WikiSky.eps}
% %http://www.daviddarling.info/encyclopedia/C/Curtis.html
%%  \includegraphics[width=11.cm]{./P2/fig1.eps}
%\vskip 0.5cm
% \caption{HST image of M87, only one jet is visible because of relativistic beaming effect}
%%\vskip -0.75cm
%\label{lab:image}
% \end{center}
%\end{figure}
%The exact process through which the jets are launched is still a question.
Recent radio observations also limit the jet generation region to be very close to the central object extending up to a distance of less than $100$ Schwarzschild radii ($r_{\rm s}$) \citep{jbl99,detal12}. This implies that the entire accretion disc does not take part in jet generation. As the inner region is hot and radiatively very active, the radiation field around it is very intense and hence it becomes very important to study the impact of thermal pressure as well as radiation driving on the dynamics of the jet.

In current work, we have investigated the dynamics of jets around black hole (BH) candidates like
X-ray binaries and AGNs under thermal and radiation driving. 
%The very first accretion model had a disc around the BH that was Keplerian in nature.  
\section{Brief Chronology and Theoretical Developments}
Ever since the emergence of the first theoretical model of accretion discs, that is, the Keplerian disc (KD) \citep{ss73}, or later disc models like the thick discs \citep{pw80}, the advection-dominated accretion flow (ADAF) \citep{nkh97} and advective discs
\citep{f87,c89}, there have been numerous attempts to understand how photons radiated
away from these discs interact with outflowing jets. 

The equations of motion (EoM) of radiation hydrodynamics (RHD)
were developed 
by many authors \citep{hs76,mm84,kfm98} in special relativistic (SR) regime. Later the general relativistic (GR) version of those equations
was also obtained \citep{p06, t07}. Many authors used these EoMs under a variety of approximations
to study radiatively driven jets. Under Thomson scattering domain, \citet{i80} studied the matter flow in the radiation field above a Keplerian disc. \citet{sw81} studied particle jets in SR regime, driven by the radiation field in the funnel of a
thick accretion disc and obtained a terminal speed of
$\vt \sim 0.4c$ ($c{\equiv}$speed of light in vacuum) for electron-proton or $\ep$ jets, although the
terminal Lorentz factor obtained was $\gamt \sim 3$ for electron-positron or $\el$ jets.  \citet{o81} found that the radiation force imparted under elastic scattering increases and 
results in to enhanced radiative driving, a phenomenon called `Compton rocket'. However, later, \citet{p82} down played the significance of Compton rocket in presence of Compton cooling.
\citet{i89} obtained a theoretical upper limit or `magic speed' $v_{\rm m}=0.45c$ above a KD using the near disc approximation for radiation field. Any speed above $v_{\rm m}$ would
invoke radiative deceleration induced by radiation drag. Around the same time, in elastic scattering domain, \cite{ftrt85} studied
radiation interaction with a fluid jet in SR regime. They mostly assumed isothermal jets
with a non-radial cross-section. A Newtonian gravitational field was added ad hoc to the EoM.
The radiation field was computed from
disc models for a variety of disc thicknesses. They obtained mildly relativistic jets and shocks induced by non-radial nature of the jet cross-section as well as the radiation field.
\citet{f96} studied radiatively driven off-axis particle jets, using the radiation field similar to Icke. The detailed radiation field around a BH was calculated by \citet{hf01} above a KD governed by a point mass gravity
using Newtonian and pseudo-Newtonian potentials (pNp) to mimic {non-rotating and rotating BH exterior}. The strength of the radiation field using Schwarzschild pNp was found to be half of the Newtonian potential, but it was about one order greater for Kerr pNp. In another attempt, \citet{fth01} considered a hybrid disc
consisting of outer KD and inner ADAF. Such a scenario produced jets with $\gamt \sim 2$, and also induced collimation. Except few studies in Compton driving of disc winds \citep{tnc86,qp85} very few attempts are made with consideration of energy transfer by interaction of radiation with jet matter. As stated before, some of the studies include Compton losses and acceleration of jets through Compton thrust up to relativistic speeds \citep{p82, o81,co81}

A large number of jet studies in recent years have relied upon numerical simulations. Most of these works investigate how special relativistic jets interact with
the ambient medium \citep{dh94,mm97}, or how magnetic fields affect them \citep{agmimaah01,kbvk07,mrb10}.
\citet{tnm11}, on the other hand, simulated magnetically arrested disc and jet launching from such a disc.  Although not a simulation, \citet{msvtt06} studied steady jets in the meridional plane in general relativistic magneto hydrodynamics (GRMHD). These kinds of studies are important because they enhance
the understanding of the system and act as test cases for numerical simulations.

In the advective disc regime, numerical simulations \citep{mrc96,dcnm14,lckhr16}
and theoretical investigations \citep{cd07,kc13,kcm14,kc17,kscc13,ck16} showed that the extra-thermal gradient force in the post-shock region automatically generates bipolar outflows.
Anticipating that the intense radiation from the accretion disc may accelerate jets, \citet{cc00a,cc00b,cc02a,cc02b}
investigated radiative driving of jets by advective disc photons.
It was noted that cold jets could be efficiently accelerated to $\vt \sim$few$\times0.1$c.
But to achieve $\vt > 0.9$c for jets, the required base temperature and injection speed was quite high, which does not match with inner accretion disc parameters. Moreover, being in the non relativistic regime, the formalism followed by \citet{cc00a,cc02a} is only correct up to the first order of the flow velocity.
In order to gauge the full extent of radiative acceleration,
investigations of radiatively driven particle jets in SR regime \citep{cdc04,c05}
were undertaken. Under such conditions, disc photons could accelerate jets up to $\gamt \gsim 2$ and
significant collimation could be achieved. The radiation field above such discs has two sources,
one being the hard radiation from the inner post-shock disc,  and the other being the soft radiation
from the pre-shock disc. A compact, hot, low-angular-momentum corona close to the BH, which produces hard radiation, and an external disc producing softer radiation, are not exclusive to shocked advective discs but are seen in many other models \citep{sl76,dwmb97,gzdjeuhp97}. Therefore, the source of radiation, that is,
the underlying accretion disc, may be an advective disc, or any other disc model which considers a compact, geometrically thick corona close to the BH and an outer disc.

%{{ Efforts have been made for study of jets 2-D 
In most of the investigations of relativistic fluid jets, the cross-section was assumed to be spherical ($\propto r^2$, $r$ being the radial distance). \citet{mstv04} considered thermally driven relativistic jets
in Schwarzschild metric, modifying an approximate equation of state (EoS) of single species relativistic gas \citep{m71}. They hid the actual acceleration process in an ad hoc adiabatic index ($\Gamma$) and obtained monotonic jets having mildly to ultra-relativistic terminal speeds.
%from mildly- to ultra-relativistic jet terminal speed.
In contrast, \citet{ftrt85} studied jet driven by radiation as well as the cross-section deviating from the spherical description. Since then, the possibility of internal shocks in outflows, except
for non-spherical solar winds
\citep{lh90}, has rarely been reported; leaving it unclear whether non-conical geometry or
the external radiation field triggers the shock in the jet. 

\section{Motivation}
\label{sec_motivation}
%In view of the development described in last section, we aim to explore the advanced aspects of radiation hydrodynamics of the jets deeper in various regimes.
As the jets originate very close to the BH, general relativistic analysis becomes important. One of the reasons that the study of relativistic jets has mostly been carried out considering Newtonian or pseudo-Newtonian potential and ignoring general relativity was the complexity of handling radiation field in curved space. Generally the relativistic flows in astrophysics are studied considering adiabatic index to be constant. While temperatures in such sources happen to be relativistic, so relativistic equation with variable adiabatic index is required. We mostly consider relativistic treatment to the thermodynamics of the matter. %These set initial motivations for the work as we plan to explore advanced aspects of radiation hydrodynamics of the jets in curved space as well as using relativistic equation of state with variable adiabatic index.
The aim of the thesis is to find answers to some questions regarding nature of the jets in extreme radiation field as well as curved space-time. The basic question that we seek to answer is, up to what speeds, the jets are accelerated in view of radiation acceleration in association with thermal driving. The significance of this question can be realized if we quote a line from \cite{ggmm02} :

\textit{``... but (radiation acceleration) is unable to explain the acceleration of the powerful jets associated with radio loud AGN..."}

This has been general consideration as radiation driving is ineffective for hot flows. Following which, most of the studies are carried out in particle regime where the thermal acceleration is absent \citep{c05,cdc04}. We wish to further examine it in this thesis and will come back to above claim when we conclude the thesis.

Other aspects that we seek to investigate in this attempt are dependence of composition of the matter on the jet dynamics, role of radiation drag along with radiation acceleration, possibility of formation of radiation driven internal shocks, impact of non-conical jet geometry along with radial cross section, comparative study of elastic and in-elastic scattering consideration between photons and jet matter. We seek to explore, where does the analysis stand when it is compared with already existing knowledge about radiation driving of jets and whether it is able to explain observational features of astrophysical sources that harbour relativistic jets. The dissertation is dedicated to explore all these areas.
\section{Thesis at Glance}
In table (\ref{tableA}), we briefly sketch the prominent features of all the chapters along with their significance and implications.
\begin{table}
\caption{Outline}
%\begin{center}
\label{tableA}
    \begin{tabular}{ | p{1.55cm} | p{4cm}|p{7.8cm}|}        \hline
    \hline
{\bf Chapter \newline No.} &		{\bf Key features}		& 	{\bf Key outcomes and significance of the study}\\
\hline
Ch. 1.

& Introduction

& -- \\
\hline
Ch. 2.

& Mathematical Structure 
& -- \\
\hline
Ch. 3.

& Methods of Analysis

& -- \\
\hline
Ch. 4.

& Radiative driving of non rotating, radial, relativistic jets under scattering regime. Flat space assumed and relativistic EoS considered. Pseudo Newtonian potential used to account for strong gravity.

& 1. Explorative study is carried out of radiatively driven astrophysical jets and their Lorentz factors are obtained to be a few. \newline 
2. Studied dependence of jet terminal speeds upon accretion disc luminosity, magnetic pressure, jet composition etc in detail. \newline 
3. Jets in extragalactic sources are obtained to be faster than that in microquasars. \\
\hline

Ch. 5.
& Thermal driving of non-radial jets in Schwarzschild space-time
& 1. It is shown that non-radial geometry of the jets result into internal shocks. \newline 
2. Internal shocks may account for observed high energy tail of spectrum of high energy radio loud sources\\
\hline
%\pagebreak
Ch. 6.
& Radiative driving of relativistic jets in curved space-time under Thomson scattering regime, general relativistic treatment in both fixed gamma and variable gamma EoS.
& 1. In GR analysis, the flow is less hotter as compared to analysis in SR regime. \newline
2. Impact of curved space-time on radiation field is non-linear. \newline
3. Lorentz factors of jets in hot and intense radiation field may reach upto $3$ for $\ep$ and $10$ for $\el$ composition. \newline
4. For transonic solutions, in Thomson scattering regime, the jets at the base are required to be in 'unbound' state with generalized Bernoulli parameter $E>1$
\\
\hline
Ch. 7.
& General relativistic radiative driving of jets under Compton scattering regime.
& 1. Compton scattering allows jets to emerge and accelerate even if the jet base is bound ($E<1$) \newline
2. In radiation field, in Compton regime, even cold and kinetically inefficient jets at the base are driven up to relativistic speeds ($\gamma>10$) and heated up to relativistic temperatures ($T>10^{11} \rm K$).\\
\hline
Ch. 8.

& Conclusions

& -- \\
\hline
    \end{tabular}
%\end{center}
\end{table}
\chapter{Mathematical Structure}
\label{CH:2}
\label{mathematical_structure}
\section{Space-time metric and unit system used}
%Considering spherical coordinates, flat spacetime metric is described by 
%\bea
%ds^2=-g_{tt} c^2dt^2+
%g_{rr}dr^2+g_{\theta \theta}d{\theta}^2+g_{\phi \phi}d\phi^2 
%=-\left(1-\frac{2G\mbh}{c^2r} \right)c^2dt^2 \nonumber \\ 
%+\left(1-\frac{2G\mbh}{c^2r}\right)^{-1}dr^2 
%+r^2d{\theta}^2+r^2\sin^2{\theta}d\phi^2
%%\eqno{(1)}
%\label{metric.eq}
%\eea
The space-time around a non-rotating black hole is described by Schwarzschild metric:
\be
ds^2=g_{\mu\nu}dx^{\mu \nu}=g_{tt} c^2dt^2+
g_{rr}dr^2+g_{\theta \theta}d{\theta}^2+g_{\phi \phi}d\phi^2  \nonumber \\ 
\ee
\be
=-\left(1-\frac{2G\mbh}{c^2r} \right)c^2dt^2
+\left(1-\frac{2G\mbh}{c^2r}\right)^{-1}dr^2 
+r^2d{\theta}^2+r^2\sin^2{\theta}d\phi^2
%\eqno{(1)}
\label{metric.eq}
\ee
Here $r$, $\theta$ and $\phi$ are usual spherical 
coordinates,
$t$ is time, $\mbh$ is the mass of the central 
black hole (BH) and $G$ is the universal constant of gravitation.\\
$g_{\mu \mu}$ are diagonal metric components
\be
-g_{tt}=g^{rr}=\left(1-\frac{2G\mbh}{c^2r}\right)
\ee

However, in chapter \ref{CH:P1}, we consider flat space and hence the metric components become independent of mass of the BH
\be 
-g_{tt}=g_{rr}=1
\ee
Hereafter, we have used geometric units (unless specified otherwise) where $G=\mbh=c=1$ with, such that the units of mass, length and time are $\mbh$, $\rg=G\mbh/c^2$ and $\tg=G\mbh/c^3$, respectively. In this system of units, the event horizon or Schwarzschild radius is at $r_{\rm \small S}=2$
\section{Equation of state}
\subsection{Fixed adiabatic index equation of state}
\label{sbsbsec2.1.1}
Equation of state (EoS) is a
closure relation between internal energy density ($e$), pressure ($p$) and mass
density ($\rho$) of the fluid. If EoS for Newtonian (non-relativistic) fluids with fixed adiabatic index equation of state is extended for relativistic fluids with $\Gamma<5/3$ then
%If we consider the jet fluid obeying polytropic EoS having fixed adiabatic index ($\Gamma$), the form of EoS is given as
\begin{equation}
e=\rho+\frac{p}{\Gamma-1}
%e=n_{e^-}m_ec^2f,
\label{eos0.eq}
\end{equation}

Expressions for adiabatic sound speed $a$ in relativistic regime and enthalpy $h$ are given by \citep{vc18a}
\begin{equation}
a^2=\frac{\Gamma p}{e+p}=\frac{\Gamma \Theta}
{1+N \Gamma \Theta}; ~~ h=\frac{e+p}{\rho}=1+\Gamma N \Theta
\label{sound0.eq}
\end{equation}
Here $N(=\frac{1}{\Gamma-1})$ is polytropic index of the flow and non-dimensional temperature is defined as
$\Theta=p/\rho$.
\subsection{Relativistic equation of state with variable adiabatic index}
As the fluid gets transition between non relativistic and relativistic temperatures, the adiabatic index of the fluid is likely to depend upon temperature \citep{vkmc15}. The value of $\Gamma$ ranges from $5/3$ to $4/3$ as the flow goes from non relativistic temperatures to relativistic temperatures. Considering this, we use EoS for multispecies, relativistic
flow proposed by \citet{c08,cr09} which is an extremely close approximation \citep{vkmc15}
to the exact one \citep{c38,s57}. The EoS is 
given as,
\begin{equation}
e=n_{e^-}m_ec^2f, \mbox{ in physical dimensions}
\label{eos.eq}
\end{equation}
where, $n_{e^-}$ is the electron number density, $m_e$ is the electron rest mass and dimensionless quantity
$f$ is given by
\begin{equation}
f=(2-\xi)\left[1+\Theta\left(\frac{9\Theta+3}{3\Theta+2}\right)\right]
+\xi\left[\frac{1}{\eta}+\Theta\left(\frac{9\Theta+3/\eta}{3\Theta+2/\eta}
\right)\right].
\label{eos2.eq}
\end{equation}
Here,
$\Theta=kT/(m_ec^2)$ is a measure of temperature ($T$), $k$ is Boltzmann constant and
$\xi (= n_{p^{+}}/n_{e^{-}})$ being the relative proportion of number densities of 
protons and electrons.
$\eta (= m_{e}/ m_{p^{+}}$) is the mass ratio of electron and proton.
In relativistic EoS, the expressions of $N$, $\Gamma$,
$a$ and $h$ (in geometric units) are obtained as \citep{cr09}
\begin{equation}
N=\frac{1}{2}\frac{df}{d\Theta} ;~~ \Gamma=1+\frac{1}{N} ; ~~
a^2=\frac{\Gamma p}{e+p}=\frac{2 \Gamma \Theta}
{f+2\Theta}.; ~~ h=\frac{f+2 \Theta}{\tau}
\label{sound.eq}
\end{equation}
Here $\tau(=2-\xi+\xi/\eta)$ is a function of composition.
\section{Radiation Hydrodynamic Equations Governing the Dynamics of Relativistic Fluids }
%\subsubsection{Equation of continuity}
The energy-momentum tensor for the matter ($T^{\mu \nu}_M$) and radiation ($T^{\mu \nu}_R$)
are given by
\begin{equation}
T^{\mu \nu}_M=(e+p)u^{\mu}u^{\nu}+pg^{\mu \nu};
~~T^{\mu \nu}_R={\int}I ~l^{\mu}l^{\nu}d{\Omega},
\end{equation}
Here,  $u^{\nu}$ are the components of four velocity
$l^{\nu}$s are directional derivatives, $I$ is the frequency integrated specific intensity
of the radiation field% where $\nu$ is the frequency of the radiation and
and $d\Omega$ is the differential solid angle subtended by a source point at the accretion disc surface on to the field point at the jet axis.\\
The combined energy momentum tensor for the system is given by 
\be 
T^{\mu \nu}=T^{\mu \nu}_M+T^{\mu \nu}_R
\ee
The equations for conservation of energy and momentum are obtained by that fact that the four divergence of the energy momentum tensor vanishes. Following this and the mass conservation, the equations of motion are given by
\begin{equation}
T^{\mu \nu}_{M;\nu}=-T^{\mu \nu}_{R;\nu}~~~~  \mbox{and} ~~~~
(\rho u^{\nu})_{; \nu}=0,
\label{eqnmot.eq}
\end{equation}

The $i^{\rm th}$ component of the momentum balance equation is obtained by projecting $(T^{\mu \nu}_M+T^
{\mu \nu}_R)_{;\beta}=0$ with the projection tensor 
${\cal P}^i_\nu=(g^{i}_{\nu}+u^iu_{\nu})$

\be
(e+p)\left(u^\nu \frac{\partial u^i}{\partial x^\nu}+\Gamma^i_{\nu \sigma}u^\nu u^\sigma\right)=-(g^{i \nu}+u^iu^{\nu})\frac{\partial p}{\partial x^\nu}+{\cal P}^i_\mu T^{\mu \nu}_R;\nu
\ee

Under assumption of non rotating jets, in steady state, considering spherical coordinates, the $r$ component of momentum balance equation becomes $\footnote{In second term of R.H.S, the multiplied factor is $\frac{\sigma}{m_ec}$ but in our unit system $c=1$, hence it is $\frac{\sigma}{m_e}$}$
\begin{equation}
u^r\frac{du^r}{dr}+\frac{1}{r^2}=-\left(1-\frac{2}{r}+u^ru^r\right)
\frac{1}{e+p}\frac{dp}{dr}+{\rho}_e\frac{{\sigma\sqrt{g^{rr}}\gamma^3}
}{m_e(e+p)}{\Im}^r
\label{eu1con.eq}
%\eqno{(6a)}
%\footnote{In physical units it is $\frac{\sigma_T}{m_ec}$ but in our unit system $c=1$} 
\end{equation}

Here, $\gamma$ is bulk Lorentz factor of the jet, $\rho_e$ is total lepton density and $\sigma$ is cross section given as \citep{by76,p83}, 
\be
\sigma= \chi_c\sigma_T=\left[\frac{1}{1+\left(\frac{T_e}{4.5 \times 10^8}\right)^{0.86}}\right]\sigma_T
\ee
$\sigma_T$ is Thomson scattering cross section. $\chi_c$ accounts for Compton opacity and is less 1. $T_e$ is electron temperature in physical units, which in dimensionless units, is given as a function of $N$ \citep{kc14,vc18b}

$$
\Theta_e=-\frac{2}{3}\pm\frac{1}{3}\sqrt{\left[4-2\left(\frac{2N-3}{N-3}\right)\right]}
$$

${\Im}^r$ is the net radiative contribution to the momentum impart onto the plasma and is given by
\be
{\Im}^r=\frac{\sigma}{m_e} \left[(1+v^2){\cal R}_1-v
\left(g^{rr} {\cal R}_0+\frac{{\cal R}_2}{g^{rr}}\right)\right]
\label{radcontrib.eq}
\ee
Three-velocity $v$ of the jet is defined as $v^2=-u_iu^i/u_tu^t=-u_ru^r/u_tu^t$, {\ie}
$u^r={\gamma}v\sqrt{g^{rr}}$ and $\gamma^2=-u_tu^t$ is the Lorentz factor. ${\cal R}_0,~{\cal R}_1$ and
${\cal R}_2$ are zeroth, first and second moments of specific intensity of the radiation and physically can be identified
as the radiation energy density, the flux and the pressure respectively.\\
We define $R_0=\sigma_T{\cal R}_0/m_e,~R_1=\sigma_T{\cal R}_1/m_e$ and $R_2=\sigma_T{\cal R}_2/m_e$ which
are terms proportional to the radiative moments
%radiation energy density (${\cal R}_0$), flux (${\cal R}_0$)and pressure (${\cal R}_0$),
but for simplicity in rest of this thesis, we call these quantities ($R_0,~R_1,~\&~R_2$) as respective radiative moments.\\

%In scattering regime, 
First law of thermodynamics, or energy equation ($u_{\alpha}T^{\alpha \beta}_{M_{;\beta}}=-u_{\mu}T^{\mu \nu}_{R_{;\nu}}$) is obtained as,
\begin{equation}
\frac{de}{dr}-\frac{e+p} {\rho}\frac{d\rho}{dr}=-\frac{\gamma \rho_e(1-\chi_c)R_t}{\sqrt{g^{rr}}},
\label{en1con.eq}
\end{equation} 
Here $R_t$ is radiative contribution representing energy exchange between imparted radiation and fluid : 
\be
R_t=\frac{\sigma}{m_e}\left[\frac{g^{rr} {\cal R}_0}{v}-\frac{v {\cal R}_2}{g^{rr}}-2{\cal R}_1\right]
\ee 
Integrating the conservation of mass flux equation ($[\rho u^\alpha]_{;\alpha}=0$),
we obtain the mass outflow rate 
\begin {equation}
{\dot M}_{\rm o}=\rho u^r \cal A
\label{mdotout.eq}
\end {equation}
Here ${\cal A}$ is the cross-section of the jet.
It should be noted that under elastic scattering assumption, $\chi=1$ (or $\sigma=\sigma_T$) and the R.H.S of equation (\ref{en1con.eq}) vanishes which means that no energy exchange takes place between radiation and the plasma and radiation interacts with the jet only through momentum exchange. Therefore, the system is isentropic \citep{mm84}. Thomson scattering assumption enables us to integrate
equation (\ref{en1con.eq}) with the help of the EoS (\ref{eos.eq}) we obtain an adiabatic relation between $\Theta$ and $\rho$
\citep{kscc13}. Replacing $\rho$ of the adiabatic relation into the equation (\ref{mdotout.eq}), we obtain 
expression of entropy-outflow rate
\begin{equation}
\mdtj=\mbox{exp}(k_3) \Theta^{3/2}(3\Theta+2)
^{k_1}
(3\Theta+2/\eta)^{k_2}u^r{\cal A},
\label{entacc.eq}
\end{equation}
where, $k_1=3(2-\xi)/4$, $k_2=3\xi/4$, and $k_3=(f-\tau)/(2\Theta)$.
This is also a measure of entropy of the jet and in elastic scattering regime, it remains constant 
along the jet except at the shock.
We integrate equations (\ref{eu1con.eq} and \ref{en1con.eq}), and obtain the generalized relativistic Bernoulli
parameter in the radiation driven regime and is given by,
\be 
\begin{split}
 E=-h u_te^{-X_f},~~~~\mbox{where,~~~~} 
 X_f=\int dr \frac{\gamma (2-\xi)}{(f+2 \Theta )\sqrt{g^{rr}}}\left[\Im-(1-\chi_c)R_t\right]
\end{split}
\label{energy.eq}
\ee

%\be 
%\begin{split}
%& E=-h u_t{\rm exp}(-X_f),~~\mbox{where,} \\
%& X_f=\int dr \frac{\gamma (2-\xi)}{(f+2 \Theta )\sqrt{g^{rr}}}\left[\Im-(1-\chi)R_t\right]
%\end{split}
%\label{energy.eq}
%\ee

The kinetic power of a jet, is defined as the energy flux at large distances and is given as:

\be 
L_j=\dot{E}={\dot M}_{\rm o}{E_{\infty}}
\label{ljet.eq}
\ee
Here, $ E_\infty=[-hu_t]_{r\rightarrow \infty}$ is the Bernoulli parameter at infinity.
Expressing $\wp^r=\sigma_T\Im^r/(m_e)$, equations (\ref{eu1con.eq}) and (\ref{en1con.eq}) can be expressed as gradients of $v$ and $\Theta$
and are given by
%$$
%\gamma^2vg^{rr}r^2\left(1-\frac{a^2}{v^2}\right)\frac{dv}{dr}=a^2\left(2r-3\right)-1+\frac{\wp^r r^2(2-\xi)}{(f+2 \Theta)\gamma^2} {\rm~~~~~ \bf remove}
%%\label{dvdr.eq}
%$$

\bea
\gamma^2vg^{rr}r^2\left(1-\frac{a^2}{v^2}\right)\frac{dv}{dr}=a^2\left(\frac{g^{rr}r^2}{\cal A}\frac{d\cal A}{dr}+1\right)-1+\frac{(2-\xi)\gamma r^2 \sqrt{g^{rr}}}{f+2 \Theta}\left[\Im-\frac{(1-\chi_c)R_t}{N}\right]
\label{dvdr.eq}
\eea

and
%$$
%\frac{d{\Theta}}{dr}=-\frac{{\Theta}}{N}\left[ \frac{{\gamma}
%^2}{v}\left(\frac{dv}{dr}\right)+\frac{2r-3}{r(r-2)}
%\right] {\rm~~~~~ \bf remove}
%$$

\begin{equation}
\frac{d{\Theta}}{dr}=-\frac{{\Theta}}{N}\left[ \frac{{\gamma}
^2}{v}\left(\frac{dv}{dr}\right)+\frac{1}{\cal A}\frac{d\cal A}{dr}+\frac{1}{g^{rr}r^2}-\frac{(2-\xi)(1-\chi_c)\gamma R_t}{2 \Theta \sqrt{g^{rr}}}
\right]
\label{dthdr.eq}
\end{equation}
Equations (\ref{dvdr.eq}) and (\ref{dthdr.eq}) are integrated to solve for $v$ and $\Theta$ of a steady jet
plying through the radiation field ($\Im^r$) of the underlying accretion disc.

\section{Radiation model of accretion disc}
\subsection{Radiative moments}
\label{sbsec2.2}
\begin {figure}[h]
\begin{center}
 \includegraphics[width=15.cm]{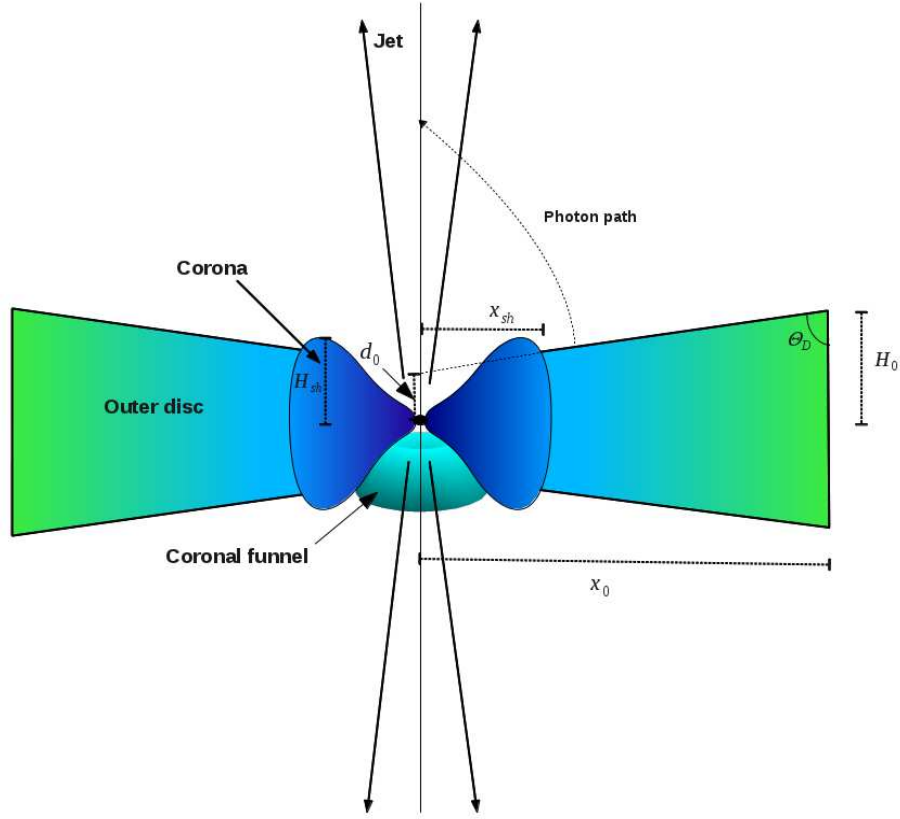}
\vskip -0.5cm
 \caption{Cartoon diagram of cross-sections of axis-symmetric
accretion disc and the associated jet in ($r,~\theta, \phi$ coordinates).
The shock location
$\xsh$, the intercept of outer disc on the jet axis ($d_0$), height of the shock $H_{\rm sh}$, the outer edge of the
disc $x_0$ are marked. Semi-vertical
angle of corona is $\theta_{\rm \small C}$ and for outer disc it is $\theta_{\rm \small D}$. The
funnel of the corona is also shown. \citep{vc18a}}
%\vskip -0.75cm
\label{lab:fig1}
 \end{center}
\end{figure}

In Fig. (\ref{lab:fig1}),
we present the schematic diagram of the accretion disc-jet system, where the jet, the corona
and the outer disc are shown. 
The outer boundary of the corona is $\xsh$ and the half height is $H_{\rm sh}$ and the outer boundary
of the outer disc is $x_0$ and the half height is $H_0$. The accretion disc plays an auxiliary role, where it is considered only as a 
source of radiation. The jet passes through this radiation field above the disc and then interacts with it. The accretion disc assumed, has a
geometrically thick, compact corona, which supplies the hard photons by inverse-Comptonization of seed photons,
and an outer disc supplying softer photons. Such a disc structure is broadly consistent with many
accretion disc models as mentioned in chapter \ref{Ch:intro}. The Keplerian component in the outer disc
is ignored, because the radiative moments computed from an outer Keplerian
disc are negligibly small compared to those from the inner corona, or from the outer advective flow
\citep{cdc04,c05,vkmc15}. 
%\subsubsection{Estimating approximate accretion disc variables}

\subsubsection{Relativistic transformation of intensities from various disc components}
%\label{sbsbsec2.2.1}
To solve equations of motion of the jet, we need to compute radiative moments on the jet axis that requires information of specific intensities from both the outer disc and the corona. The details of estimating the temperature (\ref{acctemp.eq}) and velocity (\ref{accvel.eq}) from accretion discs and thereby estimating the radiative intensity (\ref{skint.eq}, \ref{coronaint.eq}), are given in appendix A. However, the form of the intensities is in the local rest-frame of the disc surface, and therefore,
those intensities need to be transformed from the disc rest frame 
to the curved frame. After special and general relativistic transformations the specific intensities
become, 
\begin{equation}
I_{\rm i}=\frac{{\tilde I}_{\rm i}}{\gamma^4_{\rm i}\left[1+{\vartheta}_jl^j\right]^4_{\rm i}}\left(1-\frac{2}{x}\right)^2
\label{Itrans.eq}
\end{equation}
Here ${\tilde I}_{\rm i}$ is the frequency integrated specific intensity measured in the local rest frame of the accretion disc, ${\vartheta}^j$ is $j^{\rm th}$ component of 3-velocity of accreting matter, $l^j$s are directional cosines, $\gamma_{\rm i}$ is Lorentz factor and $x$ is the radial coordinate of the source point on the accretion disc. 
The suffix ${\rm i}{\rightarrow {\rm C},~{\rm D},~{\rm K}},$ signifies the
contribution from the corona, the outer sub-Keplerian disc (SKD), and the Keplerian disc (KD) respectively. The presence of $(1-2/x)^2$ in the above equation
reduces the intensity of radiation close to the horizon \citep{b02}.

\subsubsection{Calculation of radiative moments in curved space-time}
\begin {figure}[h]
\begin{center}
 \includegraphics[width=15.cm]{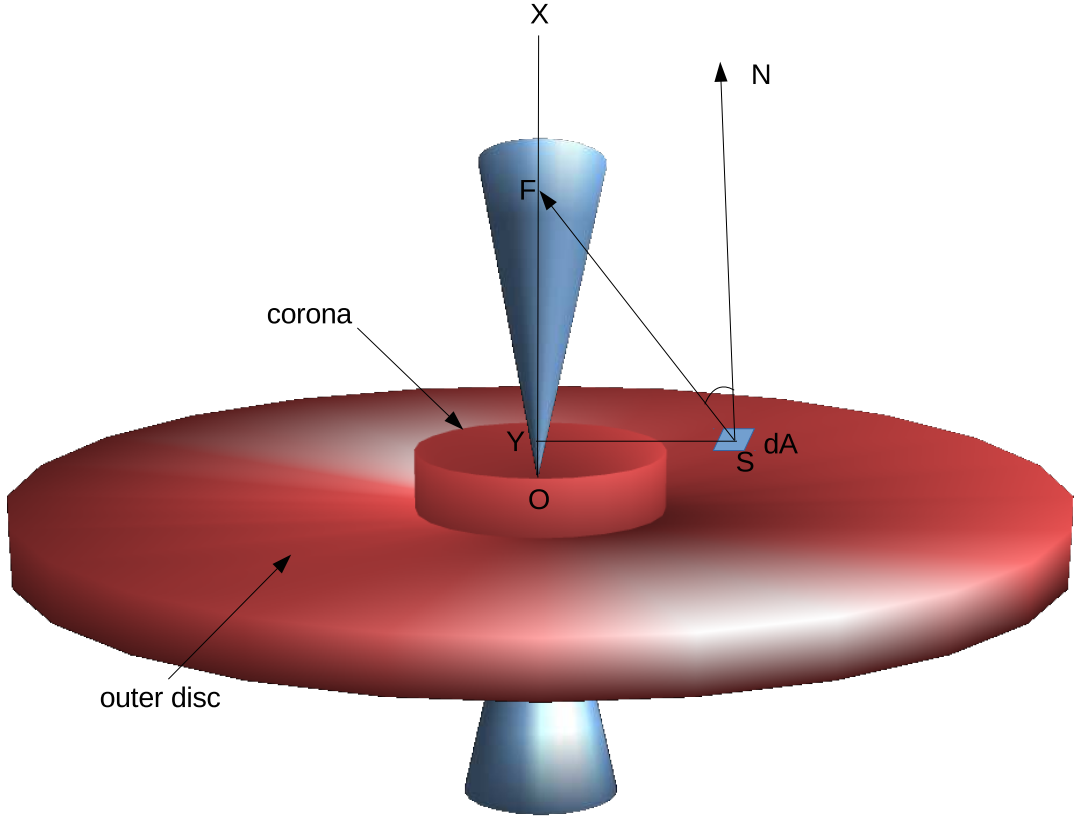}
\vskip -0.5cm
 \caption{Geometrical representation of calculation of radiative moments from the disc components .}
%\vskip -0.75cm
\label{lab:fig_rad_calc}
 \end{center}
\end{figure}
Radiative moments are defined as zeroth, first and second moments of specific intensity
i. e., $\int I d\Omega;~\int I l^j d\Omega;~\&~ \int I l^j l^k d\Omega$, respectively, which are
ten independent components \citep{mm84,c05}. However,
it was also found that for a conical narrow jet only three of the moments are dynamically important.  
%\textbf{\color{blue} Since} the space-time is curved around a black hole, therefore,
%the direction cosines need to be transformed in order to incorporate the curvature.
If $l_{\rm \small F}$ is the relevant direction cosine in the flat space-time, 
then it is related to the one in the curved space as \citep{b02}, 
\bea
l_{\rm i}=l_{\rm i \small F}\left(1-\frac{2}{x}\right)+\frac{2}{x} \nonumber \\
d \Omega_{\rm i} = \left(1-\frac{2}{x}\right) d \Omega_{\rm i \small F} 
\label{l_do_trans.eq}
\eea

%Here, as before ${\rm i} \rightarrow {\rm \small C}~\&~ {\rm \small D}$ signifies disc components.
Method of calculation of $l_F$ and $d\Omega_F$ is shown in Fig. (\ref{lab:fig_rad_calc}). S is source point on the accretion disc, F is the field point on the jet axis where radiative moments are to be determined. SN is local normal on the disc surface. The solid angle subtended by the differential area $dA$ on the disc surface on to the field point $f$ is $d\Omega_{DF}=\frac{dA {\rm cos} \angle {\rm NSF}}{{\rm SF}^2}$ and the respective direction cosine is $l_{DF}=\frac{\rm OF - OY}{{\rm SF}^2}$. Similarly they are calculated for corona and Keplerian disc also.

The expressions of flat space differential solid angle $d \Omega_{\rm i \small F}$ and direction cosines
$l_{\rm i \small F}$ reduce to
\bea
d \Omega_{\rm i \small F}=\frac{rxd\phi dx}{[(r-x \cos\theta_{\rm i})^2+x^2\sin\theta_{\rm i}^2]^{3/2}}, {~~\rm and}~~~~
l_{\rm i \small F}=\frac{(r-x \cos\theta_{\rm i})}{\sqrt{[(r-x \cos\theta_{\rm i})^2+x^2\sin\theta_{\rm i}^2]}}
\eea

We use equations (\ref{Itrans.eq}) and (\ref{l_do_trans.eq}) in the definition of various radiative moments, and express
all the radiative moments ($R_0, ~ R_1~\&~R_2$) in a compact form given by,
\bea
R_{n\rm i}=\int^{x_{\rm i 0}}_{x_{\rm ii}} \int^{2\pi}_{0}\left(1-\frac{2}{x}\right)^3\frac{{\tilde I}_{\rm i}}{\gamma^4_{\rm i}\left[1+{\rm v}_jl^j\right]^4_{\rm i}} \nonumber \\
\times \left[\frac{(r-x \cos\theta_{\rm i})}{\sqrt{[(r-x \cos\theta_{\rm i})^2+x^2\sin\theta_{\rm i}^2]}}\left(1-\frac{2}{x}\right)+\frac{2}{x}\right]^n \nonumber \\
\times \frac{rxd\phi dx}{[(r-x \cos\theta_{\rm i})^2+x^2\sin\theta_{\rm i}^2]^{3/2}}
\label{moments2.eq}
\eea
Here limits of radial integration are $x_{\rm ii}$ (inner edge) and $x_{\rm i 0}$ (outer edge) of the respective disc component.
The index $n=0, 1, 2$ gives us
$R_0,~R_1~\&~R_2$, i. e., radiative energy density, radiative flux along $r$ and the $rr$ component
of the radiative pressure.

Expanding Eq. (\ref{moments2.eq}), expressions for $R_0, ~ R_1~\&~R_2$, can be expressed as
\bea
R_{0\rm i}=\int^{x_{\rm i 0}}_{x_{\rm ii}} \int^{2\pi}_{0}\left(1-\frac{2}{x}\right)^3\frac{{\tilde I}_{\rm i}}{\gamma^4_{\rm i}\left[1+{\rm v}_jl^j\right]^4_{\rm i}} \frac{rxd\phi dx}{[(r-x \cos\theta_{\rm i})^2+x^2\sin\theta_{\rm i}^2]^{3/2}}
\label{moments2_r0.eq}
\eea

\bea
R_{1\rm i}=\int^{x_{\rm i 0}}_{x_{\rm ii}} \int^{2\pi}_{0}\left(1-\frac{2}{x}\right)^3\frac{{\tilde I}_{\rm i}}{\gamma^4_{\rm i}\left[1+{\rm v}_jl^j\right]^4_{\rm i}} \left[\frac{(r-x \cos\theta_{\rm i})}{\sqrt{[(r-x \cos\theta_{\rm i})^2+x^2\sin\theta_{\rm i}^2]}}\left(1-\frac{2}{x}\right)+\frac{2}{x}\right] \nonumber \\
\times \frac{rxd\phi dx}{[(r-x \cos\theta_{\rm i})^2+x^2\sin\theta_{\rm i}^2]^{3/2}}
\label{moments2_r1.eq}
\eea

\bea
R_{2\rm i}=\int^{x_{\rm i 0}}_{x_{\rm ii}} \int^{2\pi}_{0}\left(1-\frac{2}{x}\right)^3\frac{{\tilde I}_{\rm i}}{\gamma^4_{\rm i}\left[1+{\rm v}_jl^j\right]^4_{\rm i}} \left[\frac{(r-x \cos\theta_{\rm i})}{\sqrt{[(r-x \cos\theta_{\rm i})^2+x^2\sin\theta_{\rm i}^2]}}\left(1-\frac{2}{x}\right)+\frac{2}{x}\right]^2 \nonumber \\
\times \frac{rxd\phi dx}{[(r-x \cos\theta_{\rm i})^2+x^2\sin\theta_{\rm i}^2]^{3/2}}
\label{moments2_r2.eq}
\eea

Since there are three disc components corona, outer disc and Keplerian disc, hence at a given $r$ the net moments are,
\be 
R_n=R_{n{\rm \small C}}+R_{n{\rm \small D}}+R_{n{\rm \small KD}}
\label{moments3.eq}
\ee

or

\be 
R_0=R_{0{\rm \small C}}+R_{0{\rm \small D}}+R_{0{\rm \small KD}}
\label{moments3_0.eq}
\ee

\be 
R_1=R_{1{\rm \small C}}+R_{1{\rm \small D}}+R_{1{\rm \small KD}}
\label{moments3_1.eq}
\ee

\be 
R_2=R_{2{\rm \small C}}+R_{2{\rm \small D}}+R_{2{\rm \small KD}}
\label{moments3_2.eq}
\ee

The $x$ limit of the corona are
$x_{\rm {\small C}i}=2, x_{{\rm \small C} 0}=\xsh$. However, from a given $r$, an observer cannot see the whole of the disc because the corona
blocks a portion of the disc. Therefore the inner edge of the outer disc is given by,
$$
x_{\rm {\small D}i}=\frac{r-d_0}{(r-H_{\rm sh})/\xsh+ \cot \theta_{\rm \small C}}
$$
It is clear from above that, as $r\rightarrow \infty$,
$x_{\rm {\small D}i}\rightarrow \xsh$. Moreover, up to some radius, radiation from the outer disc will never
reach the axis of the jet. If the distance above the disc up to which outer disc radiation does not reach the axis
is $r_{\rm lim}$, then
\be
r_{\rm lim}=\frac{x_0H_{\rm sh}-H_0\xsh}{x_0-\xsh}.
\label{shadolim.eq}
\ee

%\subsection{Method of obtaining solutions}
%\label{sec:method}

\chapter{Methods of Analysis}
\label{CH:3}
\section{Overview}
The jet solutions can be obtained by integrating Eqs. (\ref{dvdr.eq}) and (\ref{dthdr.eq}), which provide information of $v$ and $\Theta$ along jet length. In this chapter we brief the major aspects of process of obtaining solutions like sonic point analysis, obtaining jet variables through integration of EoM, shock conditions, shock properties and brief discussion about stability of the shocks etc. 
\section{Sonic point conditions}
\label{Sonic_point_conditions}
Since, the jet originates from the accretion flow from a region close to the horizon, the jet speed should be small but because of hot base, the jet base is subsonic.
At large distances from the BH, the jet moves with very high speed and is cold and hence it is supersonic.
So let the jet become transonic i.e, $v_c=a_c$ at the sonic point ($r=r_c$). Here suffix $c$ denotes quantities at the
sonic point.
Further, at $r_c$, $dv/dr\rightarrow 0/0$, which enables us to write down sonic point condition as
%\be 
%v_c=a_c;\\ 
%\label{sonic1.eq}
%\ee
%and 
%\begin{equation}
%a_c^2-\frac{1}{2r_c-3}+\frac{(\frad)_c}{2r_c-3}=0.
%\label{sonic2.eq}
%\end{equation}

\be
\left| a^2\left(\frac{g^{rr}r^2}{\cal A}\frac{d\cal A}{dr}+1\right)-1+\frac{(2-\xi)\gamma r^2 \sqrt{g^{rr}}}{f+2 \Theta}\left[\Im-\frac{(1-\chi_c)R_t}{N}\right]\right|_{r=r_c}=0
\label{sonic2.eq}
\ee

Ratio of $v$ and $a$ is defined as Mach number of the flow ($M=(v/a)$). $dM/dr|_c$ is calculated by employing the L'H$\hat{\rm o}$pital's  rule at $r_c$ and solving the resulting
quadratic equation of $dM/dr|_c$. The resulting quadratic equation can admit two complex roots leading to
the so-called $O$ type (or `centre' type) or spiral type
sonic points, or two real roots. The solutions with two real roots but with opposite signs
are called $X$ or `saddle' type sonic points, while real roots with same sign produce nodal type sonic points. The
jet solutions passing through X type sonic points are physical. %and in this paper care has been taken to
%study jet solutions through X type sonic points.
So for a given set of flow variables at the jet base, a unique solution will pass through
the sonic point determined by the entropy ${\dot {\cal M}}$ of the flow. For given
values of inner boundary parameters, that is, at the jet base  $r_b$, $v_b$ and $a_b$ or constants of motion (\ie $~E$ and $\md$),
we integrate equations (\ref{dvdr.eq}) and (\ref{dthdr.eq}), while checking
for the sonic point conditions (equations \ref{sonic2.eq}). We iterate till the
sonic point is obtained, and once it is obtained we continue to integrate outwards
starting from the sonic point using Runge Kutta's $4^{th}$ order method.
To determine density, one may need to explicitly supply the outflow rates ${\dot M}_{\rm o}$ which are about
few percent of accretion rates, as has been theoretically obtained \citep{ck16,kc17}. 

%\subsubsection{Shock conditions}
%The existence of multiple 
%sonic points in the flow opens up the possibility of 
%formation of shocks in the flow. At the shock,
%the flow is discontinuous in density, pressure and velocity.
%The relativistic Rankine-Hugoniot conditions relate the flow quantities across the
%shock jump \citep{t48,cc11}
%\begin{equation}
%  [{\rho}u^r]=0,
%  \label{sk1.eq}
%\end{equation}
%\begin{equation}
%   [\dot{E}]=0
%   \label{sk2.eq}
%\end{equation}
%and
%\begin{equation}
%[T^{rr}]=[(e+p)u^ru^r+pg^{rr}]=0
%\label{sk3.eq}
%\end{equation}
%The square brackets denote the difference 
%of quantities across the shock, i.e. 
%$[Q]=Q_2-Q_1$ 
%with $Q_2$ and  $Q_1$ being 
%the quantities after and before the shock, respectively.\\
%
%Equation (\ref{sk2.eq}) states that the energy flux remains 
%conserved across the shock. Dividing 
%equation (\ref{sk3.eq}) and equation (\ref{sk2.eq}) by equation (\ref{sk1.eq}) and a little 
%algebra leads to
%\be
%\left[\left(h \gamma v+\frac{2 \Theta}{\tau \gamma v}\right)\right]=0;~\&~ [E]=0.
%\label{sk5.eq}
%\ee
%We check for shock conditions (equation \ref{sk5.eq}),
%as we solve the equations of motion
%of the jet.
\section{Shock conditions}
\label{Shock_conditions}
The existence of multiple 
sonic points in the flow opens up the possibility of 
formation of shocks in the flow. At the shock,
the flow is discontinuous in density, pressure and velocity.
The relativistic Rankine-Hugoniot conditions relate the flow quantities across the
shock jump \citep{t48,cc11}
\begin{equation}
  [{\rho}u^r]= [\dot{E}]=[T^{rr}_M+T^{rr}_R]=0,
  \label{sk1.eq}
\end{equation}
%The square brackets denote the difference 
%of quantities across the shock, i.e. 
%$[Q]=Q_2-Q_1$ 
%with $Q_2$ and  $Q_1$ being 
%the quantities after and before the shock, respectively.\\
%Equations (\ref{sk2.eq}) states that the energy flux remains 
%conserved across the shock.
The square bracket shows difference of the respective quantities from post shock to the pre shock region at the shock location. Dividing $T^{rr}$ and ${\dot E}$ conservation conditions by mass conservation equation and following a little 
algebra they lead to
\be
\left[\left(h \gamma v+\frac{2 \Theta}{\tau \gamma v}\right)\right]=0;~\&~ [E]=0.
\label{sk5.eq}
\ee
We check for shock conditions (equation \ref{sk5.eq}),
as we solve the equations of motion
of the jet.
\section*{Shock parameters}
One of the major outcomes of existence of multiple 
sonic points in the jet is the possibility of 
formation of shocks in the flow. At the shock,
the flow makes a discontinuous jump in density, pressure and velocity.
%We check for shock conditions % described in section (\ref{Shock_conditions})
%through Eqs. (\ref{sk1.eq}) and (\ref{sk5.eq})
%as we solve the equations of motion
%of the jet.
The strength of the shock is measured by two parameters  
compression ratio ($R$) and shock strength ($S$). 
$R$ and $S$ are ratios of densities and Mach numbers ($M$) across
the shock (at $r=r_{\rm sh}$). In relativistic case, according to equation
 (\ref{sk1.eq}), $R$ is obtained as,
\be 
R=\frac{\rho_+}{\rho_-}=\frac{u_-^r}{u_+^r}=\frac{\gamma_- v_-}{\gamma_+ v_+},
\label{compress.eq}
\ee
where, $+$ and $-$ stand for quantities at post-shock and pre-shock flows, respectively. 
Similarly, $S$ is defined as,
\be
S=\frac{M_-}{M_+}=\frac{v_-a_+}{v_+a_-}
\label{shokstrnth.eq}
\ee
\section{Stability analysis of the shocks}
\label{shock_stability}
In steady state analysis, at many occasions, one may end up with multiple shock transitions across the flow. But the jet can only pass through one of them and hence the other shocks are bound to be unstable under small perturbations \citep{n96,yk95,ydl96}. In chapter \ref{CH:P2}, study of thermally driven jets with non-radial cross section, we will encounter such situation. In this section, we describe the method to check the stability of shocks in a general relativistic thermal flow.

The momentum flux, $T^{rr}$ for thermally driven flow is, 
\begin{equation}
[T^{rr}]=[(e+p)u^ru^r+pg^{rr}]=0
\label{sk3.eq}
\end{equation}
$T^{rr}$ remains conserved across the shock. But if the shock 
under some perturbation, moves 
from shock location, $r_{\rm sh}$ to $r_{\rm sh}+\delta r$ then $T^{rr}$ may 
not be balanced. The resultant difference across the shock is 
\be 
\delta T^{rr}=T^{rr}_2-T^{rr}_1=\left[\left(\frac{dT^{rr}}{dr}\right)_2-\left(\frac{dT^{rr}}{dr}\right)_1\right]\delta r = \Delta \delta r
\label{delta.eq}
\ee
Here labels with subscripts `1' and `2' represent quantities in pre-shock and post-shock regions at the shock location respectively.

Now, multiplying and dividing equation (\ref{sk3.eq}) 
by $\rho$ and after rearranging the expression for 
momentum flux becomes 
\be 
T^{rr}=\rho \left(hu^ru^r+\frac{2\Theta g^{rr}}{\tau}\right)
\label{Trr2.eq}
\ee
Using equation (\ref{mdotout.eq}) and 
differentiating equation (\ref{Trr2.eq}) 
followed by some algebra, one obtains
\be 
\frac{dT^{rr}}{dr}=\frac{{\dot M_{out}}}{{\cal A \tau}}\left[-A_s \frac{dv}{dr}+B_s\right],
\label{Trr3.eq}
\ee
where, 
\be
A_s={\sqrt{g^{rr}} \gamma^3}\left[ \frac{2 \Theta g^{rr}}{u^2}+\left(f+2 \Theta \right) \right]+\frac{2 \Theta {\gamma}^2}{Nuv}\left(u^2(N+1)+g^{rr}\right)
\label{as.eq}
\ee
and
\begin{eqnarray}
B_s=\frac{\gamma v}{r^2 \sqrt{g^{rr}}}\left[ \left(f+2 \Theta \right)-\frac{2 \Theta g^{rr}}{u^2} \right]-\left((f+2\Theta) u^2+2 \Theta g^{rr}\right)\frac{1}{u {\cal A}}\frac{d{\cal A}}{dr} \\ \nonumber
+\frac{4 \Theta}{ur^2}-\frac{2 \Theta}{N u}\left[u^2(N+1)+g^{rr}\right]
\label{bs.eq}
\end{eqnarray}
Using equation (\ref{Trr3.eq}) in (\ref{delta.eq}), we obtain
\be 
\Delta=\left[-A_s v'+B_s\right]=\left(A_{s1}v'_1-A_{s2}v'_2\right)+(B_{s2}-B_{s1})
\label{delta2.eq}
\ee
Now, the stability of the shock depends on 
the sign of $\Delta$. If $\Delta<0$ for finite 
and small $\delta r$ there is more momentum flux 
flowing out of the shock than the flux flowing in 
so the shock keeps shifting towards further 
increasing $r$, and becomes unstable. On 
the other hand if $\Delta>0$, then the change 
due to $\delta r$ leads to the further decrease 
in $\Delta$, and the shock is stable.

One finds from equations (\ref{as.eq}) and (\ref{bs.eq}), 
$A_s$ has positive value. Now 
the stability of the shock can be analyzed under two 
broad conditions.\\
$\bullet$ Condition 1. The shock is significantly away 
from middle sonic point, or the absolute magnitude of 
$v'$ is significantly more than 0. We find that 
$|B_s|<<|A_s|$ and  hence the stability of the 
shock depends upon the sign of $v'$. Equation 
(\ref{delta.eq}) shows that the shock is stable 
(or $\Delta > 0$) if $v'_1>0$ and subsequently 
$v'_2 < 0 $.  Hence the inner shock is stable 
and the outer shock is unstable.\\
$\bullet$ Condition 2. If the shock is close 
to the middle sonic point, $v'_1 \approx v'_2 \approx 0$.
So only second term consisting $B_s$ contributes to the 
stability analysis and one obtains that $\Delta<0$ for 
both inner and outer shocks and the shock is always 
unstable.

Finally, the general 
rule for stability of the shock can be set. If the post shock flow is accelerated then 
the shock is unstable and if the post shock flow is 
decelerated the shock is stable unless the shock is 
very close to the middle sonic point where it is always
unstable. { It should be mentioned that we have used the balancing properties of momentum flux to examine the shock stability. However, in literature, there are more rigorous studies of stability analysis \citep{n92,n93,nh94,n96}, in which, a small perturbation is given at the shock and the time evolution of the perturbation is examined. If the perturbation grows with time, the shock is found to be unstable and if it decays then the shock is stable. The stability rules obtained above are compatible with the rules obtained by this method \citep{n96}. }For detailed explanation and further clarification with examples, see \cite{vc17}.
\chapter{Special Relativistic Study of Radiatively Driven Relativistic Jets}
\label{CH:P1}
\section{Overview}
We investigate a radiatively and thermally driven relativistic fluid jet from a shocked accretion disc around a non-rotating BH. The strong gravitational field around the BH is approximated by Paczy\'nski-Wiita potential.
The accretion rate controls the shock location and therefore, the
radiation field around the accretion disc. The general formalism of computing radiation field (presented in chapter \ref{CH:2}) corresponds to curved space-time. However, since this chapter is in special relativistic regime and hence assumes flat space, all metric components having curvature information are equal to unity ($|g_{rr}|=|g_{tt}|=1$). The calculation of moments in this chapter can be obtained by putting $2/x\rightarrow0$ in Eqs. (\ref{Itrans.eq}), (\ref{l_do_trans.eq}) and (\ref{moments2.eq})  However, we compute the radiative moments with full special relativistic
transformation and the effect of a fraction of radiation absorbed by the black hole
has been approximated \citep{vkmc15}. Further, the interaction between radiation and jet fluid is considered to be dominated by Thomson scattering (\ie $~\sigma=\sigma_T$). We show that the radiative moments
around a super massive BH are different compared to that around a stellar mass black hole. We carry out an exploratory study of dependences of jet dynamics upon various parameters like accretion rate, disc luminosity, composition of the jet, magnetic pressure in the disc etc. The dynamical behaviour of jets around stellar mass BH and super-massive BH is investigated. The results of the study are published in \citet{vkmc15, vkmc15cn}.% and \cite{vkmc15cn}.
\footnote{The units of $r$ considered in \citet{vkmc15}, Schwarzschild radius is at $r=1$. Keeping in accordance with the general terminology used in all the chapters, it is considered to be at $r=2$ in this chapter.} 
%We show that the terminal speed of jets increases with the mass accretion rates, synchrotron emission of the
%accretion disc and reduction of proton fraction of the flow composition. To obtain relativistic terminal velocities
%of jets, both thermal and radiative driving are important.
%For very high accretion rates and pair dominated flow, jets around super massive black holes are truly ultra-relativistic,
%while for jets around stellar mass black holes, terminal Lorentz factor of about $10$ is achievable.

\section{Governing Equations of Jet dynamics}
Under Thomson scattering assumption, $\sigma\rightarrow\sigma_T$ in Eq. (\ref{eu1con.eq}) and (\ref{dvdr.eq}). In absence of source term, the R.H.S of Eq. (\ref{en1con.eq}) is zero. The cross section considered is radial (${\cal A}\propto r^2$). Considering these factors, Eqs. (\ref{dvdr.eq}) and (\ref{dthdr.eq}) become :
\be%gin{eqnarray}
{\gamma}^4v\left(1-\frac{a^2}{v^2}\right)\frac{dv}
{dr}  = \frac{2{\gamma}^2a^2}{r}-\frac{1}{2(r-2)^2}+ %\\ \nonumber
%& + &
\frac
{{\gamma}^3{(2-\xi)}}{f+2{\Theta}}[(1+v^2){R_1}-v
({R_0}+{R_2})] 
\label{dvdr1.eq}
\ee%nd{eqnarray}

and
\begin{equation}
\frac{d{\Theta}}{dr}=-\frac{{\Theta}}{N}\left[ \frac{{\gamma}
^2}{v}\left(\frac{dv}{dr}\right)+\frac{2}{r}\right]
\label{dthdr1.eq}
\end{equation}
In Eq. (\ref{dvdr1.eq}), the left hand side is the net acceleration term of a steady state jet. On the right hand side, the first term is accelerating thermal term a$_t
=2{\gamma}^2a^2/r$, while the second being gravity a$_g=-0.5/(r-2)^2$, it decelerates. The third term in right hand side is radiative acceleration,
a$_r={\gamma}^3{\tau}[(1+v^2){R_1}-v
({R_0}+{R_2})]/(f+2{\Theta})$. The radiative contribution is within the square bracket and the rest represents the interaction of
matter jet with the radiation field.
The physical significance of the term in the square bracket is worth noticing.
It has form$\rightarrow$~$(1+v^2){R_1}-v({R_0}+{R_2})$.
The term proportional to $v$ comes with a negative sign and would decelerate and gives rise to radiation drag.
If the first term
$(1+v^2){R_1}$ dominates, then radiation would accelerate the flow, which means the net radiative
term would either be accelerating or decelerating depending on the velocity. Dependence of radiative term
on $v$ arises purely due to relativity. In non-relativistic domain ({\ie} $v\ll 1$), the radiative term
is just ${R_1}$. In the fast but sub relativistic domain {\ie} $v^2 \ll 1$ the radiative term is
${R_1}-v({R_0}+{R_2})$ similar to the formalism followed by \citet{cc02,kcm14}. The drag term arises
due to the resistance faced by the moving material through the radiation field, and the finite value of the speed of light.\\
Much talked about equilibrium speed $v=v_{\rm eq}$, which is defined for ${\rm a}_r=0$, {\ie}
\be
v_{\rm eq}=\Re-\sqrt{\Re-1};~\mbox{where, } \Re=\frac{{R_0}+{R_2}}{2{R_1}}.
\label{veq1.eq}
\ee
From equation (\ref{veq1.eq}), it is clear that if the relative contribution of radiative moments
or $\Re$ approaches $1$, {\ie} ${R_1}={R_0}={R_2}$, then $v_{\rm eq} \rightarrow 1$, {\ie} no radiation drag.
Therefore, the nature of the quantity $\Re$ dictates, whether a radiation field
would accelerate a flow or decelerate it.
Of course the resultant acceleration depends on the magnitude
of all moments. There is an added feature of radiatively driven
relativistic fluid, that is, the radiative term is multiplied by a term inverse
of enthalpy ($\{f+2\Theta\}/\tau$) of the flow, which actually suggests
that the effect of radiation on the jet is less for hotter flow.

\section{Analysis and Results}
\subsection{Nature of radiative moments}
\begin{figure}
\begin{center}
 \includegraphics[trim={4cm 0 4cm 1cm}, width=9.cm]{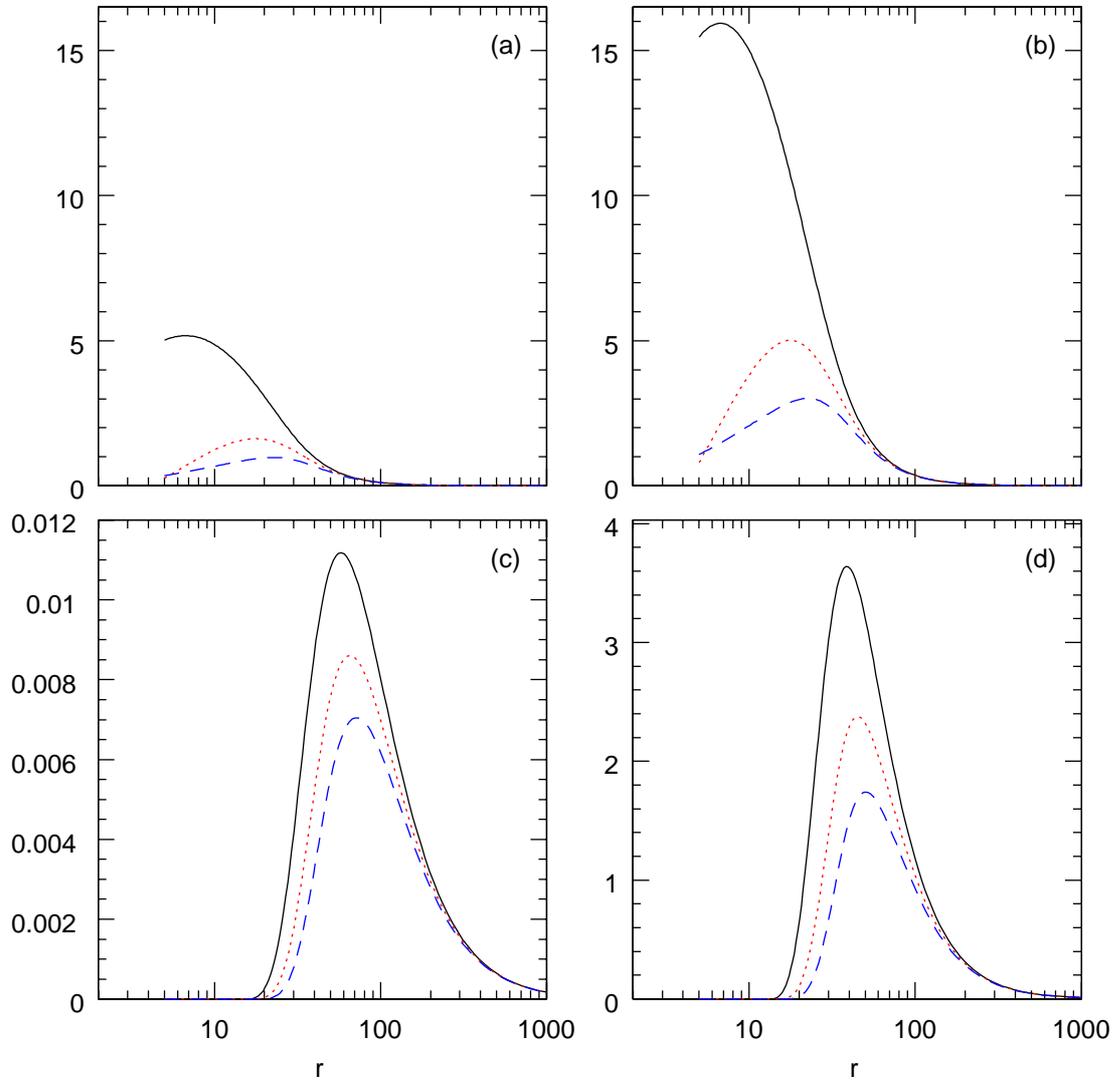}
\caption{Distribution of radiative moments  ${R_0}$ (solid black), ${R_1}$ (dotted red)
and ${R_2}$ (dashed blue) from corona for (a)
$10M_{\odot}$ and (b) $10^8M_{\odot}$ black holes.
Distribution of radiative moments from the
KD (c) and from SKD (d) is same for both types
of black holes when expressed in the geometric units. Various parameters used to compute
the moments are ${\dot m}_\sk=7$, ${\dot m}_\kd=1$ and $\beta = 0.5$. This produces
$x_s= 26.4$ and luminosities are ${\ell}_\sk=0.0265$, ${\ell}_\kd = 0.039$ and for stellar mass BH
${\ell}_\s = 0.215$ (a), 
while for larger BH $\ell_\s=0.661$
(b) \citep{vkmc15}.}
\label{lab:R2}
\end{center}
\end{figure}

\begin{figure}
\begin{center}
 \includegraphics[trim={2.8cm 0 2cm 6cm}, width=12cm]{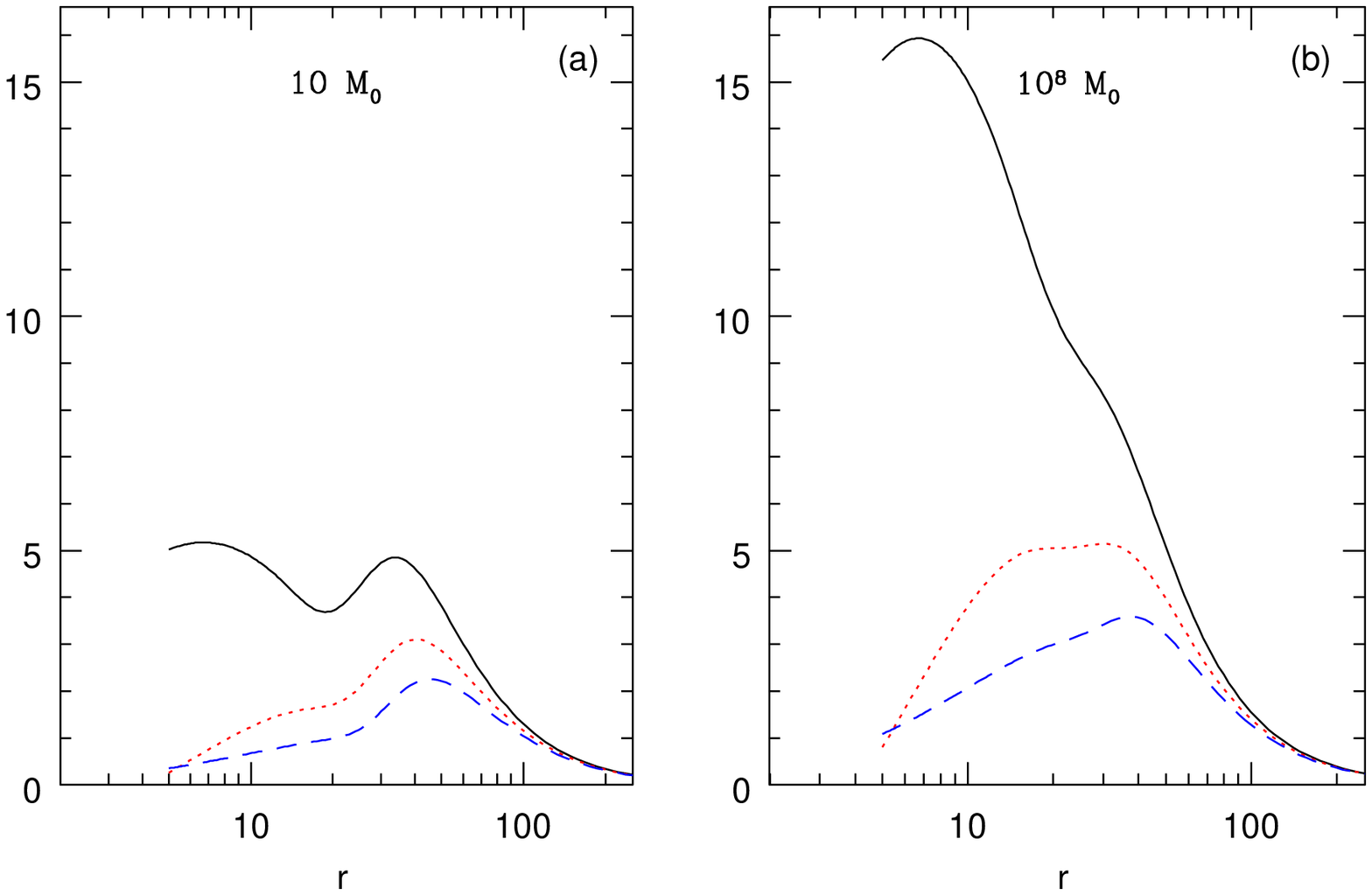}
 \caption{Combined moments from (a) $10$ and (b)
$10^8 M_{\odot}$ black holes. Various curves are ${R_0}$ (solid black), ${R_1}$
(dashed red) and ${R_2}$ (dashed blue).
Disc parameters are ${\dot m}_\sk=7$ and ${\dot m}_\kd=1$, $\beta = 0.5$
for which the shock obtained is
at $x_s= 26.41$, luminosities are $\ell_\sk=0.0265$,
$\ell_{\kd} = 0.039$. The $\ell_\s=0.215$ for $10M_\odot$ BH and $\ell_\s=0.661$ for $10^8M_\odot$ BH.
The moments are expressed in geometric units \citep{vkmc15}.}
\label{lab:R3}
\end{center}
\end{figure}

Considering the assumptions and equations described above, We numerically integrated set of Eqs.(\ref{moments2_r0.eq}-\ref{moments2_r2.eq}) to obtain radiative
moments from corona, outer disc or sub-Keplerian disc (SKD), and the Keplerian disc (KD) for the accretion disc parameters given in Table (\ref{tableP1}).

\begin{table}
\caption{Disc parameters}
\label{tableP1}
\centering
 \begin{tabular}{|c c c c c c c|} 
 \hline
 $\lambda$ & $x_0$ & $\left[\vartheta_{\rm \small D}\right]_{x_0}$ & $\left[\Theta_{\rm \small D}\right]_{x_0}$ & $\theta_D$ & $H_{\rm sh}$ & $d_0$\\ [0.5ex] 
 \hline%\hline
 $3.4$ & $3500 \rs$ & $0.001$ & $0.1$ & $85^0$ & $0.6(\xsh-1)$ & $0.4H_{\rm sh}$\\ 
 \hline
 \end{tabular}
\end{table}

From Appendix \ref{app:rad_mom} and \citet{vkmc15}, it is clear that $\ell_\s$ is different for $10M_{\odot}$ and $10^8M_\odot$
BH, for the same set of free parameters, $\dot m_\sk$ and $\dot m_\kd$ and $\beta$. This should affect the net radiation field
above the disc. 
In Fig. (\ref{lab:R2}a), we plot ${R}_0$, $R_1$ and $R_2$ with $r$ for $\dot m_\sk=7$ and $\dot m_\kd=1$ for $M_B=10M_\odot$. 
For such accretion rate the shock is at $x_s=26.41$ (see, equation \ref{xsdotm.eq}). In Fig. (\ref{lab:R2}b),
we plot radiative moments above a disc around super-massive ($M_B=10^8 M_\odot$) BH, for the same set of accretion parameters.
The radiative moments from the corona around $10^8 M_\odot$ BH are about three times than
those around $10 M_\odot$. It may be noted that for lower ${\dot m}_\sk$ the accretion shock forms farther away from the BH, and the radiative moments around stellar mass and super-massive BH are similar.
In Fig. (\ref{lab:R2}c), we plot
${R_0}_\kd$ (solid black), ${R_1}_\kd$
and ${R_2}_\kd$.
In Fig. (\ref{lab:R2}d)
we present ${R}_0$, $R_1$
and $R_2$ for SKD. The corona luminosity for $10M_\odot$ BH is $\ell_\s = 0.215$
and $\ell_\s=0.661$ for $10^8 M_\odot$ BH. The luminosities of the pre-shock disc $\ell_\sk=0.0265$ and
$\ell_\kd=0.039$ are same for discs around super massive as well as stellar mass BHs.
The moments due to corona around a super-massive BH are larger than that around stellar mass BH, however,
the SKD and KD contributions in the geometric units are exactly
same for stellar mass and super massive BH. In physical units these moments
would scale with the central mass.
Finally, we show combined
radiative moments from all the disc components for $10 M_\odot$ (Fig. \ref{lab:R3}a) and 
$10^8 M_\odot$ BH (Fig. \ref{lab:R3}b) for exactly the same disc parameters as in Fig. (\ref{lab:R2}). 
For higher ${\dot m}_\sk$, the overall radiation field (in geometric units), above a disc around a stellar mass 
BH is different than the moments around a super massive BH. This is because the for higher ${\dot m}_\sk$
the shock in accretion is located closer to the BH, which produces a cooler corona around
a stellar mass BH than a super massive BH and therefore, larger efficiency of Comptonization.
However, for lower ${\dot m}_\sk$ (equation \ref{xsdotm.eq}), the shock is located at larger distance from the BH, making the efficiency of
Comptonization similar for both kinds of BHs, and hence the moments are also similar.

\subsection{Sonic Point Properties}
\begin{figure}
\begin{center}
 \includegraphics[trim={0cm 0 0.2cm 0.5cm}, width=14cm]{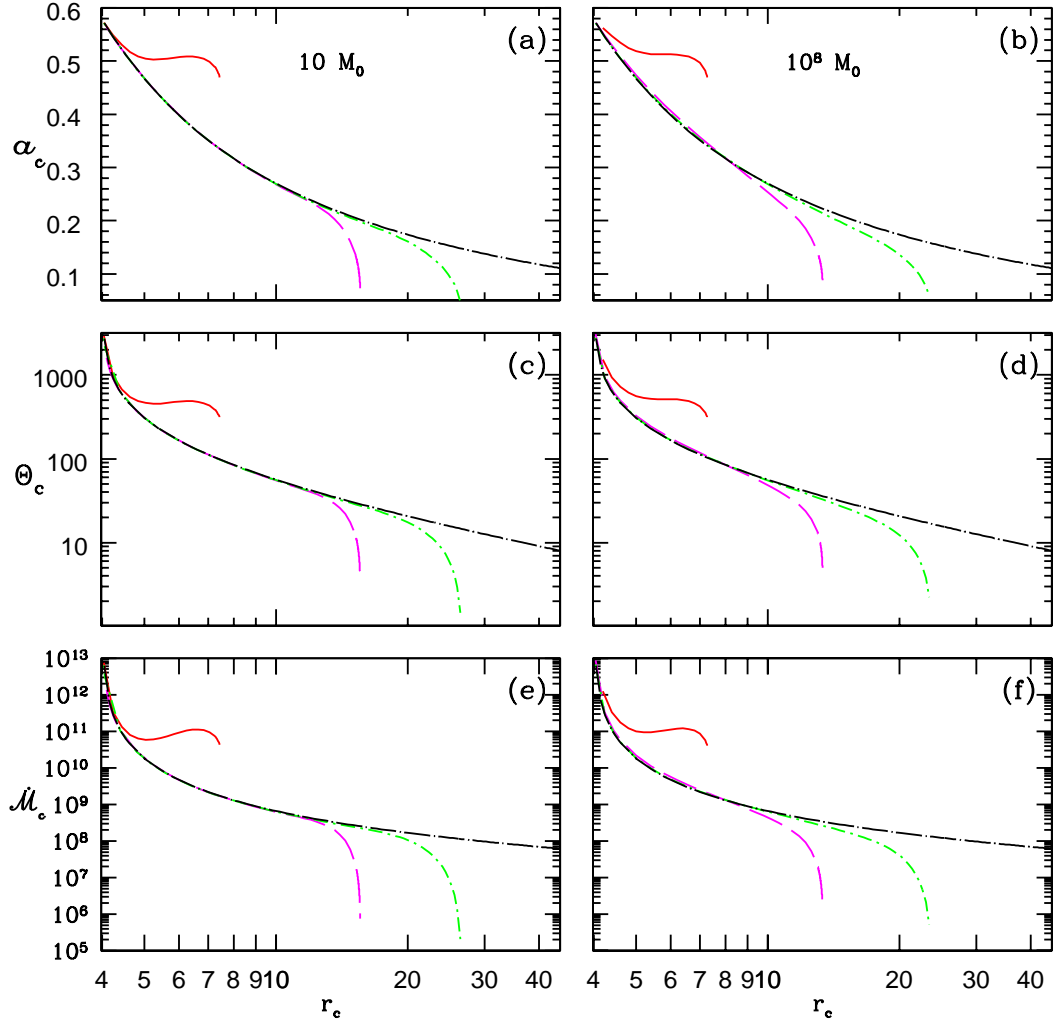}
 \caption{Variation of $a_c$ (a, b), $\Theta_c$ (c, d) and ${\dot {\cal M}}_c$ (e, f)
as a function of $R_c$ for jets are around $M1=10M_\odot$ BH (a, c, e) and $M8=10^8M_\odot$ BH (b, d, f).
Each curve corresponds
to ${\dot m}_\sk= 13$ (solid red), $10$ (long dashed magenta), $8$ (dashed-dotted green),
and only thermally driven jet (long-dashed dotted black). For all the plots ${\dot m}_\kd=1$. \citep{vkmc15}}
\label{lab:sonpt}
 \end{center}
\end{figure}
From equation (\ref{sonic2.eq}), considering assumptions in this chapter, it is clear that for a thermally driven jet,
sonic points exist in range $r_c=4~\rightarrow~ \infty$. However, radiatively driven flow may not posses
sonic point at large distance from the jet base,
because the presence of strong radiation field may render $a_c\lsim 0$ at those distances.
In Figs. (\ref{lab:sonpt}a-f), we compare the flow quantities $a_c$ (a, b), and $\Theta_c$ (c, d), ${\dot {\cal M}}_c$
(e, f)
as a functions of $r_c$. The left panels show the sonic point properties of jets around $10M_\odot$ BH (a, c, e)
and the right panels show sonic point properties of jets around $10^8 M_\odot$ BH (b, d, f). The KD accretion rate, or,
${\dot m}_\kd=1$ is kept invariant for all these plots, but various curves are for ${\dot m}_\sk=13$, $10$, $8$,
and only thermally driven jet. It is interesting to note that,
the region outside the central object available for sonic points shrinks,
as the disc luminosity increases.
For luminous discs say ${\dot m}_\sk > 10$, sonic points can only form for $r_c < 16$. This implies that
only very hot flow has thermal energy density comparable to the radiation pressure, and therefore for any
flow with less thermal energy may be considered as collection of particles rather than a fluid in such radiation
field. Moreover, for ${\dot m}_\sk=13$, multiple sonic points may form for some values of
${\dot {\cal M}}$. This possibility is further explored in next chapters.

\subsection{Jet Solutions}
\begin{figure}
\begin{center}
 \includegraphics[trim={5.5cm 0 4cm 0cm}, width=8cm]{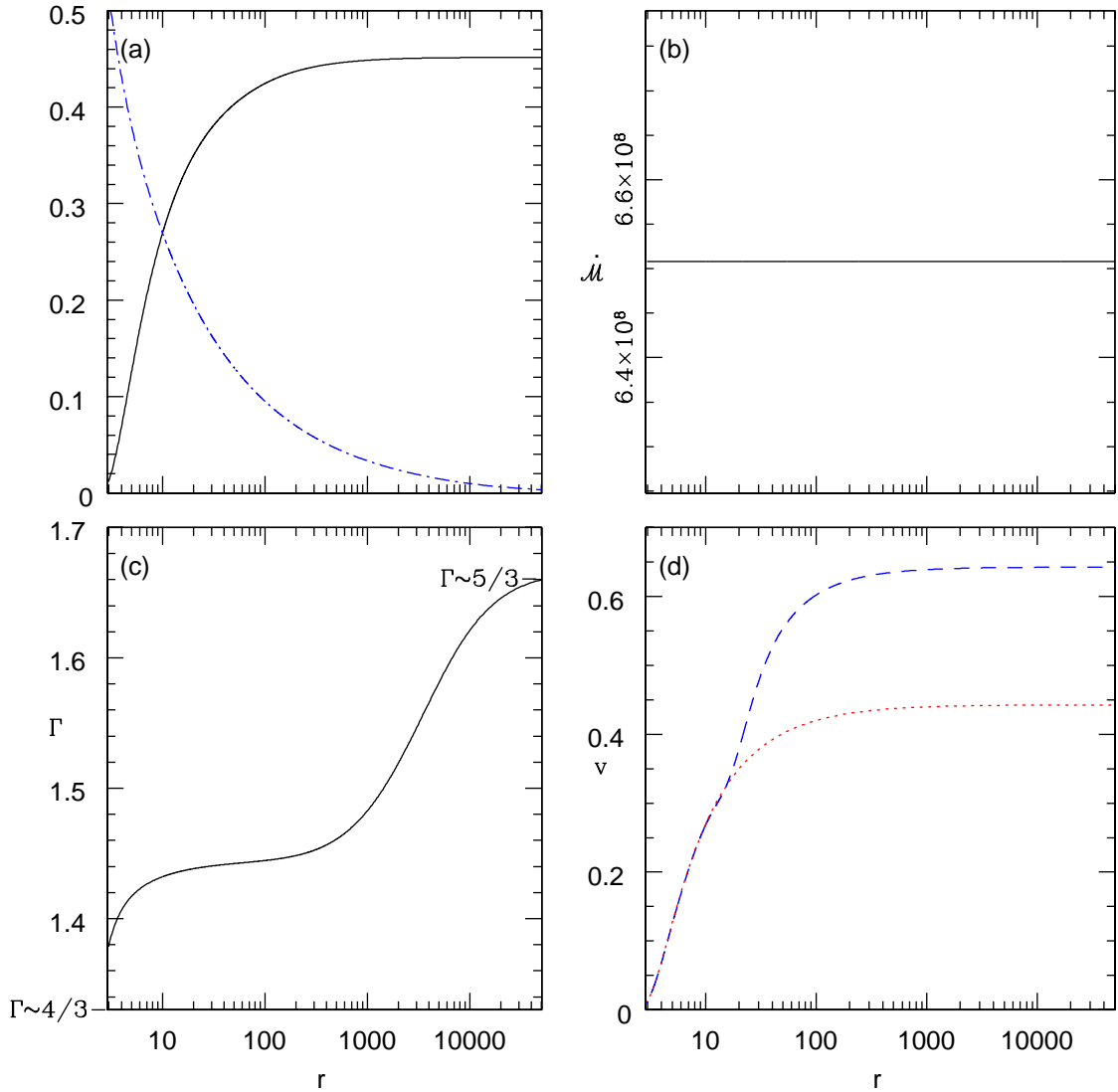}
 \caption{(a) The variation of jet 3-velocity $v$ (solid black) 
and sound speed $a$ (dashed-dotted blue) with $r$.
(b) ${\dot {\cal M}}$ is plotted as a function of $r$
and (c) the variation of $\Gamma$ with $r$. We keep ${\dot m}_\sk=5$. (d) Comparison of $v$ for a thermally driven jet (dotted red, ${\dot m}_\sk=0$) and
 radiatively plus thermally driven jet (dashed blue, ${\dot m}_\sk=10$).
The jet is launched around a $10M_\odot$ BH. \citep{vkmc15}}
\label{lab:gen}
 \end{center}
\end{figure}

A radiatively inefficient disc can only give rise to thermally driven jets, so we choose luminous disc.
%We discuss the jet properties for electron-proton jets {\ie} $\xi=1$, until specified otherwise.
We choose ${\dot m}_\kd=1$, $\beta=0.5$ and keep $\ep$ jets ({\ie} $\xi=1$) until specified otherwise.
In special relativistic study, the solutions are generated by supplying inner boundary values. The base of the jet is assumed to be at $r=r_b=3$ throughout the chapter. In Fig. (\ref{lab:gen}) we choose $v_b=0.014$ and $a_b=0.51$,
for ${\dot m}_\sk=5$ and obtain the transonic solutions. In Fig. (\ref{lab:gen}a), we plot jet 3-velocity $v$
and the sound speed $a$ as functions of $r$. This jet is from a disc
around a stellar mass BH.
The sonic point $r_c$ is at the crossing point of $v$ and $a$.
Total disc
luminosity is $\ell=\ell_\s+\ell_\sk+\ell_\kd\sim 0.14$ in units of Eddington luminosity ($L_{\rm Edd}$) and the resultant terminal speed achieved is $0.45$ in units of $c$.
${\dot {\cal M}}$ is obtained to be constant of motion (Fig. \ref{lab:gen}b).
In Fig. (\ref{lab:gen}c) we plot variation of $\Gamma$ or adiabatic index
of the jet. The base of the jet is very hot, therefore $\Gamma \rightarrow 4/3$
at the base. However, as the jet expands to relativistic velocities (at $r\rightarrow$ large), the temperature falls
such that $\Gamma \rightarrow 5/3$. In Fig. (\ref{lab:gen}d), we compare the $v$ profile of a thermally driven jet
with a radiatively plus thermally driven jet starting with the same base values. The radiatively
driven fluid jet is powered by radiation from a disc with ${\dot m}_\sk=10$. From the base to first few $r_g$, the $v$ profiles of the two flows are almost identical,
and the radiative driving is perceptible at $r>15.32$. The terminal speed of the thermally driven flow is slightly less than $0.45$
and for the radiatively driven flow it is $v_T\lsim 0.65$. The radiative driving of the jet is ineffective in regions
close to $r_b$, because the thermal driving accelerates the jet to $v \sim v_{\rm eq}$ results in a similar $v$ profile up to $r\sim 15.32$.
But beyond it, radiative driving generates a flow with a $44 \%$ increase in $v_T$.

\begin{figure}
\begin{center}
 \includegraphics[trim={6.5cm 0 7cm 0}, width=5.cm]{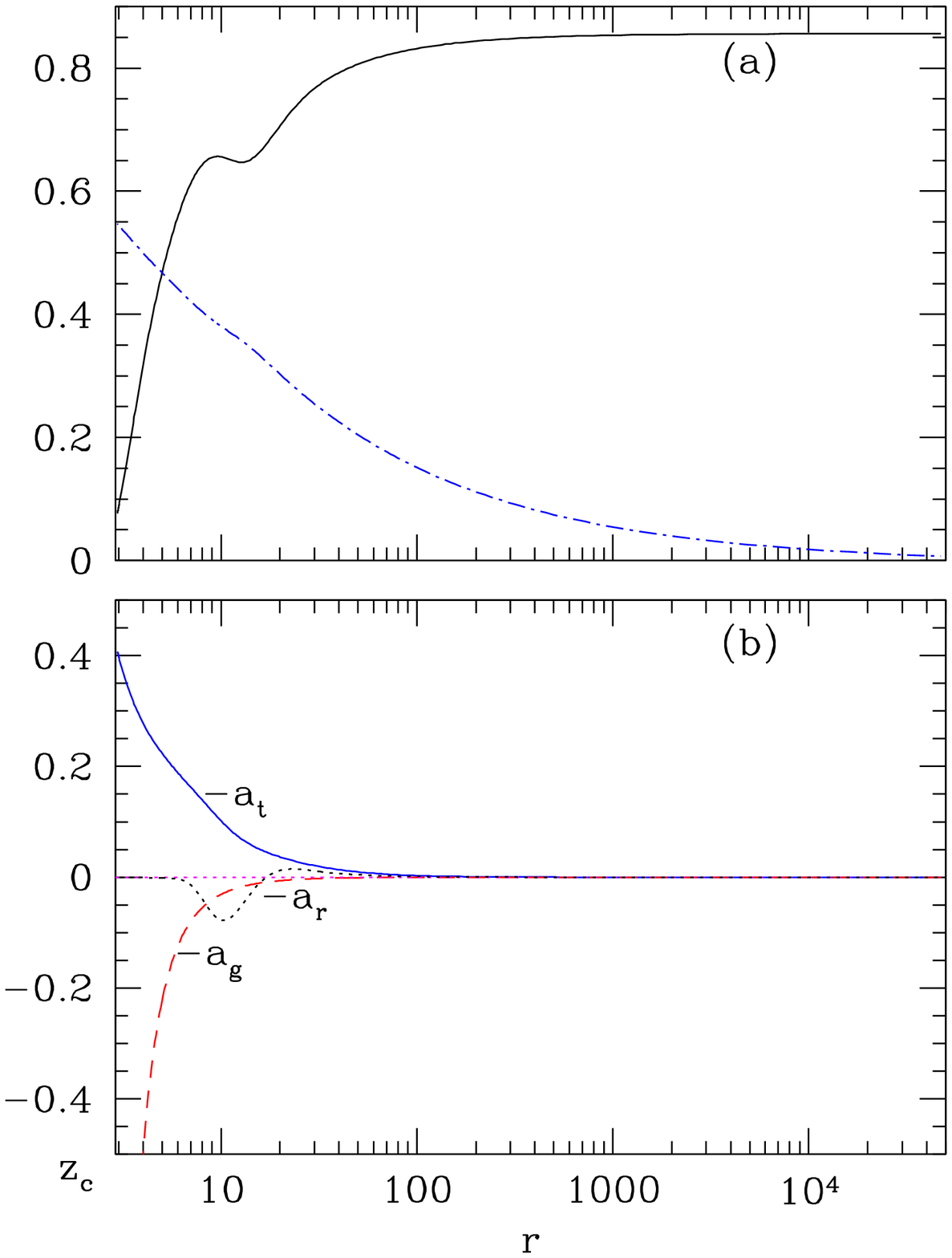}
 \caption{(a) Variation of $v$ (solid black) and $a$ (dashed-dotted blue);
(b) Variation of ${\rm a}_t$ (solid blue),
${\rm a}_r$ (dotted black) and ${\rm a}_g$ (dashed red)
with $r$. We take ${\dot m}_\sk=12$.
$x_s$ is obtained at $11.74$ \citep{vkmc15}.}
\label{lab:terms}
 \end{center}
\end{figure}

To depict the mechanism of radiation interaction with jet, in Figs. (\ref{lab:terms}a-b), we show a transonic jet from a disc around a stellar mass BH. The input parameters are chosen to be such ($v_b=0.087$ and $a_b=0.545$, ${\dot m}_\sk=12$) that, both radiation acceleration and deceleration are effective and visible.
In Fig. (\ref{lab:terms}a), we plot $v$
and $a$ with $r$. The sonic point is obtained to be at 
$r_c=5$. Luminosities of various disc components around
$10M_\odot$ BH, are $\ell_\sk=0.295$, $\ell_{\s} = 0.522$ and
$\ell_{\kd} = 0.0667$. The total luminosity turns out to be $\ell=0.884$. 
The 3-velocity $v$ increases beyond $r_c$ and up to $r\sim 8$, and then decelerates
in region $8< r \lsim 14$ and thereafter again accelerates till it reaches terminal value
$v_T \sim 0.86$. Let us analyze various terms that influence $v$.
In Fig. (\ref{lab:terms}b), we plot the 
variation of gravitational acceleration term (a$_g$), the radiative term (a$_r$), 
and acceleration due to thermal
driving (a$_t$). 
From L. H. S of equation (\ref{dvdr1.eq}), it is clear that in the subsonic
region $v$ can increases (that is, jet accelerates)
with $r$ only if the right hand side is negative. While in the supersonic region the jet accelerates if
the right hand side is positive. The gravity term or a$_g$ is always negative,
while a$_t$ is always positive. In this particular solution
a$_r<0$ for $r<17.06$.
In the sub sonic region {\ie} $r<r_c$,
$|{\rm a_r}| \ll {\rm a}_t$ and a$_t<{\rm a}_g$, therefore, R. H. S of equation (\ref{dvdr1.eq})
is negative and the jet is accelerated. At the sonic point a$_t={\rm a}_g+{\rm a}_r$.
For $r> r_c$, a$_r$ decreases to its minimum value at $r=10.30$. Gravity is less important at those
distances and a$_r\gsim {\rm a}_t$, which makes R. H. S negative. Therefore, the jet decelerates in range $9.56<r<12.68$.
For $r>12.68$, $|{\rm a}_r|$ decreases, ultimately becomes positive, making R.H.S to be positive again.
So the jet starts to accelerate at $r>12.68$ until $v\rightarrow v_T$.

\begin {figure}
\begin{center}
 \includegraphics[width=15.cm]{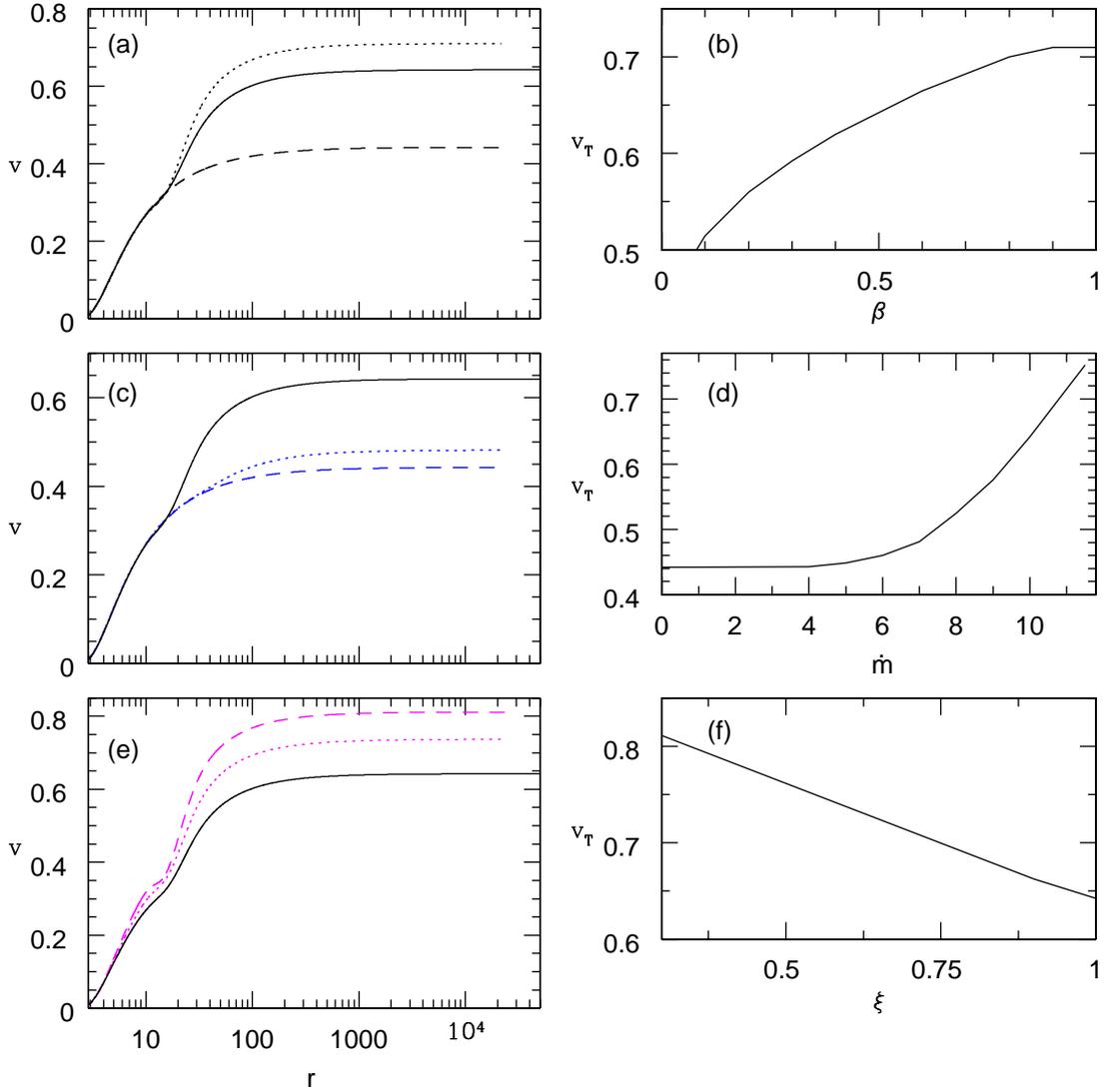}
 \caption{(a) Variation of $v$ with $r$ for jet around $10M_{\odot}$ BH. Each curve represents $\beta=1.0$ (dotted black),
$0.5$ (solid black) and
$0.0$ (dashed black).
(b) Variation of $v_{\rm T}$ with $\beta$. Other parameters same as (a).
(c) Variation of $v$ with $r$. Each curve represents ${\dot m}_\sk=1.0$ (dashed blue),
 $7$ (dotted blue) and
$10$ (solid black).
(d) Variation of $v_{\rm T}$ with ${\dot m}_\sk$. Other parameters same as (c).
(e) Variation of $v$ with $r$. Each curve represents $\xi=1.0$ (solid black),
$0.6$ (dotted magenta) and $0.3$ (dashed magenta).
(f) Variation of $v_{\rm T}$ with $\xi$. Other parameters same as (e).
Accretion parameters are ${\dot m}_\sk=10$ (in a, b, e, f), $\beta=0.5$
(in c, d, e, f), $\xi=1.0$ (in a, b, c, d) and ${\dot m}_\kd=1$. Jet base values
are $v_b=0.014$ and $a_b=0.51$ \citep{vkmc15}.
}
\label{lab:parameters}
 \end{center}
\end{figure}

Now we discuss how various disc parameters and fluid composition 
of the jet affect its dynamics. The jet is affected by the
radiation from the disc, and the radiation field above
the disc in influenced by 
$\beta$, $\dot{m}_\sk$, and $\dot{m}_\kd$. 
The base values of the jet are $v_b=0.014$
and $a_b=0.51$.
In Fig. (\ref{lab:parameters}a) 
we show comparison of $v$ as a function of $r$ for various values of ${\beta}
\rightarrow$ 0.0, 0.5 and 1.0.
The corresponding terminal speeds
($v_T$) at $r=10^4$ with $\beta$ are presented in Fig. (\ref{lab:parameters}b).
As the magnetic pressure increases in the disc, that is, $\beta$ increases, the supersonic part of the jet
is accelerated because $\ell_\sk$ increases. When magnetic pressure 
is zero ($\beta=0.0$) or when the jet is thermally and radiatively driven only by
pre-shock bremsstrahlung and thermal photons, the terminal speed is at around $0.44$. But when 
magnetic pressure is taken to be equal to the gas pressure ($\beta=1$), $v_T$ reached above
$0.7$. 
In Fig. (\ref{lab:parameters}c) we show the effect of 
SKD accretion rate on $v$ profile and the corresponding terminal speeds
are shown in Fig. (\ref{lab:parameters}d) with ${\dot m}_\sk$.
$v_T$ ranges from 0.42 to 0.72 when ${\dot m}_\sk$
is varied from $0.1$ to $11.5$. The velocity profiles of a thermally driven jet and a jet driven by radiation
acted on by ${\dot m}_\sk=1$ are similar. Only when the luminosity is close to $L_{\rm Edd}$ or $\ell \rightarrow 1$,
the radiative driving is significant. In Fig. (\ref{lab:parameters}e), we carry out 
similar analysis for variation of composition parameter 
($\xi$) in the jet, and plot $v$ profiles for jet with $\xi=0.3$, $0.5$ and $1.0$. With the lighter jet $v$ increases, and
this is also seen in the $v_T$ dependence of $\xi$ in
Fig. (\ref{lab:parameters}f). As $\xi$ increases, high proton fraction makes the jet
heavier per unit pair of particles, and the optical depth decreases
due to the decrease in total fraction of leptons. So the net radiative momentum deposited on to the 
jet per unit volume decreases, in addition the inertia also increases.
This makes the jets with higher proton fraction (high $\xi$) to be slower.
\begin{figure}
\begin{center}
 \includegraphics[trim={0 0 -2cm 1cm}, width=15.5cm]{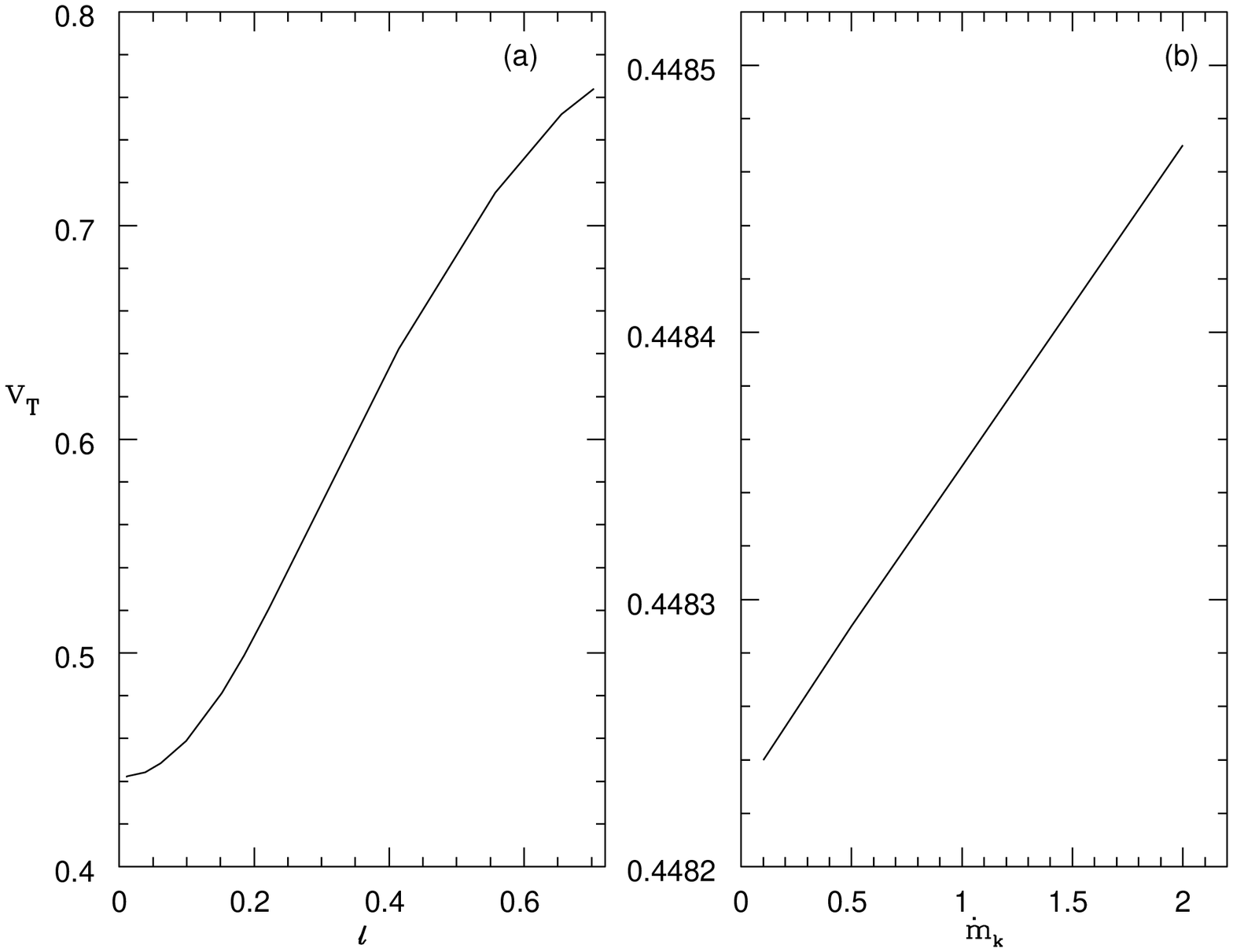}
 \caption{Dependence of terminal speeds $v_T$ on (a) $\ell$ and on
(b) $\dot{m}_{kd}$ for the similar base parameters ($v_b=0.014$ and $a_b=0.51$).
The value of ${\dot m}_\kd=1$ for (a), and ${\dot m}_\sk=5$ for (b) \citep{vkmc15}.}
\label{lab:lumkd}
 \end{center}
\end{figure}

As $x_s$ depends upon ${\dot m}_\sk$,  therefore, not only $\ell_\sk$ but also $\ell_\s$
varies. In fact since $x_s$ is the inner edge of the KD, $\ell_\kd$ will change even though ${\dot m}_\kd$
is kept constant.
In Fig. (\ref{lab:lumkd}a), we plot $v_T$ with the total luminosity $\ell~(\equiv \ell_\s+\ell_\sk+\ell_\kd)$,
by tuning ${\dot m}_\sk$. As the luminosity of the disc increases the terminal speed increases from moderate
values of $0.44$ to high speeds of $\sim 0.8$ when the disc luminosity is closer to Eddington limit.
However, $\ell_\kd$ has limited role in determining $v_T$ as has been shown in Fig. (\ref{lab:lumkd}b).

\begin {figure}
\begin{center}
 \includegraphics[width=15.cm]{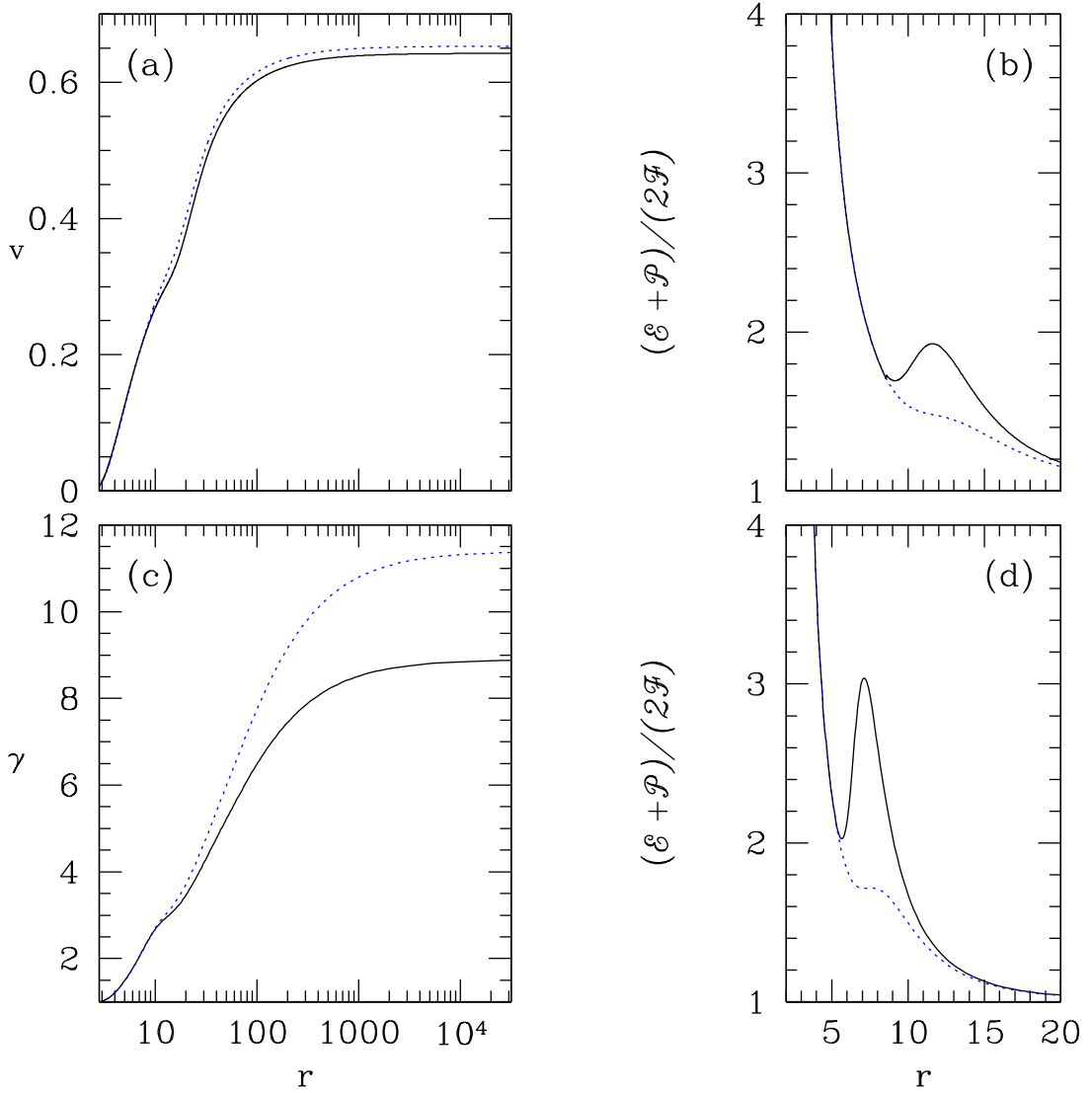}
 \vskip 0.0cm
 \caption{ Comparison of (a) $v$ profile for jets around $10 M_\odot$ (solid black)
and $10^8 M_\odot$ black holes (dotted blue) for ${\dot m}_\sk=10$.
(b) $\Re$ with $r$
from an accretion disc around $10 M_\odot$ BH (solid black)
and $10^8 M_\odot$ BH (dotted blue). 
The jet base values for (a) and (b) are $v_b=0.014$, and $a_b=0.51$.
(c) $\gamma$ profile of
a jet around $10M_\odot$ BH (solid black) and $10^8M_\odot$ (dotted blue) and
(d) $\Re$ with $r$.
The SKD accretion rate is ${\dot m}_\sk=12$. Other disc parameters are
${\dot m}_\kd=1$, $\beta=0.5$. And the jet base values for (c) and (d) are $v_b=0.19$, and $a_b=0.576$. \citep{vkmc15}}
\label{lab:m1m8}
 \end{center}
\end{figure}

In Fig. (\ref{lab:R3}) the radiative moments around a super-massive BH are shown to be significantly higher than
that around a stellar mass BH even for same accretion rates (in units of ${\dot M}_{\rm Edd}$). In order to study the effect of the mass of the central object, in 
Fig. (\ref{lab:m1m8}a), we compare the $v$ profile of the jet around $10M_\odot$ BH
with that around $10^8M_\odot$ BH. The jets are launched with the same
base parameters ($v_b=0.014$, and $a_b=0.51$), around accretion with ${\dot m}_\sk=10$. Although the radiative moments around a super-massive BH are significantly different,
yet the $v$ profiles differ by moderate amount. To ascertain the cause we plot $\Re$ or relative contribution of
radiative moments for both the jets in Fig. (\ref{lab:m1m8}b). $\Re$ is quite similar
for both the BHs close to the horizon, but in the range $8<r<20$ $\Re$ around stellar mass BH is higher than that around
super massive BH. In Fig. (\ref{lab:m1m8}c), we compare the Lorentz factor
$\gamma$ of a jet around $10M_\odot$ BH with a jet around $10^8M_\odot$ BH,
launched with hotter base ($v_b=0.19,~
a_b=0.576$) and acted by high accretion rate ${\dot m}_\sk=12$. The initial $\gamma$ ($\equiv$ $v$) is almost
same for both the jets, however, due to larger $\Re$ around a stellar mass BH, the jet around it is slower
compared to that around super-massive BH. In this case, the terminal Lorentz factor $\gamma_T$
is significantly larger for a jet around $10^8 M_\odot$ BH.

\begin {figure}
\begin{center}
 \includegraphics[width=12.cm]{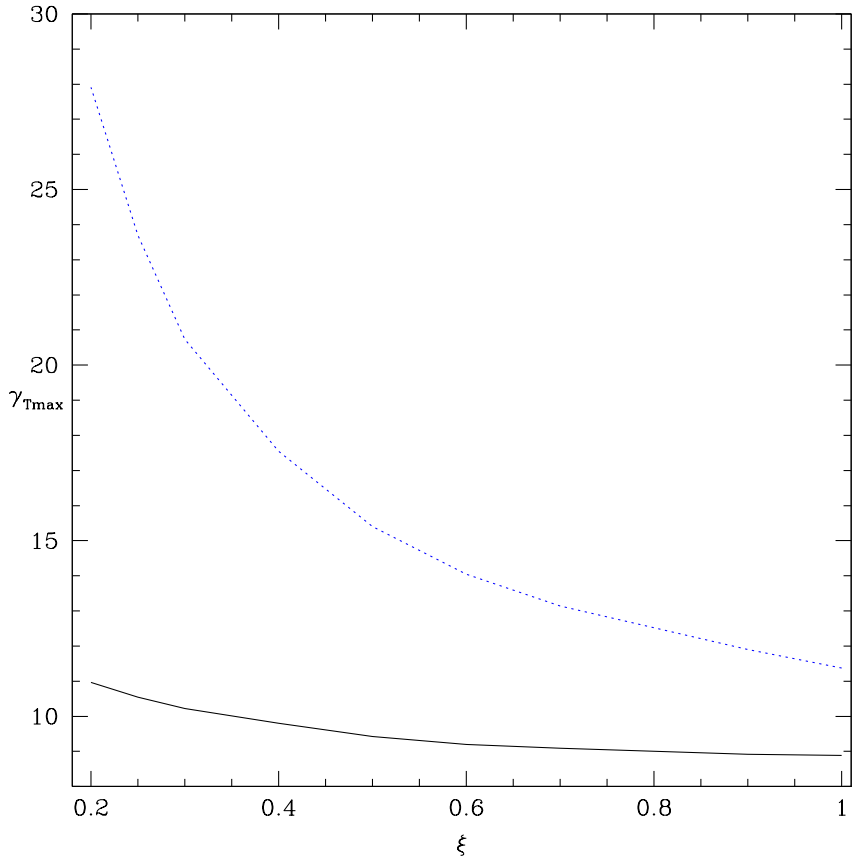}
 \caption{Variation of $\gamma_{T{\rm max}}$ with $\xi$, around $10M_\odot$ BH (solid black) and
$10^8 M_\odot$ BH (dotted blue). Accretion parameters are ${\dot m}_\sk=12$, $\beta=0.5$ and ${\dot m}_\kd=1$.
and jet base values are $r_b=3,~\&~a_b=0.5766$ \citep{vkmc15}.}
\label{lab:max1}
 \end{center}
\end{figure}

It is clear that jets around stellar mass BH are slower, and lighter jets are faster, but what is the maximum
terminal velocity possible? We choose to launch jet with maximum possible sound speed at the base and very high
accretion rate.
In Fig. (\ref{lab:max1}),
we plot the maximum terminal Lorentz factor or $\gamma_{T{\rm max}}$ possible as a function of $\xi$ for $10M_\odot$ BH
and $10^8M_\odot$, when accretion parameters are ${\dot m}_\sk=12$, $\beta=0.5$ and ${\dot m}_\kd=1$.
For jet composition $\xi=1.0$, the maximum possible terminal Lorentz factor $\gamma_{T{\rm max}}\lsim 9$
for a jet around $10M_\odot$ BH, but for $10^8 M_\odot$ BH, $\gamma_{T{\rm max}}\gsim 11$. However, for lighter jet
around stellar mass BH $\gamma_{T{\rm max}}\sim 10$, but for super-massive BH, light jets yields
$\gamma_{T{\rm max}}\sim$ few $\times 10$. So for lepton dominated composition, ultra-relativistic jets around super-massive BHs
are possible if they are driven by radiation from a luminous disc.

\section{Discussion and Concluding Remarks}
%\label{sec5}

We have investigated the interaction of a relativistic fluid-jet with the radiation
field of the underlying accretion disc. We noticed that proper relativistic transformations of the radiative intensities from the local disc frame
to the observer frame are very important and these transformations
modify the magnitude as well as the distribution of the
moments around a compact object. The corona and SKD are the major contributors in
the net radiative moments and KD contribution is much lower than either of the former.
One of the interesting facts about the moments due to various
disc components is that they peak at different positions from the disc plane giving rise to multi stage acceleration.
The jets with normal conditions at the base, produce
mildly relativistic terminal velocities $v_T\sim$ few $\times 0.1$ (Fig. \ref{lab:gen}).
The elastic scattering regime maintains the isentropic nature of the jet, and because we considered a realistic
and relativistic gas equation of state, the adiabatic index changes along the jet.
However, radiation not only accelerates but also decelerates if $v>v_{\rm eq}$. Although close to the jet base
($r_b$) the velocity is low and the radiation field should accelerate, but being hot, the effect of radiation
is not significant in the subsonic branch because of the presence of inverse of enthalpy in the radiative term ${\rm a}_r$ of the equation of motion (equation \ref{dvdr1.eq}).
Therefore, in the subsonic domain the jet is accelerated as a result of competition between thermal and the gravity term.
As the magnetic pressure is increased, the synchrotron radiation from SKD increases and it jacks up the
flow velocity in the supersonic regime. Increasing ${\dot m}_\sk$ increases both the 
synchrotron as well as bremsstrahlung photons from the SKD, which makes the SKD contribution
to the net radiative moment more dominant, and therefore increases $v$ in the supersonic
part of the flow. $v_T$ increases with $\beta$ but tends to taper off as $\beta\rightarrow 1$,
however, $v_T$ tends to increase with ${\dot m}_\sk$ and shows no tendency to taper off.
The jet tends to get faster with the decrease in protons ({\ie} decrease of $\xi$).
We do not extend our study to electron and positron or $\xi=0$ jet, since a purely
electron-positron jet is highly unlikely \citep{kscc13}, although pair dominated
jet ({\ie} $0<\xi<1$) is definitely possible. Terminal velocity $v_T$ increases with total
luminosity $\ell$, and approaches relativistic values as the disc luminosity approaches
Eddington limit. However, the KD plays a limited role in accelerating jets (Figs. \ref{lab:lumkd} a-b).

If accretion rates are moderately high, then jets around super-massive BHs are slightly faster than
those around stellar mass BHs (Figs. \ref{lab:m1m8}a-b). However, if accretion accretion rates are
high such that the moments from the central region of the disc differ significantly, then ultra-relativistic
jets around super massive BHs can be obtained compared to stellar mass BH (Figs. \ref{lab:m1m8}c-d).
Comparison of jet $\gamma_{T{\rm max}}$ around a stellar mass BH and that around super massive one
as a function of $\xi$,
shows that jet around super massive BH can be accelerated to $\gamma_T \gsim 11$ even for fluid composition
$\ep$, while that around stellar mass BH $\gamma_T \sim$ few. However, lighter jets around $10^8 M_\odot$ BH
can be accelerated to
truly ultra-relativistic speeds, compared to jets around stellar mass BH.
\chapter{General Relativistic Version of de Laval Nozzle and Non-Radial Jets with Internal Shocks}
\label{CH:P2}
\section{Overview}
In chapter \ref{CH:P1}, the space-time was assumed to be flat, while here we have considered Schwarzschild space-time around BH. The requirement of using general relativistic analysis is explained in Appendix (\ref{app:pw_gr}) where we show that apart from physical incompatibility of PW potential with special relativity, general relativity is required to properly dealing with outflow solutions. Former approach with flat space-time assumption gives greater deviation as compared to curved space consideration.

We use  two models for the jet geometry, (i) model M1 - a conical geometry and (ii) model M2 - a geometry with non-conical cross-section. The non-conical jet is taken to be similar to de Laval Nozzle where after emerging at the base, close to BH, the jet is collimated by accretion funnel (corona) and then above the funnel it again expands radially. Radiation field is ignored for simplicity. Similarly, we consider a fluid described by a relativistic equation of state. Along with thermal acceleration, we explore possibility of multiple sonic points as well as internal shocks in the jet. We discuss possible consequences and observational implications of the obtained internal shocks. The results are published in \cite{vc17}.
\section{Assumptions, governing equations and jet geometry}
\label{sec2}
The jet fluid is considered to be in steady state ({\ie} 
$\partial/\partial t=0$) and as they are collimated, we consider them to be on axis
({\ie} $u^\theta=u^{\phi}=\partial/\partial \theta=0$) and 
axis-symmetric ($\partial/\partial \phi=0$).
In advective disc model, the inner funnel or corona acts as the base of the jet \citep{cd07,dc08,kc13,kcm14,kc14,dcnm14,ck16,lckhr16}.
The shape of the corona is torus \citep[see simulations][]{dcnm14,lckhr16} and its dimension is about $\gsim$few$\times10 \rg$.
Therefore, launching the jet inside the torus shaped funnel, simultaneously
satisfies the observational requirement that the corona and the base of the jet be compact.
This also automatically satisfies that the jet is unlikely to be present if the corona is absent.
%Having said so, we must point out that, we actually do not obtain the jet input parameters ($E$ \& ${\dot M}$) from advective accretion disc solutions, but the input parameters are supplied. This implies, any accretion disc model with compact torus like hot corona will satisfy the underlying disc model. However, advective disc model with corona gets a special mention because possibility of hot electron distribution close to the
%central object, is inbuilt to the model.
The only role of the disc considered here is to confine the jet flow boundary at the base for one of the jet model (M2) considered.
Since the exact method of
how the jets originate from the discs is not being considered, the jet inputs are actually free parameters independent of the disc solutions. Hence, in short, in this chapter, we carry out an exploratory study of the thermally driven jets and role of flow geometry close to their base. 
\subsection{Geometry of the jet}
\label{sec:jet_geom}
Observations of extra-galactic jets show high degree of collimation, so it is quite
common in jet models to consider conical jets with small opening angle. As stated previously, we have considered two models of jets, the first model being a jet
with conical cross-section and we call this model M1.
However, the jet at its base is very hot and subsonic, and since 
the pressure gradient is isotropic, the jet expands in all directions.
The walls of the inner funnel of the corona provides natural collimation
to the jet flow near its base. If the jet at the base is very energetic, then it is quite likely
to become transonic within the funnel of the corona.% (see Fig. \ref{lab:fig1}).
%As the jet accelerates to the region above the funnel of corona, then radiation pressure from
%the disc would push the jet material towards the axis and then the jet expands \citep{cc02b}, resulting in
%a converging-diverging cross-section. Interestingly
%{\color{blue} In a magneto-hydrodynamic simulations}
%of jets, \citet{kms02} showed that the jet indeed flows through the open field line region, but at a certain distance above the torus shaped disc,
%the jet surface is pushed towards the axis and beyond that the jet expands again resulting into a converging-diverging cross-section. %Therefore, it is also not necessary that the jet retains the same cross-section all the time.
%In other words, radiation field from the disc, or, magnetic field structure above the disc
%would ensure a contracting-expanding geometry.
%The cross-section adopted for Model M2 in this paper,
%is inspired by these \textcolor{blue}{kinds of jet simulations}.
%We ensured that the variation of the geometry of M2 is smooth and slowly varying, such that the solutions are
%not crucially dependent on the particular shape of the flow geometry.

Considering above points, the general form of jet cross section or, ${\cal A}(r)$ is taken as
\be
{\cal A}={\cal G}r^2,~~ \mbox{where,~}{\cal G}=2\pi(1-{\rm cos}\theta);~~\theta \mbox{ is polar angle at }r
\label{jetgeom.eq}
\ee 
For the first model M1: 
\be
cos\theta=\frac{-(C m)/r+{\sqrt{{-C^2/r^2}+m^2+1}}}{(m^2+1)},
\label{con2a.eq}
\ee
where, $C$ and $m(= {{\tan {\theta_0}}})$ are the constant intercept and slope of the jet boundary with the equatorial plane, with {$\theta_0$}
being the constant opening angle with the axis of the jet.% The value of the constant
%intercept might be
%$C=0$ or $C \neq 0$, either way, the solutions are qualitatively same and we take $C=0$.

The second model M2 mimics a geometry whose outer boundary is the funnel shaped surface
of the accretion disc at the jet base i. e., $d{\cal A}/dr >0$.
%(see, Fig. \ref{lab:accretion_disc_jet_plot}a).
As the jet
leaves the funnel shaped region of the corona, the increase of the gradient of the cross-section reduces, (\ie 
$~d{\cal A}/dr \rightarrow 0$). It again expands
and finally becomes conical at large distances, where $d{\cal A}/dr \propto r$.
The functional form of the jet surface in spherical coordinates for model M2 is given by,
\be
r~{\rm sin}\theta=\frac{k_1d_1(r-n)}{1+d_2(r-n)^2}+m_{\infty}(r-n)+k_2,
\label{geomvar.eq}
\ee
where, $k_1=5\xsh/\pi$, $k_2=\xsh/2$, $d_1=0.05$, $d_2=0.0104$, $m_{\infty}=0.2$ and $n=5$.
%A schematic diagram of the geometry of the M2 jet model and disc is shown in Fig. \ref{lab:fig1},
here $\xsh$ is the shock
in accretion, or in other words, the length scale of the inner torus like region.
In Eq. (\ref{geomvar.eq}), $k_1$, $k_2$
are parameters which influence the shape of jet geometry at the base, while $d_1$, $d_2,~m_{\infty}$ and $n$
are constants, which together with $k_1$ and $k_2$, shape the jet geometry. Here, we assume at large distances
the jet is conical and $m_\infty(= {\tan {\theta_{\infty}}})$ is the gradient which corresponds to the terminal
opening angle taken to be 11$^\circ$. This geometry is shown in Figure (\ref{lab:accretion_disc_jet_plot}a) for $\xsh=40$. The details of Model M2 are discussed below.
%The size of the corona or $\xsh$ influences the jet geometry and
%the jet geometry at the base is shaped by the shape of the corona as shown in appendix (\ref{disc_structure}).
%A typical jet geometry for a given set of accretion solution is plotted in Fig. \ref{lab:accretion_disc_jet_plot}.
\subsection{Approximated accretion disc quantities}
\label{disc_structure}
The jet geometry of M2 model is taken according to
equation (\ref{geomvar.eq}). The inner part of the accretion disc or Corona shapes the jet
geometry near the base.
\begin{figure}[hp]
%\centering
%\captionsetup{justification=centering}
\begin{center}
 \includegraphics[trim={0 0 0 6.2cm},clip,width=11cm]{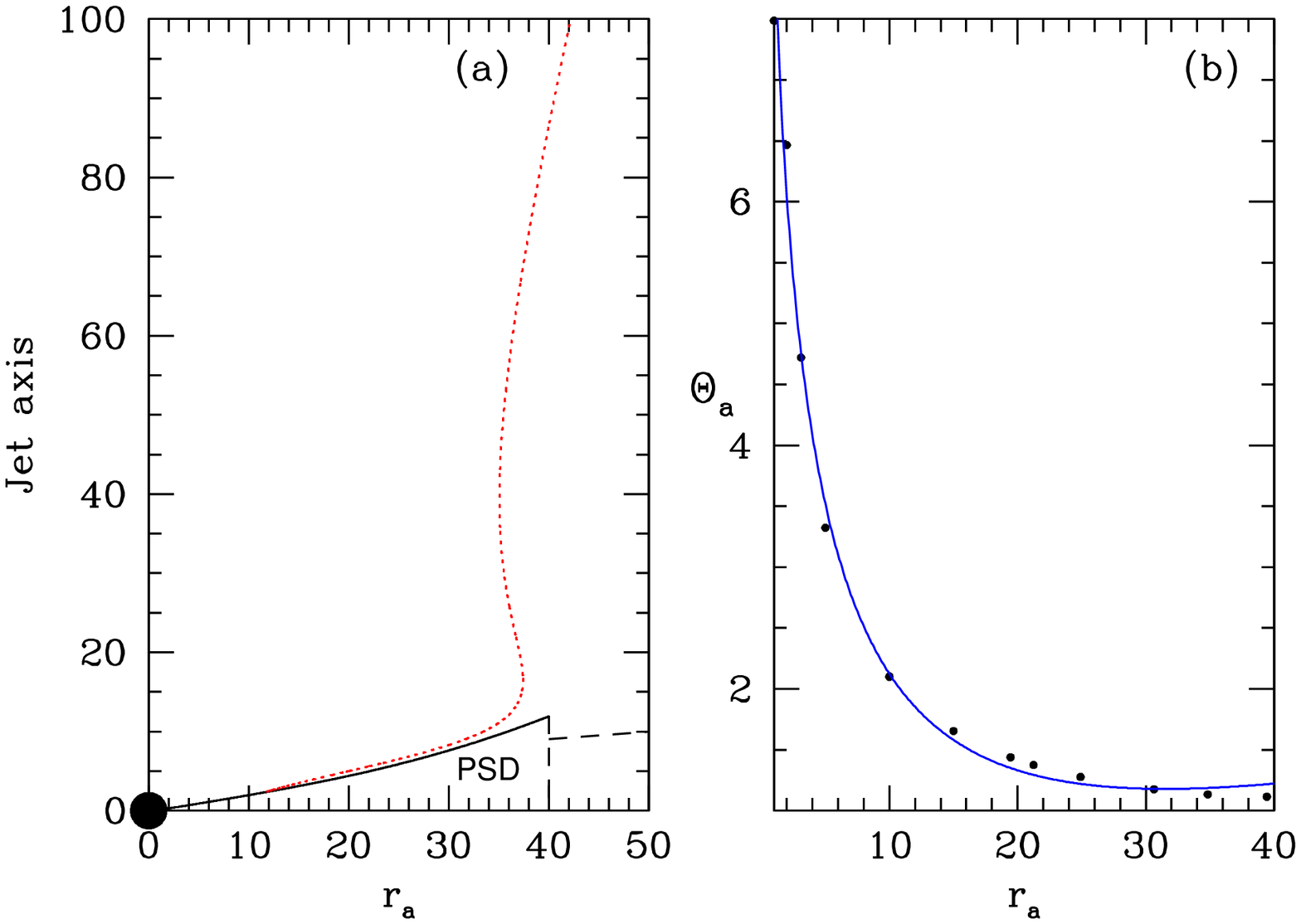}%{app3.eps}
 \caption{(a) Accretion disc height $H_{\rm a}$ is plotted with $r_{\rm a}$ with solid line. Shock location is at $x_{\rm sh}=r_{\rm a}=40$, shown by long dashed line. The jet width $r sin \theta$ is over plotted with red dotted line. Black hole resides at $r_{\rm a}=0$ with shown as black sphere with $r_s=2$ (b) Fitted $\Theta_{\rm a}$ \citep{vc17}
 }
\label{lab:accretion_disc_jet_plot}
 \end{center}
\end{figure}
If the local density, four velocity, pressure and dimensionless temperature in the accretion disc are $p_{\rm a}$, $u^r_{\rm a}$, $\rho_{\rm a}$ and $\Theta_{\rm a}$, respectively, and the angular momentum of the disc is $\lambda$, then the local height $H_{\rm a}$ of the post shock region for an advective disc is given by \citep{cc11}
\be 
H_{\rm a}=\sqrt{\frac{p_{\rm a}}{\rho_{\rm a}}\left[r_{\rm a}^3-{\lambda}^2(r_{\rm a}-2)\right]}={ \sqrt{\frac{2 {\Theta}_{\rm a}}{\tau}\left[r_{\rm a}^3-{\lambda}^2(r_{\rm a}-2)\right]}},
\label{h_a.eq}
\ee
where, $r_{\rm a}$ is the equatorial distance from the black hole.
We obtain $\Theta_{\rm a}$ in an approximate way following \citet{vkmc15}.
The $u^r_{\rm a}$ is obtained by solving geodesic equation.
Since $u^r_{\rm a}$ is known at every $r_{\rm a}$, and accretion rate is a constant, $\rho_{\rm a}$
is known. We also know that $\Theta_{\rm a}$ and $\rho_{\rm a}$ are related by the adiabatic relation. So
supplying $\Theta_{\rm a}$, $\rho_{\rm a}$ and $u^r_{\rm a}$ at the $\xsh$, we obtain $\Theta_{\rm a}$
for all values of $r_{\rm a}$. We plot $\Theta_{\rm a}$  in Fig. (\ref{lab:accretion_disc_jet_plot}b)
for $\xsh=40$ in filled dots.
An analytic function to the variation of $\Theta_{\rm a}$ with $r_{\rm a}$ is obtained as. 
\be
\Theta_{\rm a} = \exp(-r_{\rm a}^{a_t}+{b_t})c_t+d_tr.
\label{Theta_a.eq}
\ee
With $a_t=0.391623$, $b_t=2.30554$, $c_t=2.22486$ and $d_t=0.0225265$. 
This fit is shown in Fig. (\ref{lab:accretion_disc_jet_plot}-b) by solid line with 
points being actual values of $\Theta_{\rm a}$.
The fitted function is used to compute $H_{\rm a}$ and
is plotted in Fig.  (\ref{lab:accretion_disc_jet_plot}-a) with solid line. We over plot the jet structure (equation \ref{geomvar.eq}) in red dots. 
\subsection{Equations of motion of the jet}
Considering outlined assumptions above, the gradients of $v$ (equation \ref{dvdr.eq}) and $\Theta$ (equation \ref{dthdr.eq}) along jet axis reduce to
\begin{equation}
\gamma^2v\left(1-\frac{a^2}{v^2}\right)\frac{dv}
{dr}=\left[a^2\left\{\frac{1}{r(r-2)}+\frac{1}{{\cal A}}
\frac{d{\cal A}}{dr}\right\}-\frac{1}{r(r-2)}\right]
\label{dvdr2.eq}
\end{equation}
and
\begin{equation}
\frac{d{\Theta}}{dr}=-\frac{{\Theta}}{N}\left[ \frac{{\gamma}
^2}{v}\left(\frac{dv}{dr}\right)+\frac{1}{r(r-2)}
+\frac{1}{{\cal A}}\frac{d{\cal A}}{dr}\right]
\label{dthdr2.eq}
\end{equation}

All the information like jet speed, temperature, sound speed, adiabatic index and 
polytropic index as functions of spatial distance can be obtained by integrating equations (\ref{dvdr2.eq}-\ref{dthdr2.eq}).
A physical system becomes tractable when solutions are described in terms of their constants of motion.
In absence of radiation, $X_f=0$ in equation (\ref{energy.eq}) and the generalized Bernoulli parameter becomes
\be
E=-hu_t.
\label{energy_2.eq}
\ee

The system being isentropic, Equations (\ref{entacc.eq}) and (\ref{energy_2.eq}) are 
measures of entropy and energy of the flow that remain 
constant along the streamline. However, at the shock, 
there is a discontinuous jump in ${\dot {\cal M}}$.
\section{Results}

\begin{figure}
\begin{center}
 \includegraphics[trim={0.35cm 0 0 6.5cm},clip, width=14.7cm]{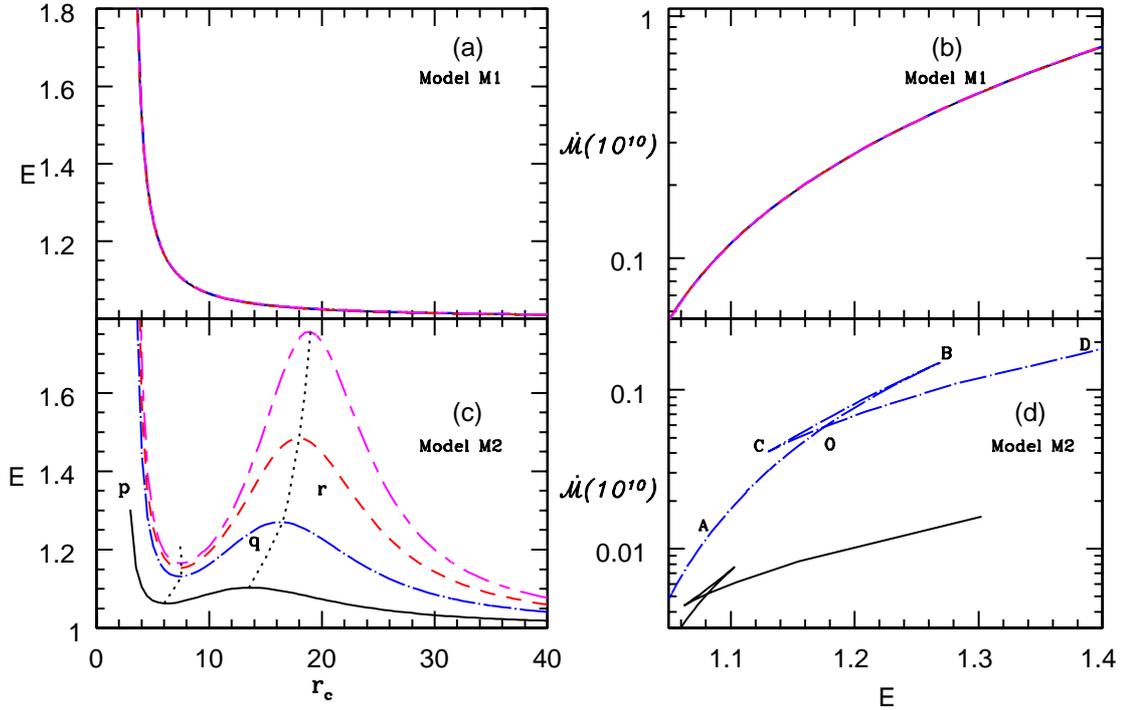}%{parameter_space.eps}
 \caption{Variation of $E$ with $r_c$ (a, c) and ${\md}$ with $E$ (b, d) for the two models M1 (a, b)
 and M2 (c, d). Each curve represent $x_{\rm sh}=12$ (long-short dash, magenta), $10$ 
 (dash, red), $7$ (dash-dot, blue),
 % 5(dotted green online), 
 $3$ (solid, black). (d) $\md~(r_c)$-$E~(r_c)$ are plotted
 for only two $\xsh=3$ (solid, black) and $\xsh=7$ (dash-dot, blue). The composition of the jet is for $\xi=1.0$ \citep{vc17}.
 }
\label{lab:P2_fig2_2}
 \end{center}
\end{figure}
\subsection{Nature of sonic points}
\label{sbsec3.1}
%Jets originate from a region in the 
%accretion disc, which is close
%to the central object, where, the jet is subsonic and very hot.
%The thermal gradient being very strong, works against the gravity and
%power the jet to higher velocities and in the process, $v$ crosses
%the local sound speed $a$ at the sonic point $r_c$, which makes the jet supersonic.
%The sonic point is also a critical point because, at $r_c$
%equation (\ref{dvdr.eq}) 
%takes the form $dv/dr\rightarrow 0/0$.
%In current regime, the sonic point conditions are obtained as
%\be 
% v_c=a_c {~~ and~~} a_c^2=\left[1+r_c(r_c-2)\left(\frac{1}{{\cal A}}\frac{d{\cal A}}{dr}\right)_c\right]^{-1}\
% \label{sonic.eq}
%\ee 
%
%Solving for $dv/dr|_c$ at $r_c$ one obtains two complex roots leading to
%the either $O$ type (or `centre' type) or `spiral' type
%sonic points, or two real roots but with opposite signs
%(called $X$ or `saddle' type sonic points), or real roots 
%with same sign (known as nodal type sonic point).\\
%So for a given set of flow variables at the jet base, a 
%unique solution will pass through
%the sonic point(s) determined by the 
%entropy ${\dot {\cal M}}$ and energy $E$ of the flow.
%Model M1 is independent of the shock location $\xsh$ in the accretion disc,
%but model M2 depends on $\xsh$.
In Figs. (\ref{lab:P2_fig2_2}a, b), we plot the sonic point properties of
the jet for model M1, and in Figs. (\ref{lab:P2_fig2_2}c, d) we plot the
sonic point properties of M2. Each curve is plotted for 
$x_{\rm sh}=12$ (long-short dash, magenta), $10$ 
(dash, red), $7$ (dash-dot, blue), $3$ (solid, black).
The Bernoulli parameter of the jet $E$ is plotted as a function of $r_c$ in Fig. (\ref{lab:P2_fig2_2}a, c).
In Fig. (\ref{lab:P2_fig2_2}b, d), ${\dot {\cal M}}$ is plotted as a function of $E$ at the sonic points.
We assumed that the jet model M1 is a conical flow. So,
curves corresponding to various values of $\xsh$ coincide with each other in Figs. (\ref{lab:P2_fig2_2}a-b).
Moreover, both $E$ and ${\dot {\cal M}}$ are monotonic functions of $r_c$. In other words, a flow with a given $E$ will have one sonic point, and the transonic solution will correspond to
one value of entropy, or ${\dot {\cal M}}$.
The situation is different for model M2. As $\xsh$ is increased from $3,~7,~10,~12$, the $E$ versus $r_c$ plot
increasingly deviates from monotonicity and produces multiple sonic points
in larger range of $E$ (Fig. \ref{lab:P2_fig2_2}c). 
For small values of $\xsh$ the jet cross-section is very close to the conical geometry and therefore, multiplicity of sonic points is obtained in a limited range of $E$.
It must be noted that, for a given $\xsh$, jets with very high and low values of $E$
form single sonic points. The range of $E$ within which multiple sonic points may form,
increases with increasing $\xsh$.
% The reason M1 has only one $r_c$, is amply clear from our discussion on DLN in section \ref{sec:eom}.
If r. h. s of equation (\ref{dvdr2.eq}) is zero, a sonic point is formed.
Since ${\cal A}^{-1}d{\cal A}/dr=2/r$ is always positive for model M1, the r. h. s 
becomes zero only due to gravity, and therefore, there is only one $r_c$ for M1.

For M2, the cross-section near the base increases faster than a conical cross-section,
therefore, the first two terms in R.H.S of equation (\ref{dvdr2.eq}) compete with gravity.
As a result, the jet rapidly accelerates to cross the sonic point within the funnel like
region of the corona
But as the jet crosses the height of the corona, the expansion is arrested and at some height
${\cal A}^{-1}d{\cal A}/dr \sim 0$. If this happens closer to the jet base
then the gravity will again make the r. h. s of the equation (\ref{dvdr2.eq}) zero,
causing the formation of multiple sonic points. 
For low values of $E$ in M2, the thermal driving
is weak and, hence, the sonic point forms at large distances. At those distances
${\cal A}$ becomes almost conical and therefore, for reasons cited above, 
the jet has only one sonic point.
If $E$ is very high, then the strong thermal driving makes the jet transonic
at a distance very close to the jet base.
For such flows, the thermal driving remains strong enough even in the supersonic domain, which
negates the effect of changing $d{\cal A}/dr$ and does not produce more sonic points.
For intermediate values of $E$, the jet becomes transonic at slightly larger distances.
For these flows,
the thermal
driving in the supersonic region becomes weaker and at the same time, the expansion of the jet cross section
term decreases \ie,  
${\cal A}^{-1}d{\cal A}/dr \sim 0$. At those distances, the gravity again becomes dominant
than the other two terms, which reduces the
r. h. s of equation(\ref{dvdr2.eq}) and makes it zero to produce multiple sonic points.
In Fig. (\ref{lab:P2_fig2_2}c), the maxima and minima of $E$ is the 
range which admits multiple sonic points.
We plotted the locus of the maxima and minima with a dotted line, and then divided the region
as `p', `q' and `r'. Region `p' harbours inner X-type sonic point, region `q' harbours O-type
sonic point and region `r' harbours outer X-type sonic points.
Figure (\ref{lab:P2_fig2_2}d) 
is the knot diagram \citep[similar to `kite-tail' for accretion, see][]{c89,kscc13} between $E$ and 
${\dot {\cal M}}$ evaluated at $r_c$, for two values of $x_{\rm sh}=3$ (solid, black) and $\xsh=7$ (dashed-dot, blue).
For $\xsh=7$ (dashed-dot, blue),
the top flat line of the knot (BC) represents the O-type sonic points from region `q'
of Fig (\ref{lab:P2_fig2_2}d). Similarly,
AB represents outer X-type sonic points from region `r'.
And CD gives the values of $E$ and ${\dot {\cal M}}$ for which only
inner X-type sonic points (region `p') exist. 
If the coordinates of points `B' and `C' in Fig. (\ref{lab:P2_fig2_2}d) be marked as ${\md}_{\rm B},~E_{\rm B}$ and so on, then it is clear that for $\xsh=7$ (dashed-dot blue), multiple sonic points
form for jet parameter range $E_{\rm C}\leq E \leq E_{\rm B}$. At the crossing point `O', the entropy of both the `x' type sonic points are same.
The plot for $\xsh=3$ (solid, black) is plotted for comparison which shows that if the shock in accretion is formed
close to the central object, then multiple sonic points in jets are formed for moderate range of $E$, a fact also
quite clear from Fig. (\ref{lab:P2_fig2_2}c).
\subsection{Jet solutions}
We classify the jet solutions for the two types of geometries considered:
(i) model M1: conical jets and (ii) model M2: jets through a variable and non-conical cross-section,
as described in section \ref{sec:jet_geom}.

\subsubsection{Model M1 : Conical jets}
\label{sec:m1}
\begin{figure}
\begin{center}
 \includegraphics[trim={0.5cm 6.5cm 0.2cm 0.2cm},clip,width=14.6cm]{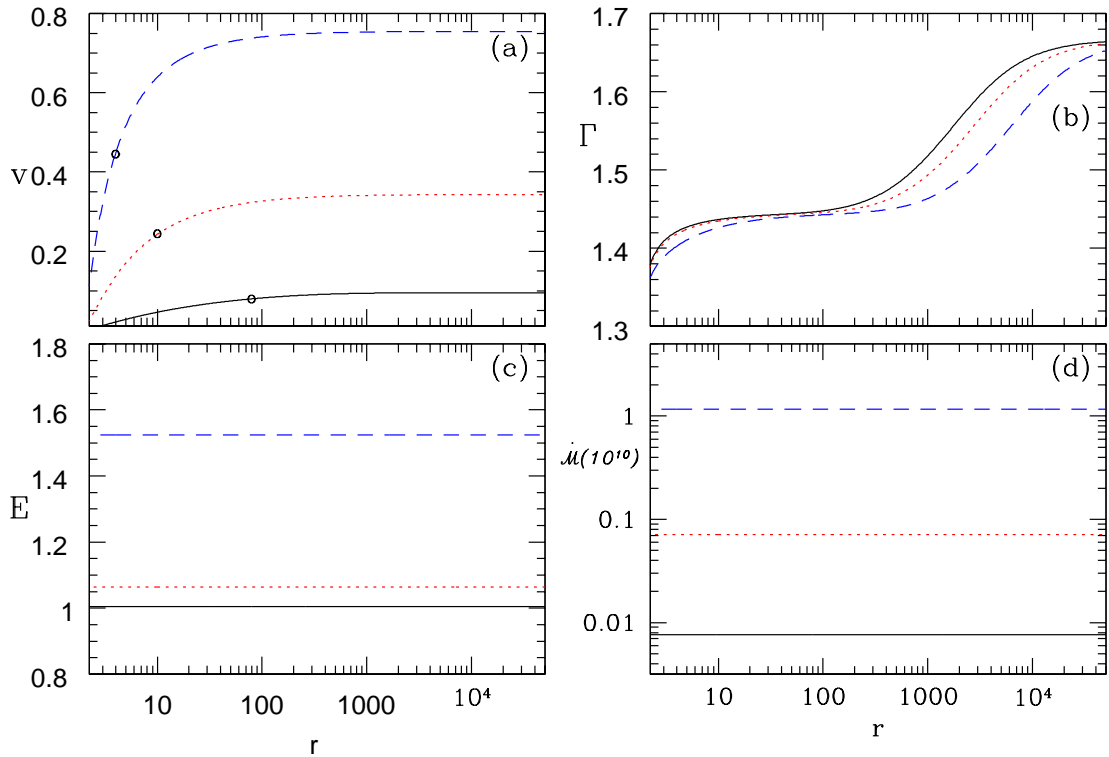}%{sph_sol_ep.eps}
 \hskip -5.0cm
 \caption{Model M1: Variation of (a) three velocity $v$, (b) 
adiabatic index $\Gamma$, (c) energy parameter $E$ and (d) entropy $\md$ as functions of $r$.
 The shock in accretion $x_{\rm sh}=12$ and flow composition $\xi=1.0$. 
 Each curve is characterized for $E=1.0045$ (solid, black), 
 $E=1.064$ (dotted, red) and $E=1.525$ (dashed, blue). 
 Black open circles in (a) show the location of sonic points \citep{vc17}.}
\label{lab:P2_fig3}
 \end{center}
\end{figure}
Inducting equation (\ref{con2a.eq}) into (\ref{jetgeom.eq}), 
we get a spherically outflowing jet, which, for $C=0$ gives a constant 
$\theta(=\theta_0=11^\circ)$.
Keeping $\xi=1$, each curve for model M1 is plotted 
for $E=1.004$, $E=1.064$ 
and $E=1.525$ and are shown in Figs. (\ref{lab:P2_fig3}a-d). For higher values of $E$, the jet terminal speed is also higher (Fig. \ref{lab:P2_fig3}a). Higher $E$ also produces hotter flow. So at any given $r$, $\Gamma$ is lesser for higher E (Fig. \ref{lab:P2_fig3}b).
Since the jet is smooth and adiabatic, so $E$ (Fig. \ref{lab:P2_fig3}c) and $\md$ (Fig. \ref{lab:P2_fig3}d)
remain constant.
The variable nature of $\Gamma$ is clearly shown in Fig. (\ref{lab:P2_fig3}b),
which starts from a value slightly above $4/3$ (hot base) and at large distance it approaches 
$5/3$, as the jet gets colder. 
As discussed in section \ref{sbsec3.1}, for all possible 
parameters, this geometry gives smooth solutions with
only single sonic point until and unless M1 jet interacts with the ambient medium.
%($\bf{Simulation*paper?}$).\\

\subsubsection{Model M2 : Non-radial Jets}
\label{sec:m2}
\begin{figure}
\begin{center}
 \includegraphics[trim={5.5cm 0.4cm 8.5cm 0.5cm},width=5.5cm]{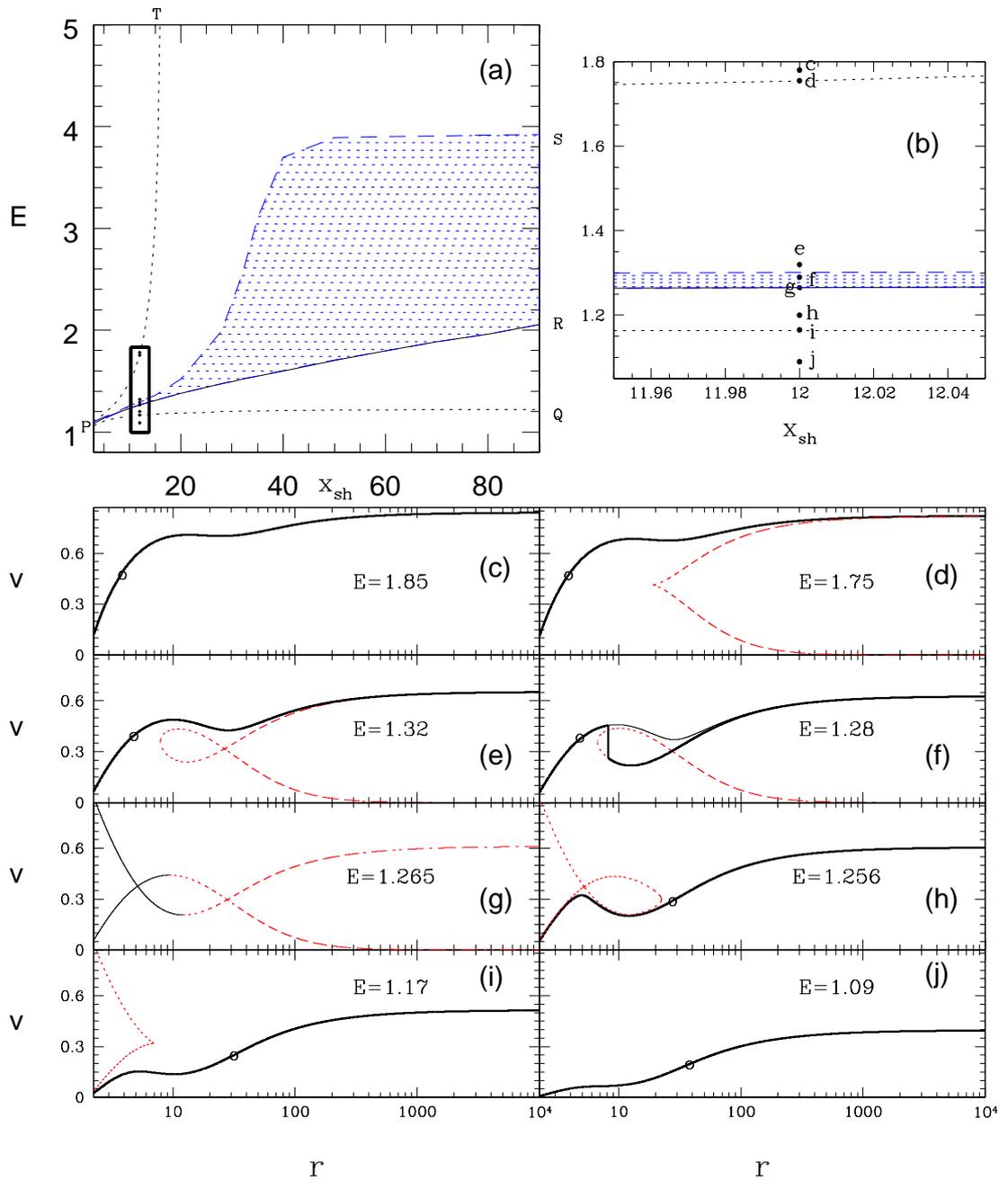}%{zc-var-1.eps}
 \caption{(a) $E$---$\xsh$ parameter space for $\xi=1$. PQRSTP is the region of multiple sonic point.
PR is the locus of points on which the two sonic points have same entropy. Shaded region PRSP permits stable shock transition. (b) Zoomed
parameter space, which shows positions `c'-`j' on $\xsh=12$ for which all typical solutions are
plotted in the panels {(c)-(j)}. Circles on global solutions (solid) and as crossings in closed solutions (dashed)
are the location of sonic points \citep{vc17}.}
% (1.256,1.32,1.30,1.28,1.27,1.256)
\label{lab:P2_fig4}
 \end{center}
\end{figure}

In this section, we discuss all possible solutions associated with the jet model M2.
The sonic point analysis of the jet model M2 showed that for larger values of $\xsh$,
multiple sonic point may form in jets for a larger range of $E$ (Fig. \ref{lab:P2_fig2_2}c). {The existence of multiple sonic points may lead to a shock transition between the solutions passing through inner and outer sonic points. We check for the Rankine-Hugoniot shock conditions in the jet, which are obtained by conserving the mass, momentum and energy fluxes across the shock front (see section \ref{Shock_conditions})}. In Fig. (\ref{lab:P2_fig4}a),
we plot the multiple sonic point region (PQRSTP) bounded by the dotted line. Dotted lines are same as those on Fig. (\ref{lab:P2_fig2_2}c),
obtained by connecting the maxima and minima of $E$ versus $r_c$ plot. 
Jets with all $E$, $\xsh$ values within PQRSTP will harbour multiple sonic points, which is similar to
the bounded region of the energy-angular momentum space for accretion disc \citep[see, Fig. 4 of][]{ck16}.
The central solid line (PR) is the set of all $E,~\xsh$ which harbour three sonic points, but the entropy is same for both the 
inner and outer sonic points. In region QPRQ, the entropy of the inner sonic
points is higher than the outer sonic point and in TPRST, it is vice versa. The zoomed part of the parameter space around $\xsh=12$ is shown in (\ref{lab:P2_fig4}b) and marked locations
from `c' to `j'. The solutions corresponding to the values of $E$ (marked in Fig.\ref{lab:P2_fig4}b) and $\xsh=12$ are plotted in panels
Fig. (\ref{lab:P2_fig4}c---j). For higher energies $E=1.85$ ($>1.75$ left side of PT), only one X-type sonic point (circle) is possible close to the
BH (Fig. \ref{lab:P2_fig4}c). Due to stronger thermal driving the jet
accelerates and becomes transonic close to the
BH. For a slightly lower $E(=1.75)$, there are two sonic points, the solution (solid) through the inner one (shown by a small circle)
is a global solution, while the second solution (dashed) terminates at the outer sonic point (crossing point) is not global
(Fig. \ref{lab:P2_fig4}d).  For lower $E~(=1.32)$, the solution through outer sonic point is $\alpha$ type
(dashed) and has higher entropy. An $\alpha$ type solution is the one which makes a closed loop
at $r < r_c$ and has one subsonic and another supersonic branch starting from $r_c$ outwards extending up to infinity (Fig. \ref{lab:P2_fig4}d).
The jet matter starting from the base,
can only flow out through the inner sonic point (solid), but cannot jump onto the higher entropy solution because
shock conditions are not satisfied (Fig. \ref{lab:P2_fig4}e). However, for $E=1.28$, the entropy difference between
inner and outer sonic points is exactly such that, the matter through the inner sonic point jumps onto the solution through outer
sonic point at the jet shock or $\rsh$ (Fig. \ref{lab:P2_fig4}f).
Solution for $E=1.265$ is on PR and produces inner and outer sonic points with the same entropy
(Fig. \ref{lab:P2_fig4}g).
Figures (\ref{lab:P2_fig4}c ---g) are parameters lying in region TPRST. For flows with even
lower energy $E=1.256$, the entropy condition of the two physical
sonic points reverses. In this case, the entropy of the inner sonic point is higher than the outer one. So, although multiple sonic points exist but no shock in jet
is possible (Fig. \ref{lab:P2_fig4}h) and the jet flows out through the outer sonic point.
In Fig. (\ref{lab:P2_fig4}i), the energy is $E=1.17$ and the solution is almost the mirror image of Fig. (\ref{lab:P2_fig4}d).
Figures (\ref{lab:P2_fig4}h, i) belong to QPRQ region. For even lower energy i. e.,
$E=1.09$ a much weaker jet flows out through the
only sonic point available at a larger distance from the compact object (Fig. \ref{lab:P2_fig4}j).
\begin{figure}
\begin{center}
 \includegraphics[width=10cm]{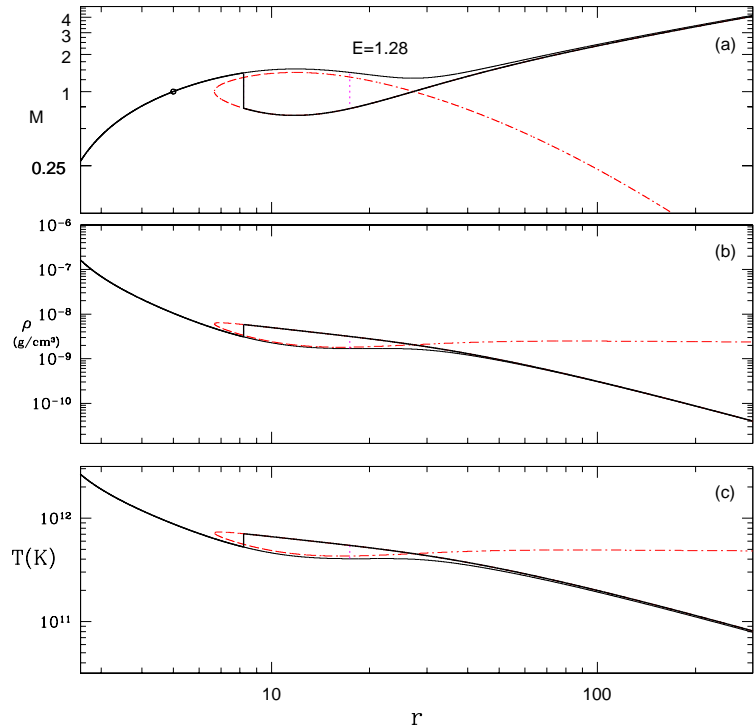}%{shock_analysis.eps}
 \caption{Variation of Mach number $M$ (a), $\rho$ (b) and $T$ (c) of a shocked jet solution. The density profile is plotted
 by assuming $M_{\rm BH}=10M_\odot$ and ${\dot M}=0.01{\dot M}_{\rm Edd}$. All the plots are obtained for $E=1.28$ \citep{vc17}.}
\label{lab:P2_fig5}
 \end{center}
\end{figure}

\begin{figure}
\begin{center}
 \includegraphics[trim={0 -0.8cm 0 -1cm},clip,width=14.8cm]{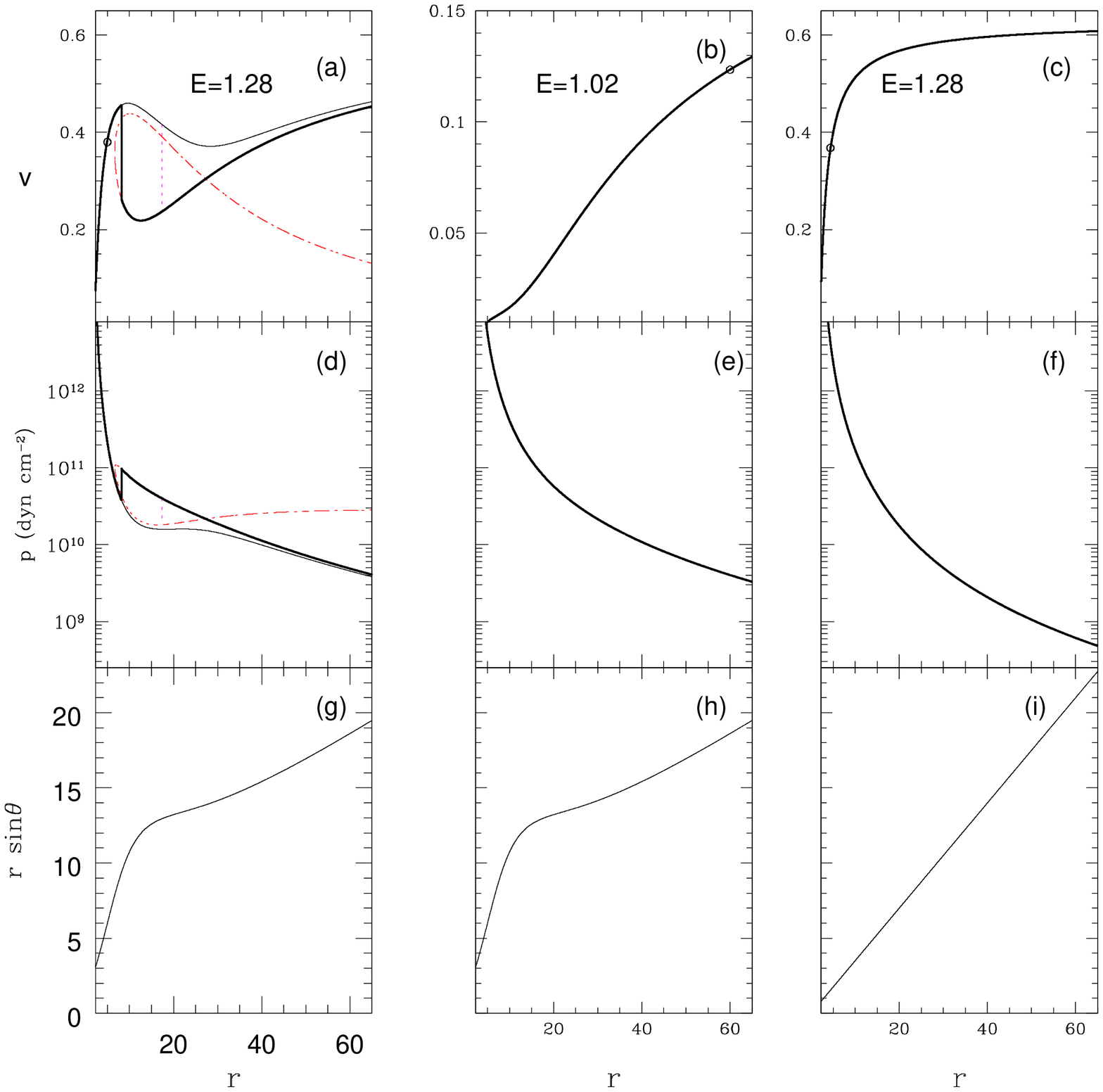}%{shock_analysis.eps}
 \caption{Three velocity $v$ (a, b, c), pressure $p$ in physical units (d, e, f)
 and jet cross section (g, h, i) are plotted
 as a function of $r$. Jet M2 are presented in the left and middle panels.
 Jet M1 is presented in the right panel. Thin-solid curve is a physical solution
jet do not follow due to shock.
Vertical lines show
stable (solid) and unstable (dotted) shocks \citep{vc17}.}
\label{lab:P2_fig6}
 \end{center}
\end{figure}

In Figs. \ref{lab:P2_fig5}a-c, we plot the solution of the inner region of a shocked jet, where the inner boundary
values are clearly seen for a particular solution. The solid curve represents the physical solution, and the
dotted curve is a multi valued solution which can only be accessed in presence of a jet shock.
The base temperature and density are quite similar to the ones obtained in advective accretion disc solutions.

%%\begin{figure}
%%\begin{center}
%% \includegraphics[width=13cm]{./P2/fig6.eps}%{shock_analysis.eps}
%% \caption{Three velocity $v$ (a, b, c), pressure $p$ in physical units (d, e, f)
%% and jet cross section (g, h, i) are plotted
%% as a function of $r$. Jet M2 are presented in the left and middle panels.
%% Jet M1 is presented in the right panel. Thin-solid curve is a physical solution
%%jet do not follow due to shock.
%%Vertical lines show
%%stable (solid) and unstable (dotted) shocks.}
%%\label{lab:P2_fig6}
%% \end{center}
%%\end{figure}

In the literature, many authors have studied shock in accretion discs \citep{f87,c89,cc11,ck16} and
the phenomena 
have long been identified as the result of centrifugal barrier developing in the
accreting flow. Shocks may develop in jets due to the interaction with the ambient medium,
or inherent fluctuation of injection speed of the jet.
But why would internal shock develop in steady jet flow, where the role of angular momentum
is either absent or very weak?
In Fig. \ref{lab:P2_fig6}(a), 
we plot velocity profile of 
a shocked jet solution
with parameters 
$E=1.28$, $\xi=1.0$ and $x_{\rm sh}$=12. 
The jet, 
starting with subsonic speeds at base, 
passes through the sonic point (circle) 
at $r_c=5$ and becomes transonic. 
This sonic point is formed due to gravity 
term (third term in the r. h. s of equation \ref{dvdr.eq})
which is negative and equals other two terms 
at $r_c$ making the r. h. s. of 
equation (\ref{dvdr.eq}) equal to zero.
It is to be remembered that, jets with higher values of $E$ implies
hotter flow at the base, which ensures greater thermal driving which makes
the jet supersonic within few $\rg$ of the base. 
However, once the jet becomes supersonic ($v>a$), it accelerates but within a short distance
beyond the sonic point the jet decelerates (thin, solid line).
This reduction in jet speed occurs due to
the geometry of the flow. In Fig. (\ref{lab:P2_fig6}g),
we have plotted the corresponding cross section of the jet.
The jet rapidly expands in the subsonic regime, but the expansion gets arrested and the expansion
of the jet geometry becomes very small ${\cal A}^{-1}d{\cal A}/dr\sim 0$. Therefore the positive contribution
in the r. h. s of equation (\ref{dvdr.eq}) reduces significantly which makes $dv/dr\leq 0$. Thus
the flow is decelerated resulting in higher pressure down stream
(thin solid curve of Fig. \ref{lab:P2_fig6}d). This resistance causes the jet to under go
shock transition at $\rsh=8.21$. The shock condition is also satisfied at $\rsh=17.4$,
however this outer shock can be shown to be unstable \citep[see, Appendix A, \citet{vc17} and also][]{n96,yk95,ydl96}.
We now compare the shocked M2 jet in Fig. (\ref{lab:P2_fig6}a, d, g) with two other jet flows,
(i) a jet of model M2 but with low energy $E=1.02$ (Fig. \ref{lab:P2_fig6}b, e, h); and
(ii) a jet of model M1 and with the same energy $E=1.28$ (Fig. \ref{lab:P2_fig6}c, f, i).
In the middle panels, $E=1.02$ and therefore the jet is much colder. Reduced thermal driving causes
the sonic point to form at large distance (open circle in Fig. \ref{lab:P2_fig6}b).
The large variations in the fractional gradient of ${\cal A}$
occurs well within $r_c$. At $r>r_c$ ${\cal A}^{-1}d{\cal A}/dr \rightarrow~2/r$, which is
similar to a conical flow. Therefore, the r. h. s of equation (\ref{dvdr.eq}) does not become
negative at $r>r_c$. In other words, flow remains monotonic. The pressure is also a monotonic
function (Fig. \ref{lab:P2_fig6}e) and therefore no shock transition occurs.
In order to complete the comparison, in the panels on the right
(Fig. \ref{lab:P2_fig6}c, f, i),
we plot for jet model of M1, with
the same energy as the shocked one ($E=1.28$). Since fractional variation of
the cross section is monotonic i. e., at $r>r_c$, ${\cal A}^{-1}d{\cal A}/dr=2/r$ (Fig. \ref{lab:P2_fig6}i),
all the jet variables like $v$ (Fig. \ref{lab:P2_fig6}c) and pressure
(Fig. \ref{lab:P2_fig6}f) remain monotonic and no internal shock develops.

Therefore to form such internal shocks in jets, the jet base has to be hot in order to make
it supersonic very close to the base. And then the fractional gradient of
the jet cross section needs to change rapidly, in order to alter the effect of gravity, so that the jet beam starts resisting the matter
following it and form a shock.

\begin{figure}
\begin{center}
 \includegraphics[trim={0 -0.8cm 0 -3cm},clip,width=14cm]{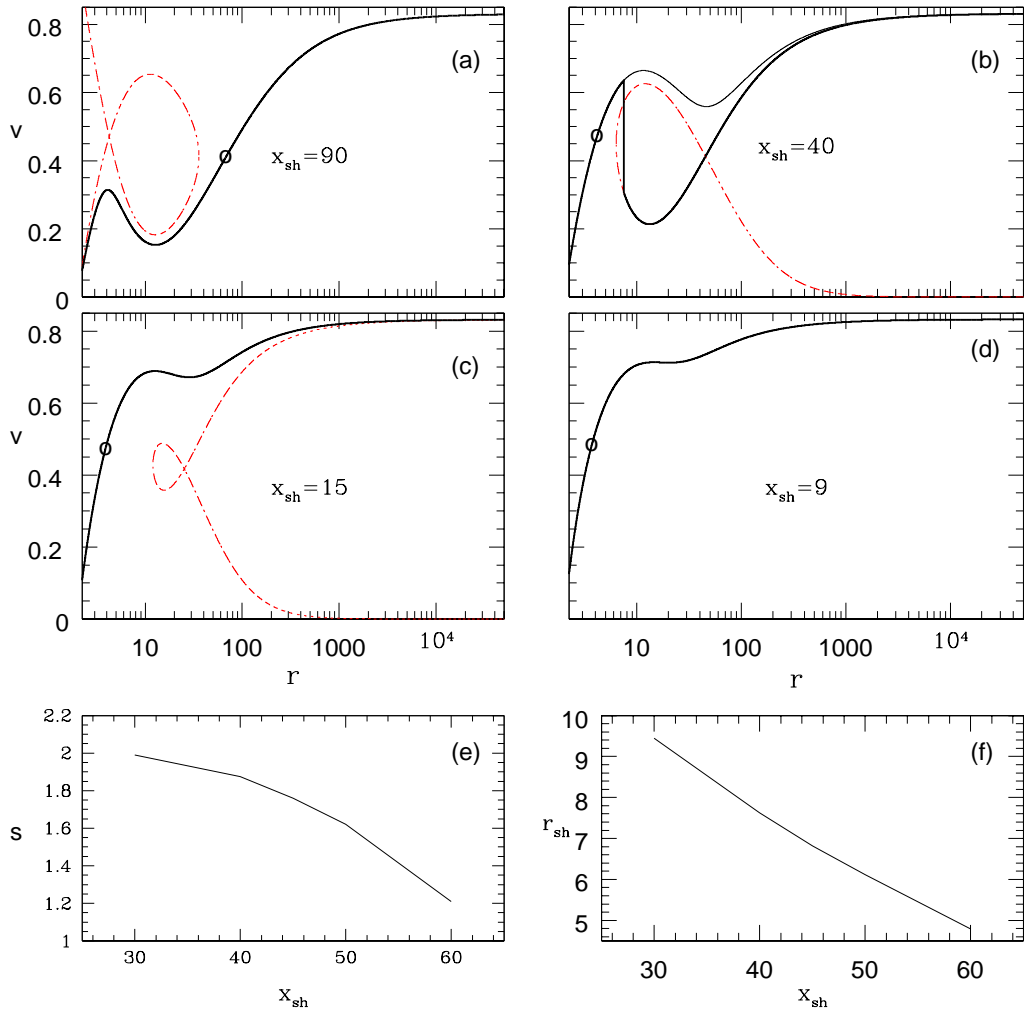}%{shk-var-1.eps}
\caption{Variation of $v$ along $r$ for different values of $x_{\rm sh}$ 
as labeled in (a)-(d). The dotted (red) curves are closed solutions,
while thin black solid curves are
possible transonic solutions. 
Vertical line show stable shock 
transition (Thick solid). Thick solid curves are the realistic 
transonic solutions of the jet;
(e) Shock strength $S$ as a function of  
$\xsh$; and (d) $\rsh$ 
as a function of $\xsh$. Here $E=1.8$, $\xi=1.0$ \citep{vc17}.}
\label{lab:P2_fig7}
 \end{center}
\end{figure}

\begin{figure}
\begin{center}
 \includegraphics[trim={0 0 0cm 0cm},clip,width=16cm, height=16.cm]{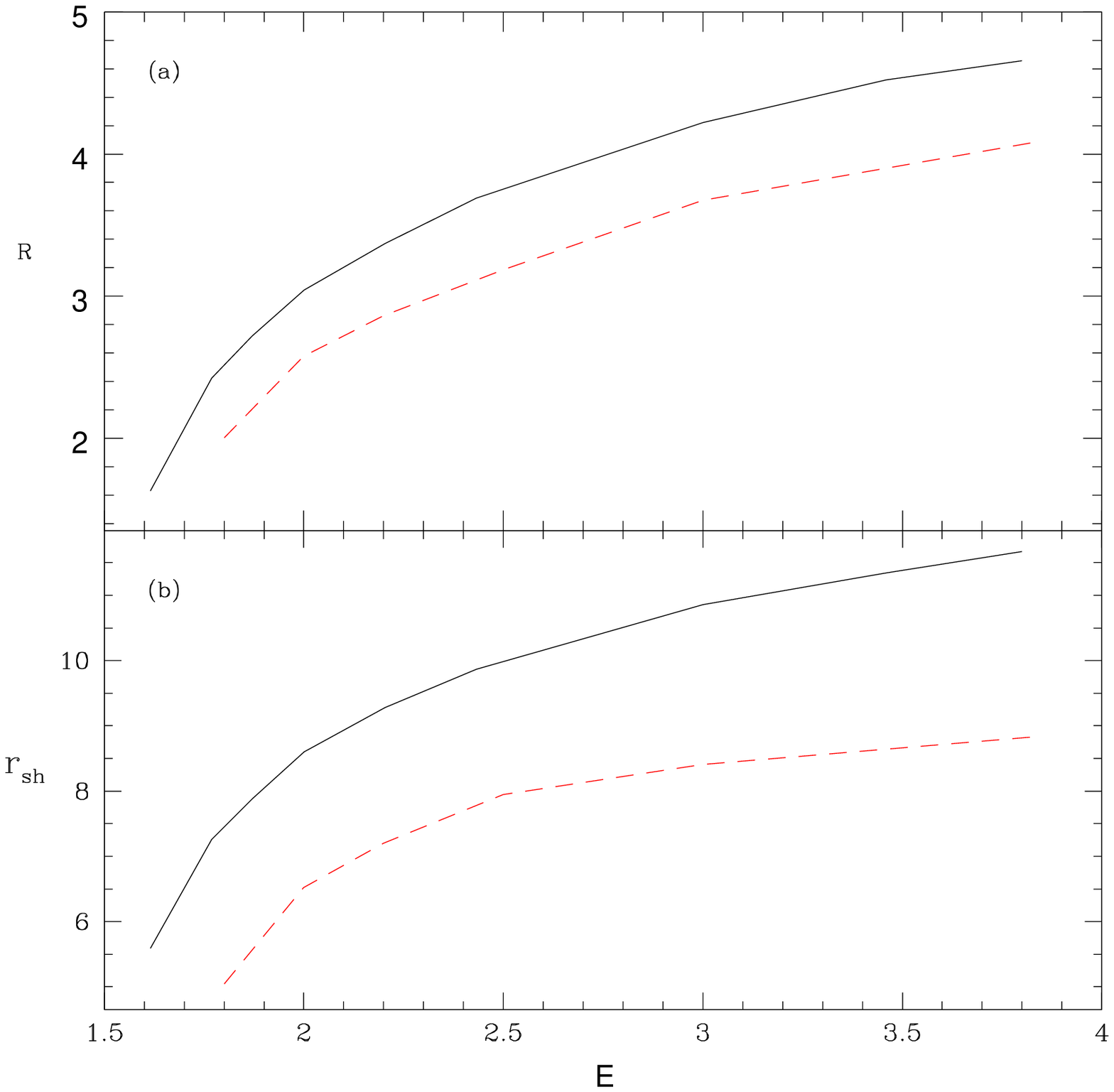}%{term_zc-var-1.eps}
 \caption{{Compression ratio ($R$)} (a) and {shock location ($\rsh$)} (b) as functions 
 of $E$ for $\xsh=40$ (solid) and 
 $\xsh=60$ (dashed, red) \citep{vc17}}
\label{lab:P2_fig8}
 \end{center}
\end{figure}

Figures (\ref{lab:P2_fig6}a-i) showed that departure of the jet
cross section from conical geometry is not enough to drive shock in jet.
It is necessary that the jet becomes transonic at a short distance from the base
and a significant fractional change in jet cross-section occurs in the supersonic regime
of the jet. Since the departure of the jet cross-section from the conical one, depends
on shape of the inner disc, or in other words, the location of the shock in accretion,
we study the effect of accretion disc shock on the jet solution.
We compare jet solutions (i. e., $v$ versus $r$)
for various accretion shock locations for e. g.,
$\xsh=90$ (Fig. \ref{lab:P2_fig7}a), $\xsh=40$ (Fig. \ref{lab:P2_fig7}b), $\xsh=15$
(Fig. \ref{lab:P2_fig7}c) and $\xsh=9$ (Fig. \ref{lab:P2_fig7}d).
In Figs. (\ref{lab:P2_fig4}c-i), all possible solutions were obtained by keeping $\xsh$
constant for different values of $E$.
In Figs. (\ref{lab:P2_fig7}a-d) we show how the jet solutions change for different values of
$\xsh$ but for same $E=1.8$ of the jet. For a large value of $\xsh=90$ (Fig. \ref{lab:P2_fig7}a),
the jet cross section near the base
diverges so much that
the jet looses the forward thrust in the subsonic regime and the sonic
point is formed at large distance. The geometry indeed decelerates
the flow, but being in the subsonic regime such deceleration do not accumulate enough pressure to break the
kinetic energy and therefore no shock is formed. As the expansion of the cross-section is 
arrested, the jet starts to accelerate and eventually becomes transonic at large distance from the BH. At relatively smaller value of $\xsh~(=40)$, the thermal term remains strong enough
to negate gravity and form the sonic point in few $\rg$. For such values of $\xsh$, the fractional
expansion
of the jet cross-section drastically reduces or, ${\cal A}^{-1}d{\cal A}/dr\sim 0$,
when the jet is supersonic. Therefore, in this case
the jet suffers shock (Fig. \ref{lab:P2_fig7}b). In fact, for $E=1.8$ the jet will under go shock transition, if the
accretion disc shock location range is from $\xsh=30$---$60$. 
For even smaller value of accretion shock location $\xsh=15$, because the opening angle of the jet is less,
the thermal driving is comparatively more than the previous case. The jet becomes supersonic at an even shorter distance. The outer sonic point is available, but because the shock condition
is not satisfied, shock does not form in the jet (Fig. \ref{lab:P2_fig7}c).
As the shock in accretion is decreased to $\xsh=9$, the thermal driving is so strong that
it forms only one sonic point, overcoming the influence of the geometry (Fig. \ref{lab:P2_fig7}d).
Although, due to the fractional change in jet geometry, the nature of jet solutions have changed, but
jets launched with same Bernoulli parameter achieves
the same terminal speed independent of any jet geometry. This is because at $r\rightarrow \infty$,
$h\rightarrow h_\infty \rightarrow 1$ $\Rightarrow ~u_t\rightarrow u_{t \infty}\rightarrow \gmt$, so  
$$
E=-hu_t=-h_{\infty}u_{t\infty}= \gmt=(1-v^2_\infty)^{-1/2}.
$$
or,
\begin{equation}
v_\infty=\left(1-\frac{1}{E^2}\right)^{1/2}	.
	\label{vterm.eq}
\end{equation}
In Figs. (\ref{lab:P2_fig7}e, f), the jet shock strength (equation \ref{shokstrnth.eq}) and 
the jet shock location $\rsh$ are plotted as functions of accretion shock location $\xsh$. It can be seen that the jet shock is pushed outward as the corona becomes smaller and subsequently the shock becomes weaker.

In Figs. (\ref{lab:P2_fig8}a, b), the shock compression ratio $R$ (equation \ref{compress.eq}) of the jet and the
jet shock location $\rsh$ as a function of $E$ are plotted. Each curve represents the accretion shock location $\xsh=40$
(solid) and $\xsh=60$ (dashed). It shows that for a given $E$, the jet shock $\rsh$
and strength $S$ decrease
with the increase of $\xsh$. From Fig. (\ref{lab:P2_fig4}a) it is also clear that for larger
values of $\xsh$,
jet shock may form in larger region of the parameter space.
The compression ratio of the
jet is above $3$ in a large part of the parameter space, therefore shock acceleration would be more efficient
at these shocks.
It is interesting to note the contrast in the behaviour of the jet shock $\rsh$ with the accretion disc shock.
In case of accretion discs, the shock strength and the compression ratio increases with decreasing shock location $\xsh$ \citep{kc13,ck16}. But for the shock in jet, the dependence of $R$ and $S$ on $\rsh$
is just the opposite i. e., $R$ and $S$ decreases with decreasing $\rsh$. 

\begin{figure}
\begin{center}
 \includegraphics[trim={0 -2cm 0 2cm},clip,width=14cm]{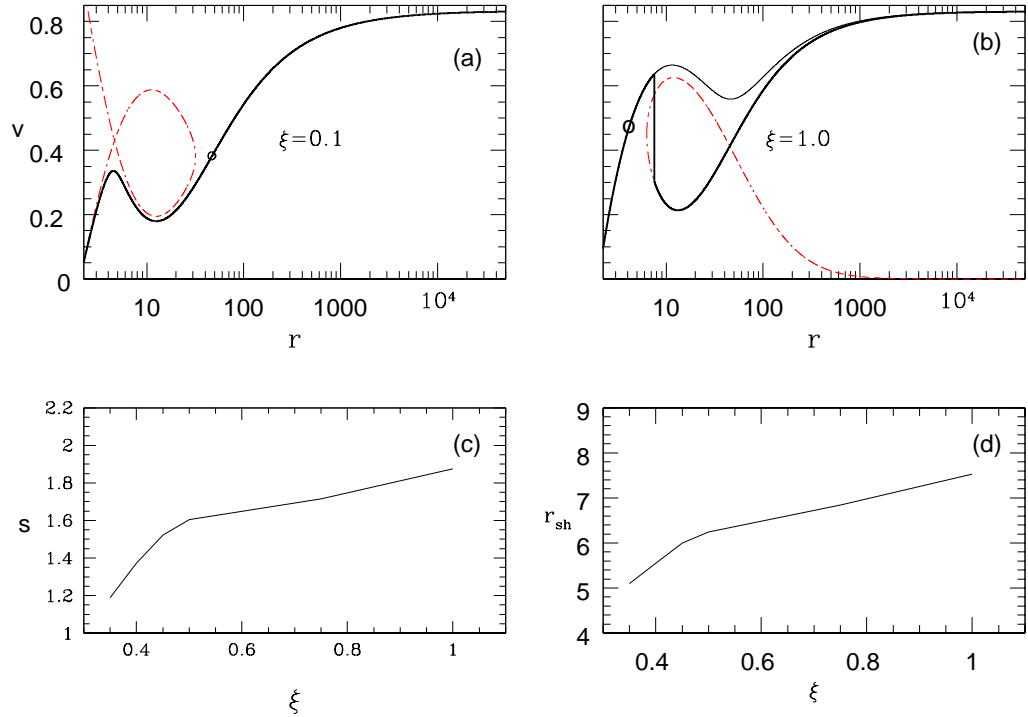}%{xi-var-1.eps}
 \caption{Variation of $v$ with $r$ for different values of $\xi=0.1$ (a) and $\xi=1.0$ (b), also
 labeled on the plots. Closed (dash) and global (solid) solutions. Thick  solid curves show the transonic physical trajectory 
 of the jet and the vertical line is the shock;
(c) $S$ and (d) $\rsh$ 
with $\xi$ plotted. For all panels
$x_{\rm sh}=40$ and $E=1.8$ \citep{vc17}.}
\label{lab:P2_fig9}
 \end{center}
\end{figure}

So far we have only studied the jet properties for $\xi=1.0$ or, electron-proton ($\ep$) flows. In Newtonian flow,
composition would not influence the outcome of the solution if cooling is not present.
But for relativistic flow composition
enters into the expression of the enthalpy and therefore, even in absence of cooling,
jet solutions depend on the composition. In Figs. (\ref{lab:P2_fig9}a) we plot the velocity
profile of a jet whose composition corresponds to $\xi=0.1$ i. e., the proton number density is $10\%$
of the electron number density, where the charge neutrality is restored by the presence of positrons. 
The jet energy is $E=1.8$ and jet geometry is defined by $\xsh=40$. In Fig. (\ref{lab:P2_fig9}b), we plot
velocity profile of an $\ep$ jet ($\xi=1.0$), for the same values of $E$ and $\xsh$.
While $\ep$ jet undergoes shock transition, but for the same parameters $E$ and $\xsh$, for the flow with $\xi=0.1$ composition, there is no shock. In fact, $\xi=0.1$ jet solution is similar to the one with lower energy.
Although, the jet solutions of different $\xi$ are distinctly different at finite $r$, but the terminal
speeds are same as is predicted by equation (\ref{vterm.eq}). For these values of $E$ and $\xsh$,
shock in jets are obtained for $\xi=0.32 \rightarrow 1.0$.
In Fig. (\ref{lab:P2_fig9}c) and (\ref{lab:P2_fig9}d) 
 we plot shock strength $S$ and the shock location
 $\rsh$ respectively, as a function of $\xi$ for the same values of $E$ and $\xsh$. Shock
 produced in heavier jet ($\ep$) is 
 stronger and the shock is located at larger distance form the BH. Figures (\ref{lab:P2_fig9}c-d) 
 again show that the shocks forming farther away from 
 BH are stronger.  
 %\subsection{Jet kinetic power and terminal speeds}

\section{Discussion and Concluding Remarks}
\label{sec5}

We see that while thermally driven radial jets
showed monotonic smooth solutions, but for model M2 (non-radial jets), depending on the jet energy $E$ and the
accretion disc parameter $\xsh$, we obtained very fast and smooth jets flowing out through one
X-type sonic point; and for some combinations of $E~\&~\xsh$, we obtained jets solutions with multiple sonic points and even shocks.
Interestingly, for low values of $\xsh$, the jet geometry of M2  differs slightly from the conical one.
Therefore, the jet shock is obtained in
a very small range of $E$. For higher $\xsh$, the range of $E$ which can harbour jet shocks
also increases. However, at same $E$, the jet shock $\rsh$ is formed at larger distance from the BH, if $\xsh$ is formed closer to the BH. This is very interesting, because in accretion disc
as the shock moves closer to the BH, the shock becomes stronger. In addition, smaller value of $\xsh$
implies higher values of $\rsh$ and higher $\rsh$ means stronger jet shock, leading to possibility of producing intense and high energy radiation field close to the BH.\\
Consideration of radiation driving for intense radiation field surely affects the jet solutions. It accelerates the jets (previous chapter), but in addition radiation drag in presence of an intense and isotropic radiation field may drive shocks, which means even M1 jet may harbour shocks (next chapter). \\
It might be fruitful to explore, whether these internal shocks satisfies some observational features.
One may recall that the charged particles 
oscillates back and forth across a shock with horizontal width $L$
 and in each cycle, its energy keeps increasing. After successive oscillations,
 the particle escapes the shock region with enhanced  energy known as 
e-folding energy and is given by \citep{bi87}
\be
E_T=\frac{L(v_+-v_-)\bsh}{\pi c}\left(\frac{\rm erg}{\rm statC}\right),
\label{E_T.eq}
\ee
where $\bsh$ is the magnetic field at the shock and $L=2GM_Br_{\rm sh}\tan\theta/c^2$.
The typical magnetic field estimates near the horizon vary from $10mG$ \citep{l11} to 
$10^4G$ \citep{kl97}. For $M_B\sim 14\msol$(Cygnus X-1), the e-folding energy for the shock obtained in
Fig. \ref{lab:P2_fig7} is obtained to be {$16$ MeV-$1.6$ TeV} (per electron charge in statC) depending on different magnetic field estimates
mentioned above. The energy associated with high energy tail of {$400$ KeV-$~2$ MeV} in Cygnus X-1 \citep{l11} can easily be explained by the internal jet shocks discussed in this chapter. 
The spectral index of obtained particles is 
 \be 
q=\frac{R}{R-1}
\ee
And for the same set of jet parameters we obtained $R=1.87$ or $q=2.15$,
therefore, even the estimated spectral index of $2.2 \pm 0.4$ of such observational estimates \citep{l11}
can be given a theoretical basis.
%\end{itemize}
 
%\begin{figure}
%\begin{center}
% \includegraphics[trim={0 0 0 0cm},clip,width=14cm]{fig9.eps}%{particle_energies.eps}
% \caption{$E_T$ as a function of $E$ (a), $\xsh$ (b) and $\xi$ (c) for $M_B=10^8M_\odot$ (solid) and $M_B=10M_\odot$ (dot)}
%\label{lab:E_T}
% \end{center}
%\end{figure}

\chapter{Radiatively Driven Relativistic Jets in Curved Space-time}
\label{CH:P3_4}
\section{Overview}
In chapter \ref{CH:P1}, we considered flat metric in special relativistic analysis of radiation driving in jets with Paczy\'nski-Wiita (PW) potential and in chapter \ref{CH:P2}, general relativistic study for thermally driven jets was carried out. Here, we carry out general relativistic study of radiatively driven conical fluid jets around non-rotating black holes.
%The requirement of using general relativistic analysis is explained in Appendix (\ref{app:pw_gr}) where we show that apart from physical incompatibility of PW potential with special relativity, general relativity is required to properly dealing with outflow solutions. Former approach with flat space-time assumption gives greater deviation as compared to curved space consideration.
We apply general relativistic equations of motion in curved space-time around a Schwarzschild black hole for axis-symmetric one-dimensional jet in steady state, plying through the radiation field of the accretion disc. Radiative moments are computed using information of
curved space-time. Thick disks are considered with $\Hsh=2.5\xsh$. All required relativistic transformations are implemented on radiation field. We use the methods laid down by  \citet{b02,bgjs15} to incorporate effects of photon bending in computing radiative moments. This work is published in \citet{vc18a}, \citep{vc18b} considering fixed $\Gamma$ EoS and variable $\Gamma$ relativistic EoS respectively. Also see \cite{vc18c1,vc18d1, vc19b1}.
% The terminal speed of the jet depends on how much thermal energy is converted into jet momentum
% and how much radiation momentum is deposited onto the jet. 
% Many classes of jet solutions with single sonic points, multiple sonic points, as well as
%those having radiation driven internal shocks are obtained. Variation of all flow variables along the jet-axis has been studied.
%Highly energetic electron-proton jets can be accelerated by intense radiation to terminal Lorentz factors $\gamt\sim 3$. Moderate terminal speed $v_{\rm \small T} \sim 0.5$ is obtained for moderately luminous discs. Lepton dominated jets may achieve $\gamt \sim 10$.
\section{Equations of Motion}
Under elastic scattering assumption considered here, $\sigma=\sigma_T$ in Eqs. (\ref{eu1con.eq}) and (\ref{dvdr.eq}). As before, right hand side of Eq. (\ref{en1con.eq})  is zero. Further, the cross section considered is radial (${\cal A}\propto r^2$). Following these assumptions, Eqs. (\ref{dvdr.eq}) and (\ref{dthdr.eq}) become :
 \bea
\gamma^2vg^{rr}r^2\left(1-\frac{a^2}{v^2}\right)\frac{dv}{dr}=a^2\left(2r-3\right)-1+\frac{W \wp^r r^2}{\gamma^2}
\label{dvdr3.eq}
,\eea
and
\begin{equation}
\frac{d{\Theta}}{dr}=-\frac{{\Theta}}{N}\left[ \frac{{\gamma}
^2}{v}\left(\frac{dv}{dr}\right)+\frac{2r-3}{r(r-2)}
\right]
\label{dthdr3.eq}
\end{equation}
As this study is carried out using both fixed $\Gamma$ EoS and variable $\Gamma$ relativistic EoS, $W$ depends upon choice of EoS:
\bea
W= (1-Na^2)/m_p{\rm ~~[Fixed~~} \Gamma {~~\rm EoS]~~~~}		\nonumber \\
{\rm ~~ and~~~~~~~~~~~~~~~~~~~~~~~}\nonumber \\
W=(2-\xi)/(f+2\Theta) {\rm ~~[Variable~~} \Gamma {~~\rm EoS]}
\eea

$W$ is a thermal term coupled with radiative term $\wp$. The difference in $W$ comes because of using Eqs. (\ref{eos0.eq}) and (\ref{eos.eq}) for fixed $\Gamma$ and variable $\Gamma$ EoS respectively.
As before, Eqs (\ref{dvdr3.eq}) and (\ref{dthdr3.eq}) are integrated to solve for $v$ and $\Theta$ of a steady jet
plying through the radiation field ($\Im^r$) of the underlying accretion disc.

The last term on the right-hand side of Eq. (\ref{dvdr3.eq}) is the radiation momentum deposition term,
{{
\be
\frad=\frac{\wp^r r^2(2-\xi)}{\tau h \gamma^2}=\frac{\wp^r r^2(2-\xi)}{(f+2 \Theta)\gamma^2}
\label{radterm3.eq}
,\ee
with
$$
\wp^r=\sqrt{g^{rr}}\gamma^3\left[(1+v^2){R_1}-v
\left(g^{rr} R_0+\frac{R_2}{g^{rr}}\right)\right].
$$
}}
%\be
%\frad=\frac{\wp^r r^2(2-\xi)}{(f+2 \Theta)\gamma^2}=\frac{r^2(2-\xi)}%{f+2\Theta}\sqrt{g^{rr}}\gamma\left[(1+v^2){R_1}-v
%\left(g^{rr} R_0+\frac{R_2}{g^{rr}}\right)\right]
%\label{radterm.eq}
%\ee

As before, here we have defined $R_i=\sigma_T{\cal R}_i/m_e$. Equation (\ref{radterm3.eq}) shows that presence of enthalpy in the denominator makes the radiation driving more effective for colder jets. The presence of the metric term $g^{rr}$ in
$R_d$ {{ implies} that gravity also affects radiation driving. 

Within the funnel for a geometrically thick corona, $R_1<0$ as is shown below, and therefore,
within the funnel $\frad <0$ for outward moving jet, that is, $v>0$. But even in regions where $R_1>0$,
$\frad \leq 0$, for any $v\geq v_{\rm eq}$ [also see \cite{cdc04}], where,
\be
v_{\rm eq}=\frac{(g^{rr}R_0+R_2/g^{rr})-\sqrt{(g^{rr}R_0+R_2/g^{rr})^2-4R^2_1}}{2R_1}
\label{equilbmv3.eq}
.\ee
{It is clear from Eq. (\ref{equilbmv3.eq}) that  the effect of radiation drag
is more effective in optically thin medium (radiation penetrates the medium) and for distributed source.
The negative terms in $\frad$
depend on $v$ and hence it is termed as `drag term'.
%\LEt{There appears to be a word missing. `What' is termed as a drag term? }
One may compare the GR version of $v_{\rm eq}$ with the special relativistic and Newtonian versions
\citep{cc02a,cdc04}.

As the system is isentropic, $E$ and $\mdtj$ Eq. (\ref{entacc.eq}) remain conserved along $r$. Expression for relativistic Bernoulli parameter (\ref{energy.eq}) in current assumptions becomes

\be 
\begin{split}
& E=-h u_t{\rm exp}(-X_f),~~\mbox{where,} \\
& X_f=\left(\int dr \frac{\gamma W}{\sqrt{g^{rr}}}\left[(1+v^2){R_1}-v
(g^{rr} R_0+\frac{R_2}{g^{rr}})\right]\right).
\end{split}
\label{energy3.eq}
\ee
\section{Analysis and Results}
\label{sec3}
\subsection{Nature of Radiative Moments}
We generate radiative moments for accretion disc parameters given in table (\ref{tableP3_4}).
\begin{table}
\caption{Disc parameters}
\label{tableP3_4}
\centering
 \begin{tabular}{|c c c c c c c|} 
 \hline
 $\lambda$ & $x_0$ & $\left[\vartheta_{\rm \small D}\right]_{x_0}$ & $\left[\Theta_{\rm \small D}\right]_{x_0}$ & $\theta_D$ & $H_{\rm sh}$ & $d_0$\\ [0.5ex] 
 \hline%\hline
 $1.7$ & $5500 \rs$ & $1.5 \times 10^{-3}$ & $0.2$ & $85^0$ & $2.5\xsh$ & $0.4H_{\rm sh}$\\ 
 \hline
 \end{tabular}
\end{table}
In Figs. (\ref{lab:fig2}a-{c}), we plot radiative energy density $R_0$
, flux $R_1$ , and radiative pressure $R_2$  as functions of $r$.
The components of the radiation field presented in all the panels are for $\dot{m}=10$ which corresponds to a corona of
size $\xsh=12.31$ (see Eq. \ref{xsdotm.eq}). The luminosity of such an accretion disc is $\ell=0.8$ around a BH of $\mbh=10 \msol$.  {  In Fig. (\ref{lab:fig2}a) we plot coronal moments $R_{n\rm \small C}$ (in compact notation) from discs around $\mbh=10 M_\odot$. 
%\begin {figure} %[h]
%%\begin{center}
% \includegraphics[width=7.3 cm, trim=0 0 180 150,clip]{./P3/fig2.eps}%
%%  \includegraphics[width=7.5cm, trim=0 0 100 180,clip, angle=0]{./P3/fig4.eps}
%\vskip -0.2cm
% \caption{Distribution of radiative moments-energy density $R_0$ (long dashed blue line), flux $R_1$ (solid black line) and pressure
% $R_2$ (dashed red line) with $r$ above an accretion disc with $\dot{m}=10$. Radiative
% moments produced by various components of the disc, for example, {(a) from corona $R_{n \rm \small C}$; (b) from outer disc $R_{n \rm \small D}$ and (c) total radiative moments $R_n.$}
%\label{lab:fig2}
%% \end{center}
%\end{figure}

\begin {figure} %[h]
\begin{center}
 \includegraphics[width=13cm, trim=1cm 0 180 150,clip]{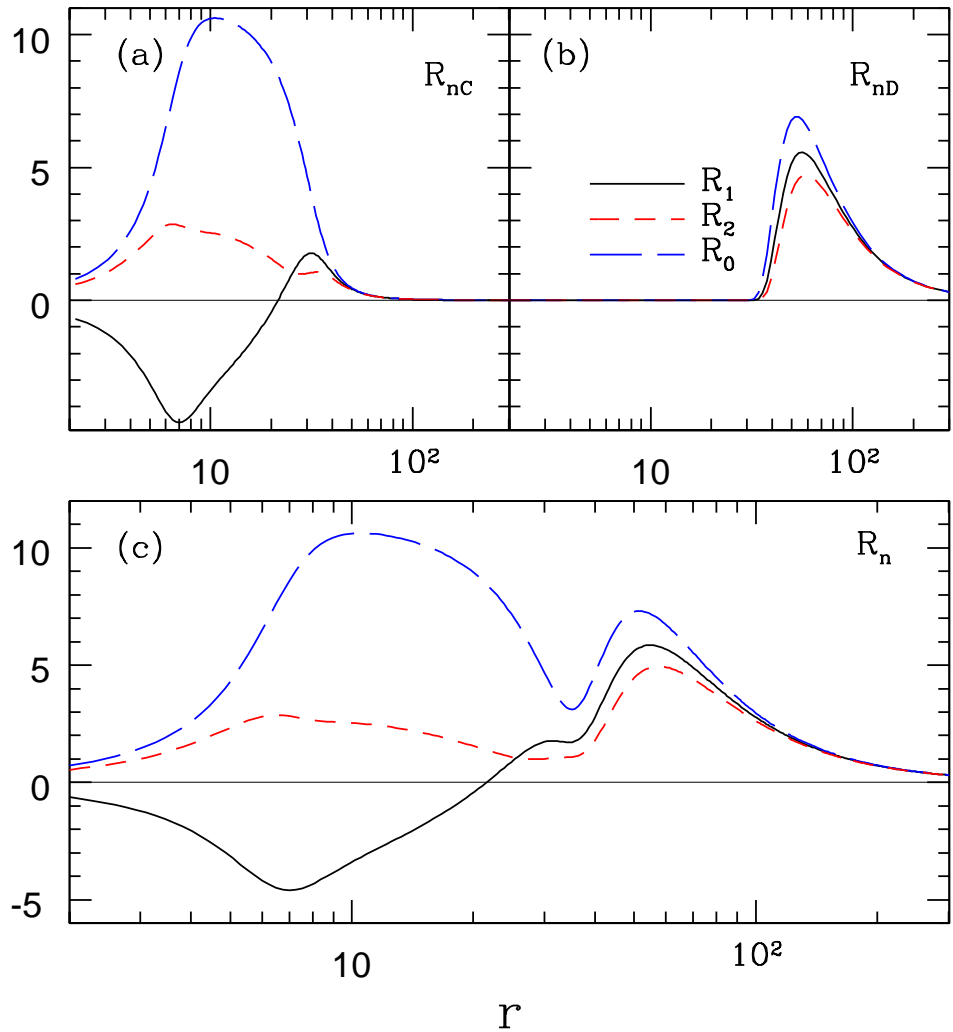}%
\vskip -0.2cm
 \caption{Distribution of radiative moments-energy density $R_0$ (long dashed blue line), flux $R_1$ (solid black line) and pressure
 $R_2$ (dashed red line) with $r$ above an accretion disc with $\dot{m}=10$. Radiative
 moments produced by various components of the disc, for example, (a) from corona $R_{n \rm \small C}$; (b) from outer disc $R_{n \rm \small D}$ and (c) total radiative moments $R_n$ \citep{vc18a}.}
\label{lab:fig2}
 \end{center}
\end{figure}
%%%%%
\begin {figure} %[h]
\begin{center}
  \includegraphics[width=12cm, trim=0 0 150 180,clip, angle=0]{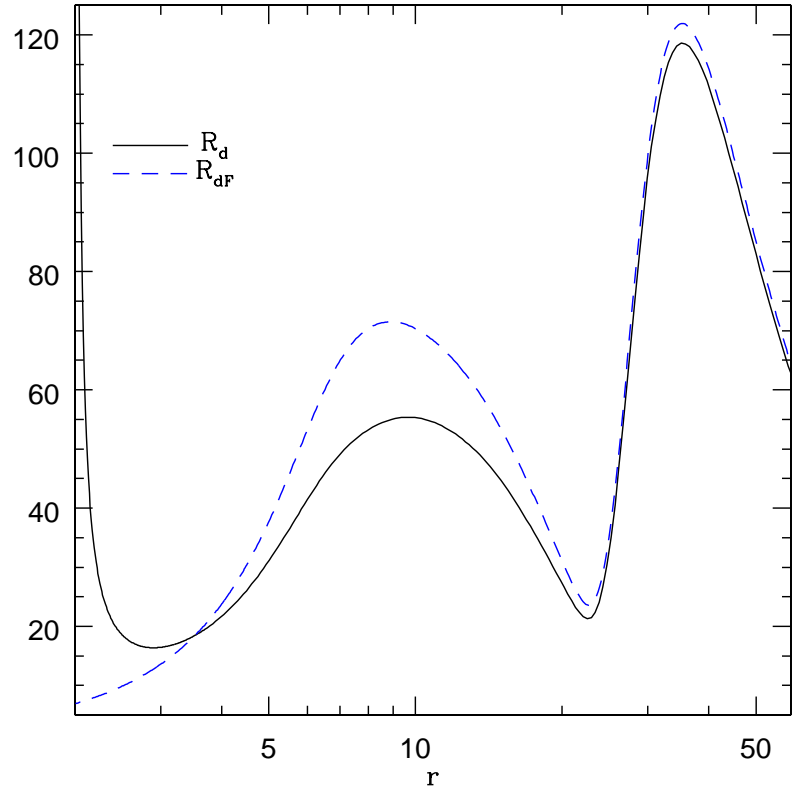}
\vskip -0.2cm
 \caption{Comparison of radiation drag term $R_d$ computed in curved space (solid black line) and flat space (dashed blue line) for the moments shown in Fig. (\ref{lab:fig2}a) \citep{vc18a}}
\label{lab:fig2B}
 \end{center}
\end{figure}
%%%%%%%%%%%%%%%%%%%%%%%%%%%%%%%%%%%
\begin {figure} %[h]
\begin{center}
\vskip -1.5cm
 \includegraphics[width=13.cm, trim=0 0 200 100,clip]{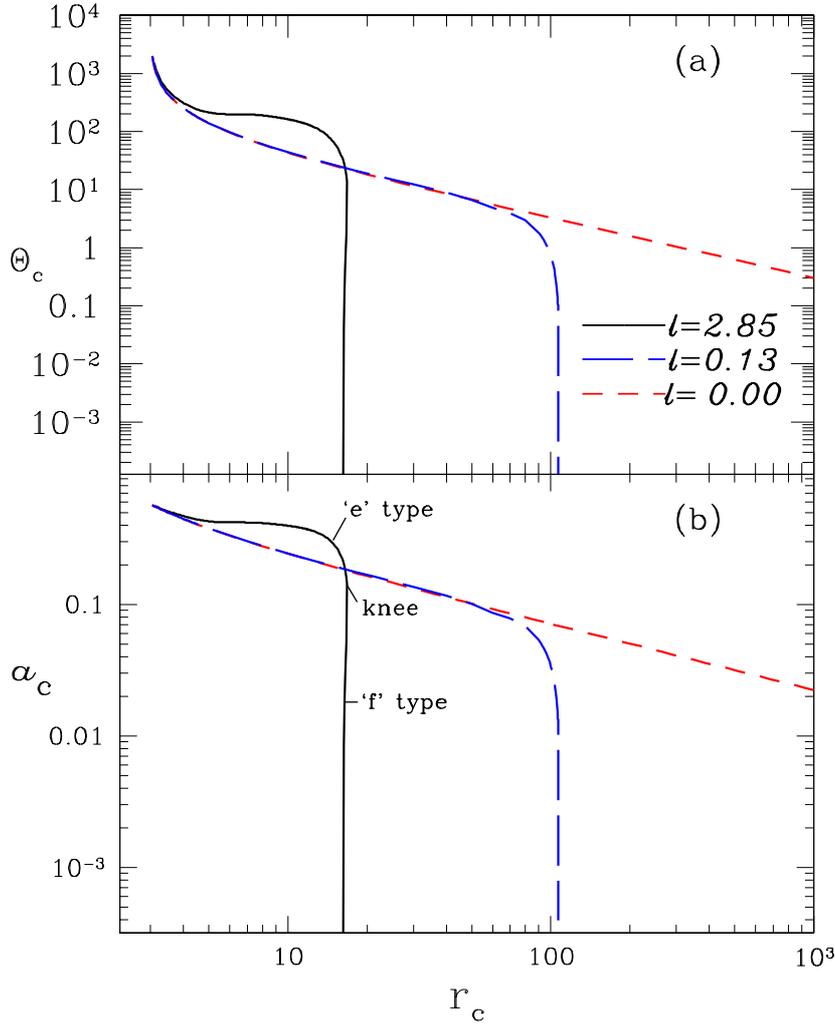}%trim=190 0 0 0,clip
\vskip -0.5cm
\caption{Variation of (a) $\Theta_c$ and (b) $a_c$ with $r_c$ for a jet acted on by $\ell=2.85$
(solid, black), $0.13$ (long dashed, blue) and thermal jet (dashed, red). The jet is composed of electrons and protons ($\xi=1$) \citep{vc18a}.}
\label{lab:fig5n}
 \end{center}
\end{figure}

\begin {figure} %[h]
\begin{center}
\vskip -1.5cm
 \includegraphics[width=11cm, trim=0 0 200 150,clip]{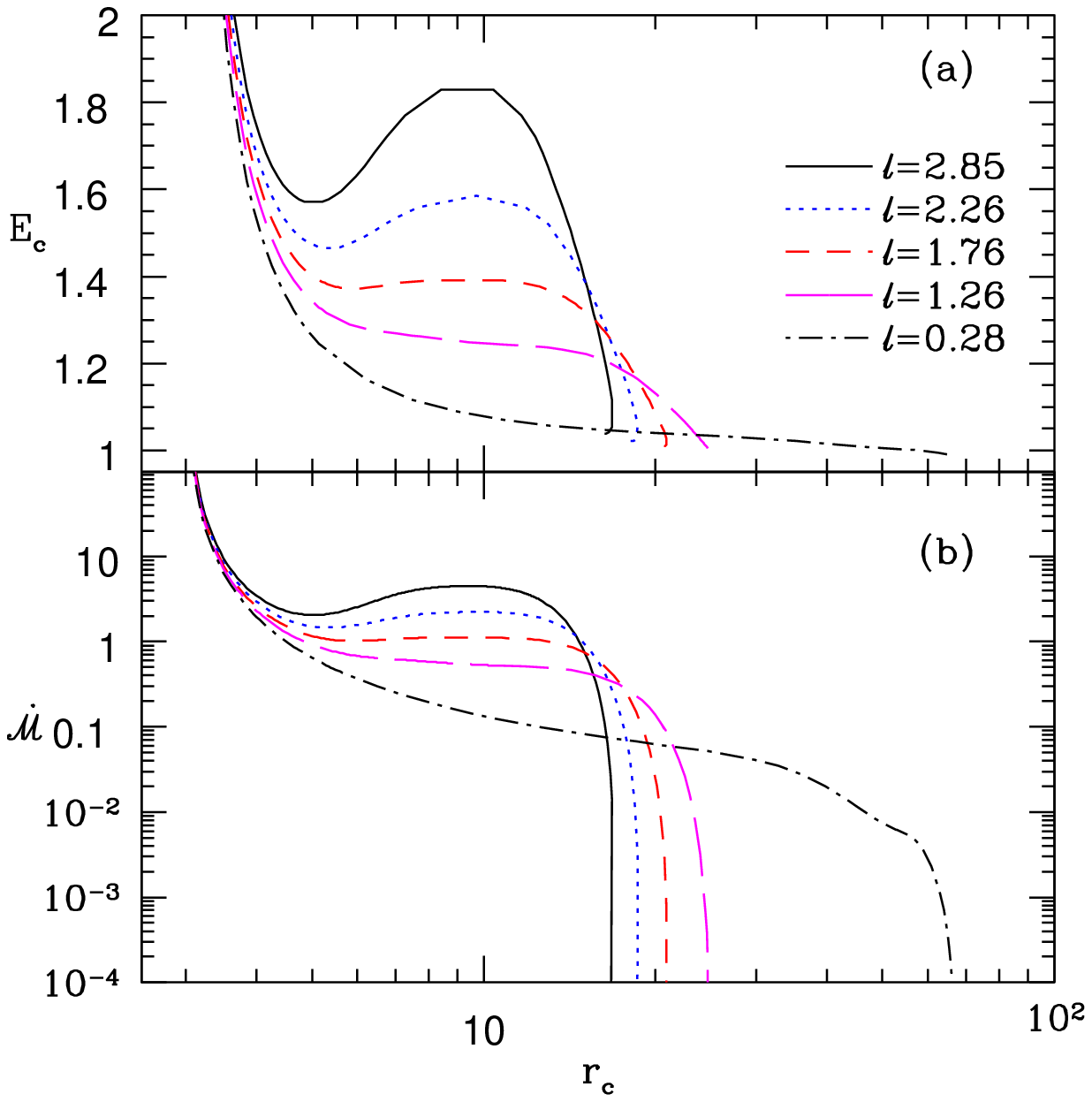}
\vskip -0.5cm
\caption{(a) $E_c$ and (b) ${\dot {\cal M}_c}$ as functions of $r_c$. Various curves represent $\ell=2.85$ (solid, black), $\ell=2.26$ (dotted, blue), $\ell=1.76$ (dashed, red), $\ell=1.26$ (long-dashed, magenta) and $\ell=0.28$ (dash-dotted, black) \citep{vc18a}}
\label{lab:fig6n}
 \end{center}
\end{figure}
%%%%%%%%%%%%%%%%%%%%%%%%%%%%%%%%%%
The moments from the corona dominate the radiation field close to the BH.
Further, because the corona is geometrically thick, the radiation flux ($R_{1\rm \small C}$) is negative
within the funnel like region and is therefore likely to oppose the jet flowing out, along with the
radiation drag terms (negative terms in right-hand side of Eq. \ref{radcontrib.eq}). Figure (\ref{lab:fig2}b) shows moments (presented in compact notation $R_{n\rm \small D}$) from the outer disc.
Because of the shadow effect from the corona, all moments of the outer disc are zero
for $r\leq r_{\rm lim}(=30)$
obtained from Eq. (\ref{shadolim.eq}). The moments of the outer disc for ${\dot m}$
peak around $r=55$.}
In Fig. (\ref{lab:fig2}c), we plot the total radiative moments from the outer disc and the corona.
Far away from the BH ($r >$few$\times 100$), the jet sees the disc like a point source and all moments fall like the inverse square of the distance, and at such distances $R_0 \sim R_1 \sim R_2$.
%In Figs. (\ref{lab:fig2}d), we compare total moments (in compact form $R_n$) computed in curved space time
%and in flat space-time. In presence of space-time curvature,
%a photon has to traverse greater distances than in a flat space-time, therefore leading to decrease in the %magnitude of the moments, as is shown in this panel. Moments with lower magnitude {{ are} computed in curved space
%and have been used in this paper. 
\subsubsection{Effect of Curved Space-time on Radiation Drag}
\label{sec:rad_curved}

The radiation field in chapter \ref{CH:P1}; \citet{vkmc15} was calculated assuming flat space as pNp do not take care of the impact of gravity in radiation fields.
Moments in curved space, $R_{n}$ , are different from that in flat space, $R_{nF}$ , due to the presence of metric components.
The metric components related to the accretion disc coordinates enter
inside the integral while calculating radiative moments (Eq. \ref{moments2.eq}).
The appearance of $n$ as a power in Eq. (\ref{moments2.eq}) shows that the curvature effects are different for different moments. 
Further, the metric component $g^{rr}$ appears inside the radiative term while determining $\wp$. This means that the curvature affects the radiative term
in a very complicated way. In order to quantify the difference curvature has on the radiative terms, we compare the
radiation drag term $R_d=g^{rr} {\cal R}_0+\frac{{\cal R}_2}{g^{rr}}$  in the curved space with its version in the flat space $R_{dF}={\cal R}_{0F}+{\cal R}_{2F}$  in Fig. (\ref{lab:fig2}d), for the similar luminosity.
%($n=0,1,2$ represent ${\cal R}_{0}, {\cal R}_{1}, {\cal R}_{2},$ respectively). As we explained in paper that one contribution comes for transformation of specific intensity (equation 27) which cuts down the radiative contribution, while others appear for transformation of solid angle and direction cosines (equation 28). The cubic term of redshift factor alone says that even at $r=25$ the moments are cut down by 22$\%$. 
The difference is clearly visible. At $r\gsim 2$ the drag term $|R_d|>>|R_{dF}|$, but at $r>3.5$ the curvature
effect changes in an opposite manner, that is, $R_d < R_{dF}$. At $8\sim r \sim 9$, $R_{dF} \sim 1.3R_d$ , which is the maximum deviation from the curved space values. However, the most interesting thing is that even at a distance of about one hundred gravitational radii, the drag term computed in the flat space is about three percent more than that computed in the curved space. In other words, not only does the curvature affect the radiative moments at moderately large distance, but since deviation varies with distance, one cannot use a scale factor to co-opt
%\LEt{please verify that this, co-opt, is a widely accepted term.}
the curvature effect on radiation in flat space.

\subsection{Nature of sonic points}
\label{sec:sonic_points}

For the choice of relativistic EoS, we present $\Theta_c$ (Fig. \ref{lab:fig5n}a) and $a_c$ (Fig. \ref{lab:fig5n}b) as { functions} of $r_c$. 
Each plot represents sonic point properties of jets in a radiation field of an accretion disc
with luminosities $\ell=2.85$ , $\ell=0.13$ , and $\ell=0.0,$ or a thermally driven jet.
%{\LEt{Repetitions of figure legend have been removed where noticed but please verify that all repetitions are taken out of the body text where not required.}}.
Physically, different values of $r_c$ imply different choices of boundary conditions that give different transonic solutions. 
In the absence of radiation, Eq. (\ref{sonic2.eq}) reduces to sonic point condition for thermal jets [$a_c^2=1/(2r_c-3)$]. This implies, for the physical values of $a_c$, that is, $1/{\sqrt3}>a_c>0$, that the range of
sonic points is $3\rg<r_c<\infty$.  
In the presence of radiation, the range of sonic points reduces to $3<r_c<$ $r_l$ ($r_l=~$some finite distance),
%{\LEt{Might it be clearer to use a symbol to express `some finite distance' and then explain the symbol in the following text?.}}%
as shown in
Figs. (\ref{lab:fig5n}a,b). The case with $\ell=0.13$  almost follows the
curve for thermal jets  until about $50\rg$ , where it deviates and terminates at a distance of
$\sim 100\rg$. { The sonic point properties (i.e. $\Theta_c$ and $a_c$) for $\ell=2.85$  are significantly different from the thermal jet  and terminate at $14\rg$.}

%For $\ell=2.85$ the curve deviates from the thermal one in few $\rg$ and terminates  
It is worth mentioning that in chapter \ref{CH:P1}, there were no sonic points in the range $3$ -- $4\rg$. Hence, solutions in the present chapter in which sonic
points are in the range $3\rg<r_c<4\rg$, cannot be found in chapter \ref{CH:P1}
because of incompatibility of pNp to mimic strong gravity close to the BH.
As a result there is enhanced thermal acceleration in all the solutions compared to the analysis. This highlights one of the drawbacks of combining
%\LEt{Please verify that this, gluing, is a widely recognised term.}
special relativistic analysis with Paczy\'nski-Wiita potential. 

The $a_c-r_c$ curve { in Fig. (\ref{lab:fig5n}b) forms} a `knee'-like structure and rapidly decreases such that
at some $r_c\rightarrow r_{c\rm f}$, $a_c \rightarrow 0$. { At the} `knee', $da_c/dr_c \rightarrow \infty$ and the curve
bulges slightly, although not perceptible in the figure. 
Truncation of $r_c$ was also seen in special relativistic \citep{vkmc15} and pseudo-Newtonian studies \citep{cc00a} of radiatively driven jets. The estimation of $r_{c\rm f}$ can be obtained from sonic point condition (Eq.
\ref{sonic2.eq}) by imposing $a_c<<1$,
\be
\frac{(2-\xi)r_{c\rm f}^{3/2}(r_{c \rm f}-2)}{\tau}R_{1c\rm f}=1.
\label{sonic3.f}
\ee
In this chapter, all the solutions corresponding to the sonic points under the `knee' are called `f'-type solutions while solutions above the `knee' are referred to as `e'-type solutions, as marked in Fig. (\ref{lab:fig5n}b).

In Fig. (\ref{lab:fig6n} a and b) we plot $E_c$ and $\mdtjc$ as functions of $r_c$ , respectively. Various curves correspond to $\ell=2.85,$  $\ell=2.26$, $\ell=1.76$, $\ell=1.26$  and $\ell=0.28$. \citet{vc17} showed that for thermal flows with conical jet geometry, $E_c$ and $\mdtjc$ were found to be monotonic functions of $r_c$. Figs. (\ref{lab:fig6n} a-b) show that { $E_c$ and $\mdtjc$ of radiatively driven conical jets are non-monotonic functions of $r_c$}. Above a certain value of $\ell$ (Fig. \ref{lab:fig6n} a), each curve has a maximum and a minimum. For a given
$E=E_c$ and $\ell$ within the two extrema, there is a possibility of forming three sonic points
(for curves with parameters $\ell=2.85,~2.26,~1.76$), where inner and outer sonic points are saddle-type, while middle sonic points are of spiral type. %The spiral nature of the sonic point is evident because of the thermal energy parameter ($-hu_t$) is not conserved along the flow.
Each of the sonic points for a given $E~\&~\ell$ has different entropy ($\mdtjc$).
Similarly, for a given choice of $\mdtj=\mdtjc$ and $\ell$ (Fig. \ref{lab:fig6n} b),
there is a possibility of three sonic points, differentiated by $E_c$.

\subsection{Jet Solutions}
\subsubsection{Solutions for Relativistic EoS with Variable $\Gamma$}
\begin {figure}
\begin{center}
 \includegraphics[width=14cm, trim=0 0 0.5cm 110,clip]{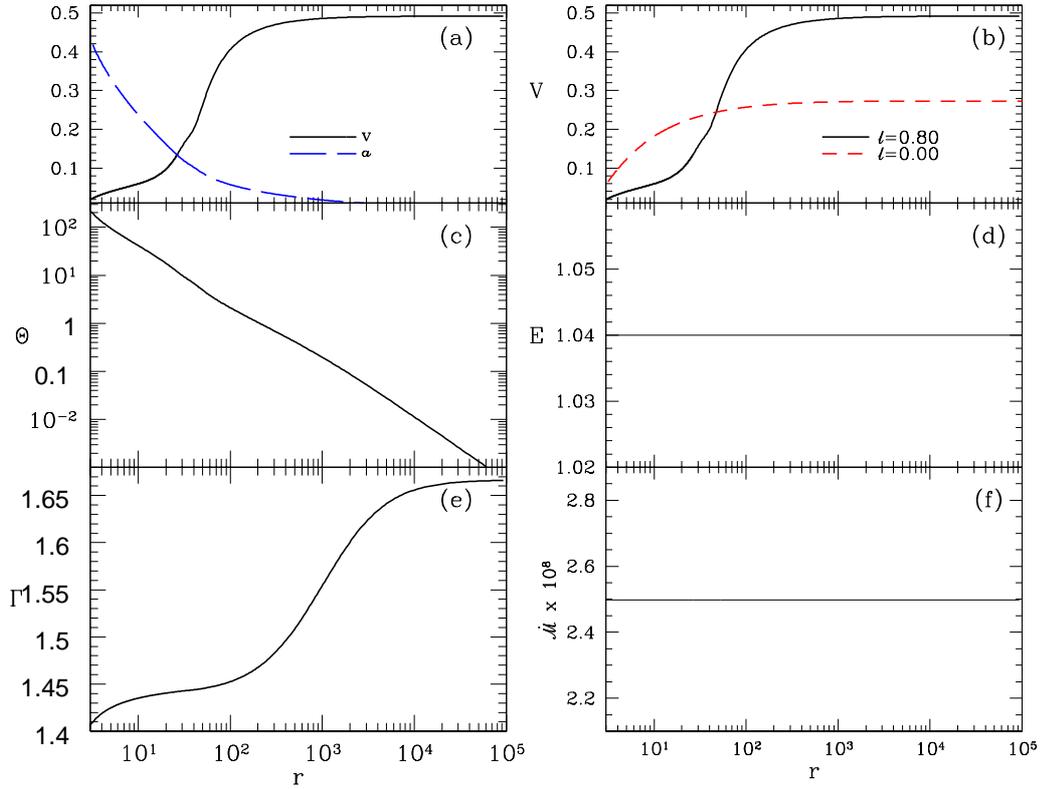}%
%\vskip 0.5cm
 \caption{(a) Three-velocity $v$ (solid black line) and sound speed $a$ (long dashed blue line) as functions of $r$.
(b) Comparison of velocity distribution of thermally driven or $\ell=0$ jet
(dashed red line) and the radiatively driven jet (solid black line);
(c) $\Theta$; (d) $E$; (e) $\Gamma$ and (f) ${\dot {\cal M}}$
as a function of $r$. All the plots are for $E=1.04$ and radiatively driven jet
is for $\ell=0.8$.\citep{vc18a}} 
%\vskip -0.75cm
\label{lab:fig7n}
 \end{center}
\end{figure}

\begin {figure}%[h]
\begin{center}
 \includegraphics[width=15.cm, trim=50 20 0 10,clip]{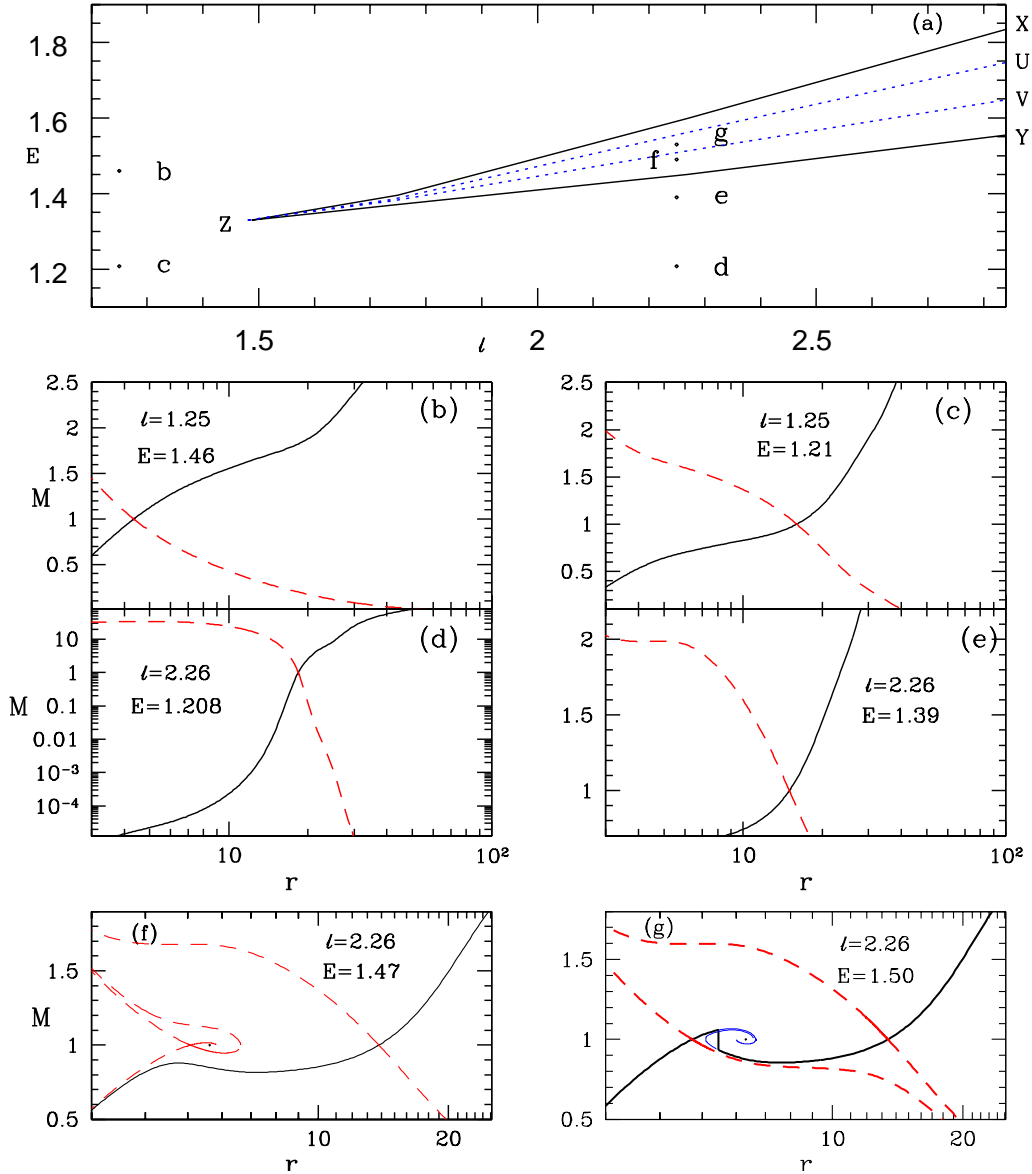}%
\vskip 0.5cm
 \caption{(a) $E-\ell$ parameter space: bounded region XZY signifies parameters for multiple sonic points in jet and region UZV within blue dotted lines represents parameters for which flow goes through shock transition.
 Filled circles named `b-g' are the flow parameters $E$ and $\ell$, for which the jet solutions
 are plotted in panels (b)-(g). Mach number $M=v/a$ is plotted as a function of $r$ for (b) $E=1.46,~\ell=1.25$
 corresponding to point `b' in panel (a); (c) $E=1.208,~\ell=1.25$ corresponding to point `c' in panel a; (d) $E=1.208,~\ell=2.26$
 corresponding to point `d' in panel (a); (e) $E=1.39,~\ell=2.26$ corresponding to point `e' in panel (a); (f) $E=1.47,~\ell=2.26$ corresponding to point `f'
 in panel (a); and (g) $E=1.5,~\ell=2.26$ corresponding to point `g' in panel (a).
 Each panel shows physical jet solutions (solid black line) and
 corresponding inflow solutions (dashed red line). Sonic points are shown by the crossing of inflow and jet
 solutions. All solutions are for $\ep$ flow \citep{vc18a}.
% \LEt{Even with my additions, the clarity of this paragraph could be improved; e.g. the meaning of  "
% or the point `b' in panel (a)", and similar phrases herein,  is not clear. Please consider rewording.}
 } 
\label{lab:fig8n}
 \end{center}
\end{figure}

In Figs. (\ref{lab:fig7n}a-d) we present 
a typical jet solution characterized by generalized Bernoulli parameter $E=1.04$ for $\xi=1$ ($\ep$ flow). In Fig. (\ref{lab:fig7n}a), three-velocity $v$  and sound speed
$a$  are plotted. The jet is transonic, starting with low $v$ and high $a$ and ending with the opposite, respectively. Interestingly, $R_1>0$ for $r>20$ above a disc and the jet starts
to accelerate significantly above that distance. The radiation { field} is for
$\ell=0.8$. 
In Fig. (\ref{lab:fig7n}b), we compare $v$ of a thermally driven jet  with $v$   of a radiatively driven
jet, where $\vt$ of the radiatively driven jet is about twice greater than that of the thermal jet.
The temperature of the { radiatively driven} jet decreases by five orders
of magnitude over a distance scale of five orders of $\rg$ (Fig. \ref{lab:fig7n}c) and consequently $\Gamma$
increases from a relativistic value to a non-relativistic one (Fig. \ref{lab:fig7n}e). The constant of motion
$E$ is plotted in Fig. (\ref{lab:fig7n}d) and since the flow is isentropic, ${\dot {\cal M}}$ is also constant (Fig. \ref{lab:fig7n}f). 

%The $E_c$---$r_c$ relation (Fig. \ref{lab:fig6n}a) show that for a luminous disc,
%the radiation may resist the flow at some regions, but drive the flow in some other.
{ Radiation from a luminous disc resists the jet within some distance above the funnel of the corona, but drives the flow beyond it.} 
As a result, multiple sonic points { are formed in jets} at high $\ell$ { for all values of $E$ within the} maxima and minima
in each of the {$E_c$---$r_c$ curves (Fig. \ref{lab:fig5n}c)}. Therefore, the loci of the maxima and the minima marks the range of
$E$ and $\ell$ {{ 
for which the flow harbours multiple sonic points demarcated by XYZ in Fig. (\ref{lab:fig8n}a).
The region UZV (dotted, blue) represents flow parameters for which a jet has stable shock solution.}}
In Figs. (\ref{lab:fig8n}b-g) we plot the Mach number $M=v/a$ as a function of $r$,
where each panel corresponds to the coordinate points marked as `b'--`g' in $E$---$\ell$ parameter space
in Fig. (\ref{lab:fig8n}a). Here all possible jet solutions are presented, but for the sake of completeness, we have also plotted the inflow solutions. The crossing points denote the locations
of the { X-type} sonic points. If the jet is illuminated by low-luminosity radiation, then it flows out through
only one sonic point (Figs. \ref{lab:fig8n}b and c). If the jet is driven by high-luminosity radiation,
then for lower energies, it will pass through a single outer-type sonic point (Figs. \ref{lab:fig8n}d and e).
However, for higher $\ell$ and $E$, the jet may possess multiple sonic points (Figs. (\ref{lab:fig8n}g and f).
In Fig. (\ref{lab:fig8n}g) { the jet undergoes shock transition}, but in Fig. (\ref{lab:fig8n}f)
it flows out only through the outer sonic point. The inner and outer sonic points are X type
and the middle one is spiral type (Figs. \ref{lab:fig8n}f and g).
Figure (\ref{lab:fig8n}d) is of special importance,
since these are `f'-type jets which {{ start}} with very low velocities but {{ achieve}} relativistic terminal {{speeds}}.

It is interesting to note that {the radiation effect is more perceptible for low-energy jets than for
higher-energy ones. To elaborate, }
we once again invoke the $E_c$---$r_c$ curve in Fig. (\ref{lab:fig9n}a) for jets acted on by three disc luminosities
$\ell=2.85$, $\ell=0.8$  ,
 $0.035$ , and mark three energy values as `b' at $E=2.71$, `c' at $E=1.04$, and
 `d' at $E=1.7$. { We compare the jet solutions at each of these values of $E$ in
 panels (b), (c), and (d) of Fig. (\ref{lab:fig9n}).}
 At high energies {(i.e. Fig. \ref{lab:fig9n}b)}, radiation has no driving power due to the presence of enthalpy
 in the denominator of the radiation term (Eq. \ref{radterm3.eq}). The thermal gradient term
 in such cases is so strong that it accelerates the jet close to its local $v_{\rm eq}$ (Eq. \ref{equilbmv3.eq}).
Therefore, shining radiation will only increase the radiation drag term and reduce the speed, as is seen
in this panel. Near the base, jets for all three $\ell$ achieve almost the same $v$. As the temperature falls
and $\frad$ starts to become effective, jets plying through higher radiation field are slower.
%In  Fig. (\ref{lab:fig9n}c), we plot jets driven by radiation fields of $\ell=2.85$ (solid, black),
%$0.8$ (long dashed, blue) and $0.035$ (dashed, red), but all the jets have same $E=1.04$.
Radiation is
quite effective for low-energy jets {(Fig. \ref{lab:fig9n}c)}. Within the funnel $R_1$ is negative, therefore, the higher the luminosity of the disc, the greater the deceleration of jets inside the funnel. 
But above the funnel where $R_1>0$, radiation from luminous disc will drive
jets to higher terminal speeds. 
For middle energies, for example, $E=1.71$ (Fig. \ref{lab:fig9n}d), the effect of radiation is even more intriguing.
In the presence of low-luminosity radiation field, jets with moderate energies are thermally driven to achieve relativistic terminal speeds which are similar to the value achieved by purely thermally driven jet. Increasing
$\ell$ increases radiation drag and the jet speeds are suppressed, reducing the terminal speed. But for even higher $\ell$, the negative $R_1$ is strong enough to cause a shock transition
in the jet. In the post-shock flow, because $v$ is significantly less than $v_{\rm eq}$, there is significant acceleration and the jet achieves approximately the same terminal speed as the thermally driven jet.
Therefore, for fluid jet, the role of radiation momentum deposition has multiple consequences
with distinctly different outcomes, which underlines the importance of this study.

\begin {figure}[h]
\begin{center}
 \includegraphics[width=16cm, trim=0 5 0 150,clip]{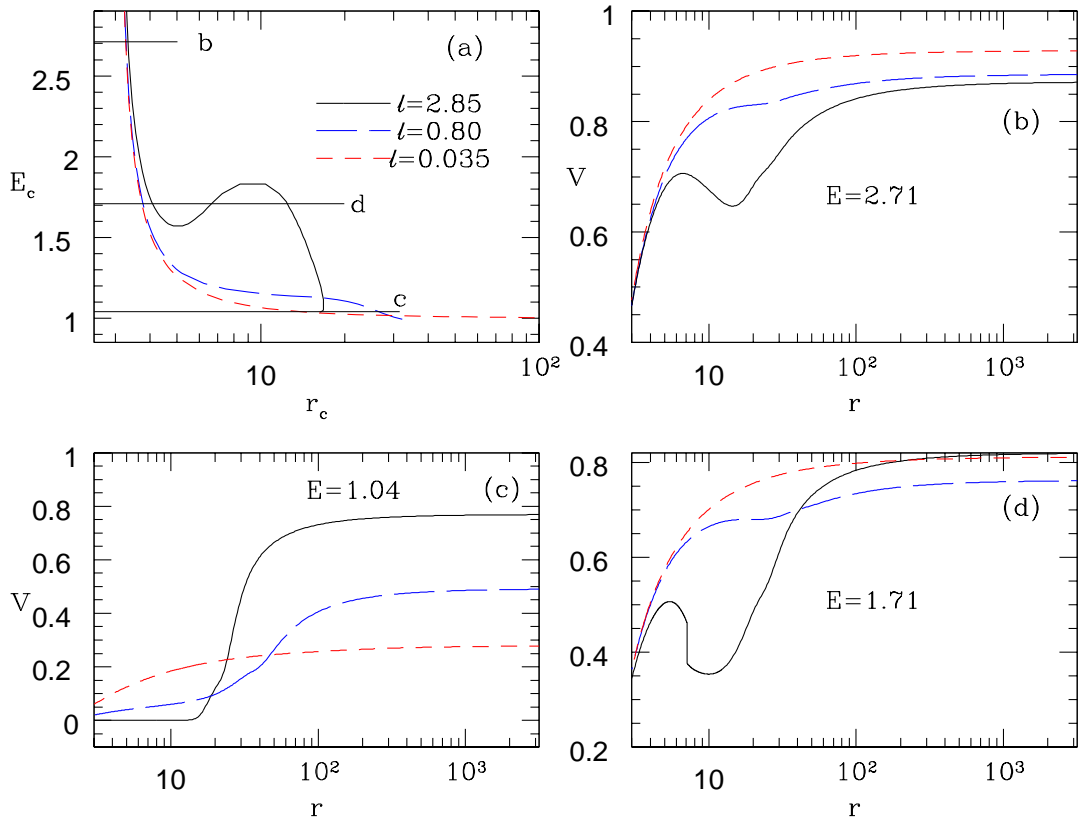}%
\vskip -0.5cm
 \caption{(a) $E_c$---$r_c$ plot with energy levels marked as `b' at $E=2.71$, `c' at $E=1.04$ and 
 `d' at $E=1.7$.
 (b) Comparison of three-velocity $v$ as a function of $r$ of jets starting with $E=2.71$.
 (c) Comparison of $v$ for `f'-type jets with $E=1.04$;
 and,
 (d)  Comparison of $v$ for jets with $E=1.7$.
 Each curve corresponds to $\ell=2.85$ (solid black line), $0.8$ (long dashed blue line) and
 $0.035$ (dashed red line).
 The composition of the jet is $\xi=1$ or $\ep$ \citep{vc18a}.} 
%\vskip -0.75cm
\label{lab:fig9n}
 \end{center}
\end{figure}
The definition of terminal speed or $\vt$ is the asymptotic jet speed, that is,
at $r\rightarrow$large, $v \rightarrow \vt$ where $dv/dr\rightarrow 0$. In Fig. (\ref{lab:fig10n}a),
we plot $\vt$ of jets with $\ell$ for three energies $E=2.71$ , $E=1.71$ 
and $E=1.04$ . For low-energy jets, terminal speed increases with $\ell$ . While for very high-energy jets,
radiation drag is more effective and $\vt$ decreases with $\ell$ .
For moderate values of $E$, radiation
decelerates the jet when $\ell$ is low, but for higher $\ell$, $R_1$ within the funnel opposes the
outflowing jet to such an extent that it triggers a shock transition. In the post-shock jet, $v$ is significantly less than $\veq$ and $R_1>0$, therefore radiation accelerates the jet efficiently to achieve high $\vt$.
In Fig. (\ref{lab:fig10n}b), we plot $\vt$ as a function of $E$, where each curve represents $\ell=2.85$  and $\ell=0.8$. 
{Similar to the previous panel, we find $\vt$ increases with $\ell$ for lower $E$ and
decreases for higher $E$.
It is interesting that for high $E$, $\vt$ is greater for lower $\ell$.}
We also define an amplification parameter A$_{\rm m}=\vt/v_b$ as a measure of acceleration of the jet,
where $v_b$ is the base speed with which the jet is launched. 
In Fig. (\ref{lab:fig10n}c), we plot A$_{\rm m}$ as a function of $E$ for $\ell=0.8$. The dotted part of the
curve represents `f'-type solutions and the solid curve represents `e'-type solutions. It is clear from
the plot of the amplification parameter that
radiation driving is more effective for `f'-type solutions, compared to the `e'-type jets.

Since the jet also contains radiation driven shock, {so we plot the shock location $R_{\rm sh}$ (Fig. \ref{lab:fig10n}d), compression ratio $R$ (Fig. \ref{lab:fig10n}e), and shock strength $S$
(Fig. \ref{lab:fig10n}f)} as a functions of $E$ with each curve plotted for constant values of $\ell$.
The composition of the jet is $\xi=1.0$ and each curve
is for $\ell=2.26$  and $\ell=2.85$. 
{ In general, $R_{\rm sh}$ increases with $E$, because higher $E$ implies higher thermal energy at the base which pushes the shock front outwards. In jets, as the shock moves outwards, the jump condition becomes steeper
and hence the shock becomes stronger. obtained shocks in chapter \ref{CH:P2} \citep{vc17} were consistent with the above fact. However, the crucial difference between \citet{vc17} and the present venture is the agent that drives the shock. In \citet{vc17}, the shock is driven by the geometry of the flow, that is, the non monotonic nature of $\cal A$ is the cause of shock formation. The geometry was coupled with the thermal term \citep[Eq. (17) of][]{vc17}. %and it becomes stronger with
%$E$.
In the present chapter, the shock is driven by the radiation that opposes
the jet flow within the funnel of the disc. In addition, the radiation term $\frad$ is more effective for flows with
lower thermal content, that is, with lower $E$. Therefore, increasing $E$ would negate the effectiveness of radiation,
and would weaken the shock. So $R$ and $S,$ which measure shock strength, initially increase but eventually decrease with increasing $E$, maximizing at some value of $E$ in stark contrast with \citet{vc17}.}
It is also quite clear that for higher $\ell$, the shock generally becomes stronger (long-dashed and solid curves).
\begin {figure}[h]
\begin{center}
 \includegraphics[width=7.5cm, trim=0 0 180 10,clip]{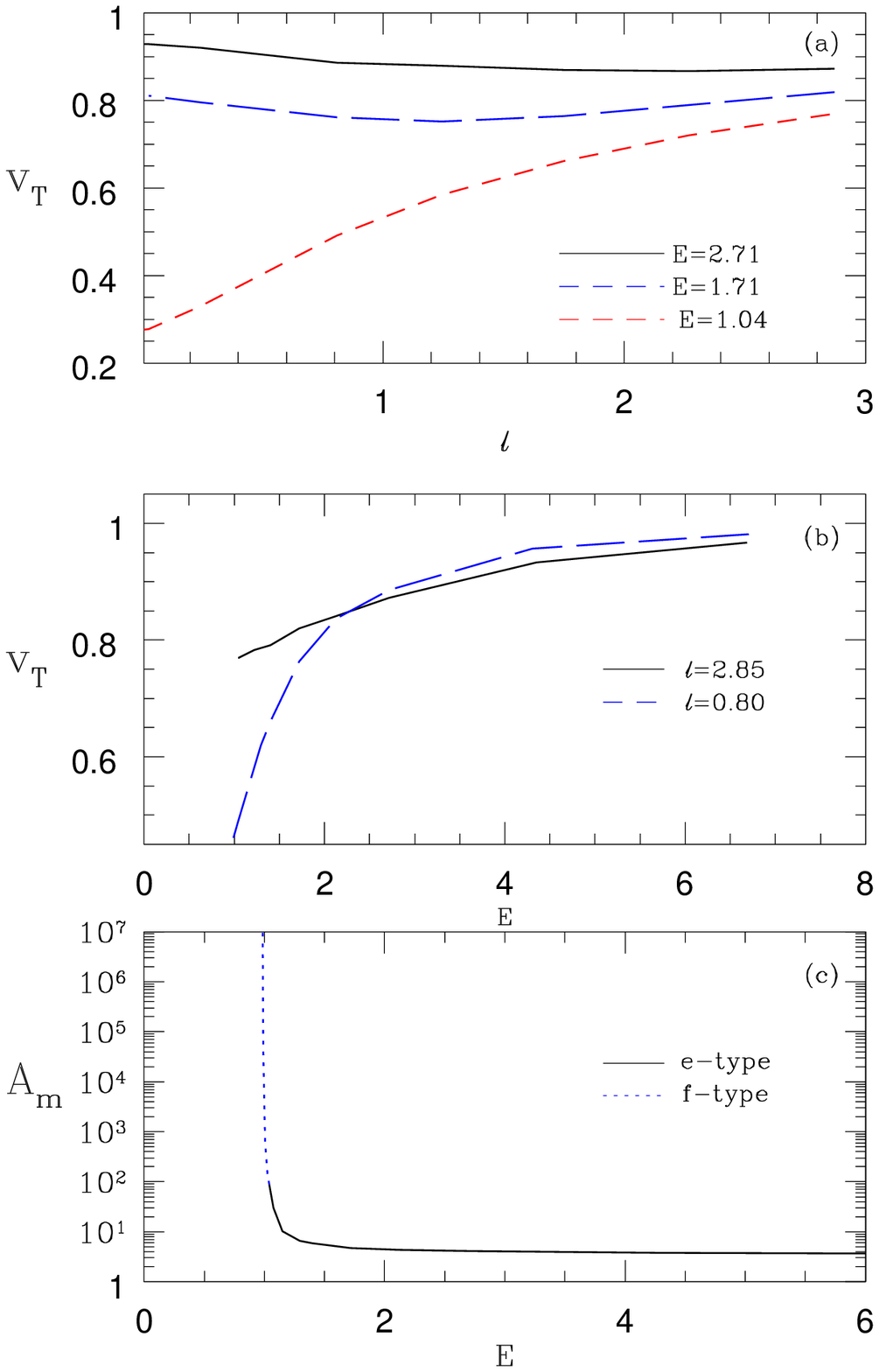}%
  \includegraphics[width=6.65cm, trim=0 0 220 10,clip]{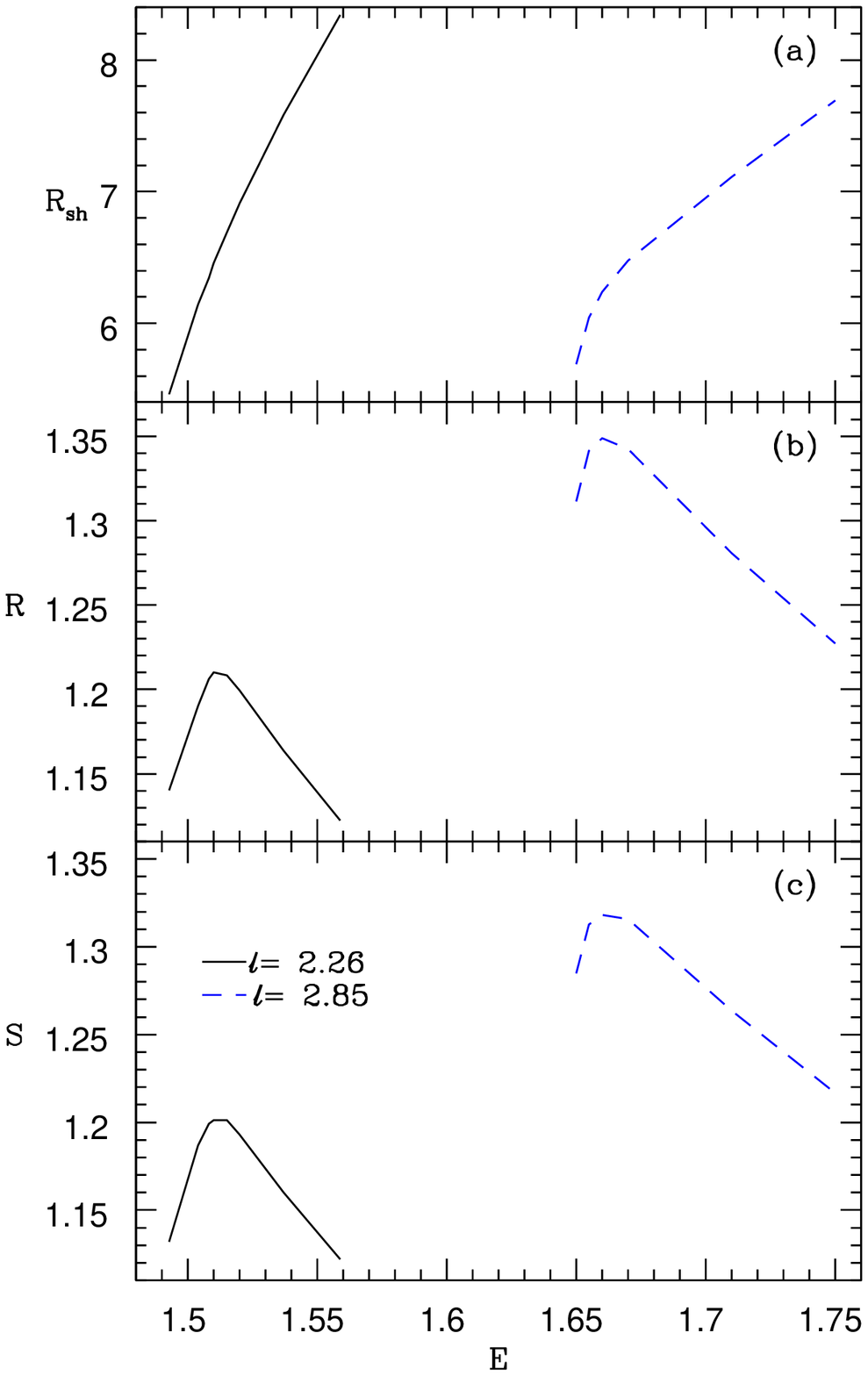}
\vskip -0.5cm
\caption{Left : (a) Variation of $\vt$ with $\ell$. Various curves represent $E=2.71$ (solid, black)
$E=1.71$ (long dashed, blue) and $E=1.04$ (dashed, red).
(b) $\vt$ as a function of $E$ for different $\ell=2.85$ (solid black) and $0.8$ (dashed, red).
(c) Amplification factor A$_{\rm m}$ with $E$ of jets flowing out through a radiation field of $\ell=0.8$. All panels have $\xi=1$. The `e'-type (solid) and `f'-type (dotted) jets are marked too. Right (a) Jet shock location $R_{\rm sh}$; (b) Compression ratio $R$
 and (c) shock strength $S$ as a function of $E$ for jets with composition $\xi=1.0$.
 Each curve is for $\ell=2.26$ (solid) and $\ell=2.85$ (long-dashed) \citep{vc18a}.}
%\vskip -0.75cm
\label{lab:fig10n}
 \end{center}
\end{figure}
\begin {figure}[h]
\begin{center}
 \includegraphics[width=14cm, trim=10 0 0 250,clip]{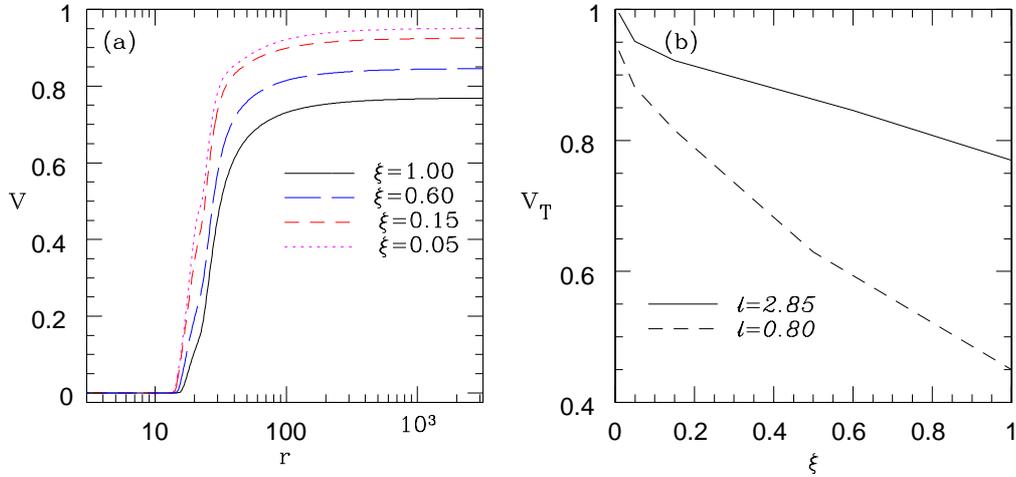}%
\vskip -0.5cm
 \caption{(a) Plot of $v$ as a function of $r$ for jets acted on by disc radiation marked by $\xi=1.0$ (solid black line); $\xi=0.6$ (long-dashed blue line), $\xi=0.15$ (dashed red line) and $\xi=0.05$ (dotted magenta line).
 The disc radiation is for $\ell=2.85$. (b) Variation of $\vt$ with $\xi$ for `f'-type jets, driven by radiation
 quantified by $\ell=2.85$ (solid line) and $\ell=0.8$ (dashed line) \citep{vc18a}.} 
\vskip -0.75cm
\label{lab:fig12n}
 \end{center}
\end{figure}

A closer look into Eq. (\ref{radterm3.eq}) reveals that $\frad$ is twice as large
for $\el$ jets as for $\ep$ jets for the same values of $\Theta$ and $v$.
Lepton-dominated flows have been shown to be colder than $\ep$ flows \citep{cr09,cc11},
which means that the term $f+2\Theta$ is lower for lepton dominated flows.
In other words, $\frad$
will be more effective for lepton-dominated jets. However, one cannot compare jets with same $E$ across a
range of compositions. If one considers Eq. (\ref{energy.eq}), then it can be easily seen that a slight change in $X_f$
will affect the value of $E$ by a large amount. Since for low $\xi$ flow, $\Theta$s are quite different
from those of $\ep$ flow,  {jets with different $\xi$, starting with similar temperature and velocity,
will have widely differing $E$}.
In Fig. (\ref{lab:fig12n}a) we compare jets launched with the same velocity at the base and driven
by radiation of the same luminosity ($\ell=2.85$), each curve corresponds to $\xi=1.0$, $\xi=0.6$, $\xi=0.15$  and $\xi=0.05$ . Jet speeds are higher for flow with lower $\xi$. In Fig. (\ref{lab:fig12n}b), we plot $\vt$ of the jet with the flow composition $\xi$;
each curve corresponds to super-Eddington luminosity ($\ell=2.85$) and sub-Eddington luminosity
($\ell=0.8$). For lepton dominated flow, the terminal speed can easily go above 90\% the speed of light.
%{{
%\subsection{On the nature of shocks}
%The shocks presented in this paper are manifestation of radiation field. The supersonic flow is resisted by %negative flux from the corona and induces shock transition in the jet. Jets in VC17 were thermal in nature. The shocks in VC17 were manifestation jet geometry where pinch off effect in supersonic branch enhances pressure in supersonic region and jet goes through shock transition. VC17 concluded that in absence of radiation, radial jets cannot form shocks. While here we show that even radial jets go through shock transition under the impact of radiation. It is to be noted that these shocks are different in nature as the reported shocks in the jets are either generated due to mechanical interaction between relatively moving plasma blobs or because of interaction of jets with ambient medium.
%The radiation driven shocks obtained in this paper are weaker compared to VC17. 
%}}
\subsubsection{Solutions for Fixed $\Gamma$ EoS}
For fixed $\Gamma$ EoS, we choose $\Gamma=1.5~~(i.e.~~ N=2)$. 
\begin {figure}[h]
\begin{center}
 \includegraphics[width=16cm, trim=10 0 00 230,clip]{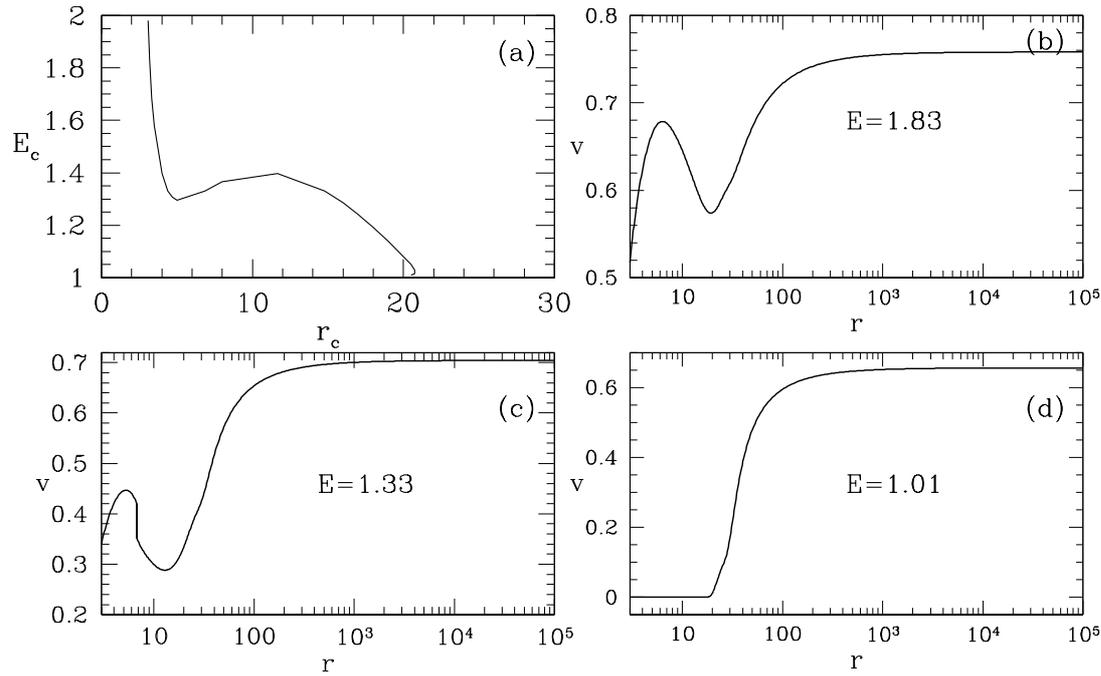}
\vskip -0.5cm
 \caption{(a) Variation of $E_c$ with $r_c$; Variation of $v$ with $r$ for (b) $E=1.83$, (c) $E=1.33$ and
 (d) $E=1.01$. For all the curves, $\ell=1.76$ \citep{vc18b}. }
%\vskip -0.75cm
\label{lab:fig5}
 \end{center}
\end{figure}
In figure (\ref{lab:fig5}a-d) we investigate behaviour of jet speeds with different boundary conditions (or for different choices of $E$). We choose $\ell=1.76$ and plot $E_c$ with $r_c$ in Fig. (\ref{lab:fig5}a). For very high value of $E(=1.83)$ the flow is hot, and radiation is in-effective {\color{blue}(Fig. \ref{lab:fig5}b)}. The jet accelerates due to the thermal gradient term and becomes transonic at $r_c=3.2$. As the jet expands, the temperature decreases and radiation becomes effective. The combined effect of negative flux in the
funnel and radiation drag term decelerates the speed. Above the funnel radiation flux become positive and starts
to accelerate the jet and it achieves terminal speed of $v_{\rm \small T}=0.76$.
If one chooses lower values of {\color{blue}$E=1.33$} (Fig. \ref{lab:fig5}c), the jet passes through inner sonic point at $r=4.4$. Because the energy is low, radiation is more effective. Radiation flux opposes the
outflowing jet inside the funnel more vigorously and causes a shock transition --- a discontinuous transition from supersonic branch to subsonic branch at $r=6.78$ and then after coming out of the accretion funnel, it again accelerates under radiation push and becoming transonic forming an outer sonic point at $r=14.77$. The terminal speeds achieved for this case is $\sim 0.7$. Here the jet crosses two sonic points with $\md$ to be higher for outer sonic point ($\md=0.291$) than the inner one ($\md=0.287$). 
%Vyas \& Chattopadhyay (2017) showed that conical jets without radiation do not form shock, but here we see that radiation field is able to induce shocks in radially outflowing jets. 
For even lower energy ($E=1.01$, Fig. \ref{lab:fig5}d),
the radiation is even more effective, and the jet speed is drastically reduced within the funnel. 
However, above the funnel it is accelerated very efficiently, becomes transonic through a single sonic point
and achieves terminal speed of about $v_{\rm \small T}\sim 0.65$.
 
\subsubsection*{Effect of corona geometry, magnetic pressure in disc ($\beta$) and $\Gamma$}
As we have been considering thick discs with corona height being $2.5$ times its width. One may wonder how the results would behave if discs are less thicker. To compare the effects of different geometries, we choose $\Hsh=0.6\xsh$ as in chapter \ref{CH:P1} \citep{vkmc15} and generate velocity profiles of the jet for $\ell=1.76$ and $E=1.33$ (same parameters as in Fig. \ref{lab:fig5}c). The profiles are plotted in Fig (\ref{lab:fig7}a) for thicker ($\Hsh=2.5\xsh$, solid, black) and thinner corona ($\Hsh=0.6\xsh$, dashed, red). It is clear that radiation from geometrically thick corona is more capable to produce shock, as the jet faces negative flux after being launched. For thinner corona, the radiation resistance is relatively less and the jet is unable to form shock inside the funnel.
Further, lesser resistance inside the funnel of thinner disc, makes radiative acceleration more effective and as a result the terminal speed is greater.

Choice of the value of $\Gamma$ is a tricky issue. This is because the base of the jet
is hot and $\Gamma$ should be closer to, but not exactly $4/3$ (Chattopadhyay \& Ryu, 2009).
And it should be lower than $5/3$, therefore, we took the median value
of $1.5$ in the previous sections. If one considers different values of $\Gamma$, then the behaviour of the jet changes because
by different choice of $\Gamma$, the net heat content of the flow changes.
In Figure (\ref{lab:fig7}b) we plot $v_T$ as a function of $E$ for $\ell=0.80$ and $\Gamma$($=1.4$, solid black), $\Gamma$($=1.5$, red dashed) and $\Gamma$($=1.6$, long dashed magenta). 
Smaller value of $\Gamma$ results in higher thermal driving and produces faster jets. 

In this study $\beta$ parameter is introduced to compute the synchrotron cooling from stochastic magnetic field.
Therefore, in steady state it is most likely that $\beta < 1$ otherwise steady disc will not form.\
We took $\beta=0.5$ as an ad hoc value. Increasing $\beta$ would increase synchrotron radiation, but would not
increase bremsstrahlung because $\dot{m}$ is not being changed. Moreover, the number of hot electrons which
inverse-Comptonize soft photons also do not change much, so although increasing $\beta$ amounts to increasing
$\ell$, but the distribution of $\ell$ is different and therefore, the response of $v_T$ to $\beta$
is different than $\dot{m}$ or $\ell$,
as was shown in Vyas et. al. (2015). 
%In Fig. \ref{lab:fig8} shows the variation of $v_T$ as a function of $\beta$ for a given $\dot{m}$ and
%$\Hsh=2.5\xsh$.
\begin {figure}[h]
\begin{center}
 \includegraphics[width=15.cm, trim=0 0 00 250,clip]{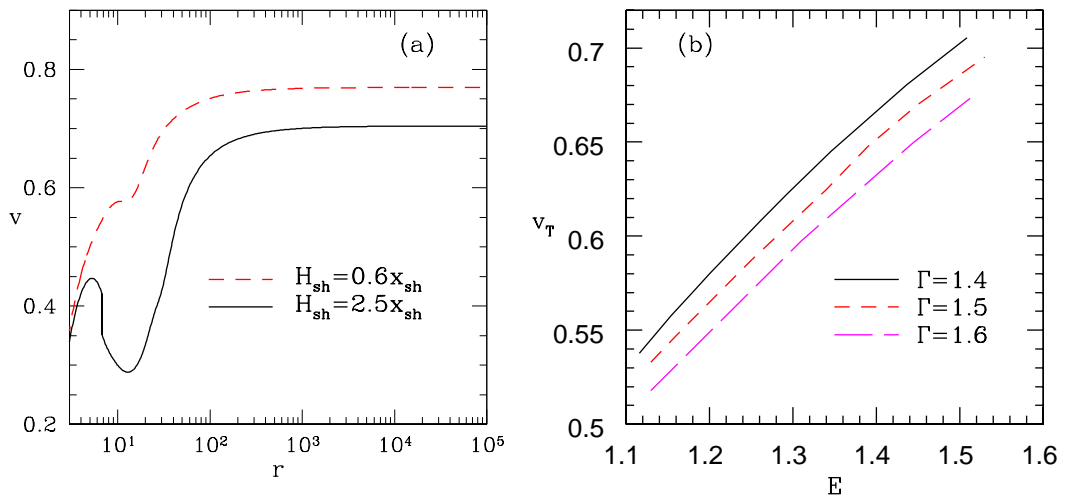}
\vskip -0.5cm
 \caption{ (a) Variation of $v$ with $r$ for various disc height ratios, $\Hsh=2.5\xsh$ (solid black) and $\Hsh=0.6\xsh$ (red dashed) for $\ell=1.76$ (b) $v_T$ as a function of $E$ for varying $\Gamma$ keeping $\ell=0.80$ \citep{vc18b}.}
%\vskip -0.75cm
\label{lab:fig7}
 \end{center}
\end{figure}

\section{Discussion and Conclusions}
\label{sec4}

Here we study radiatively and thermally driven jets with spherical cross section having a small opening angle around BH. Since the flow is hot enough to be fully ionized, the momentum transferred from radiation to the jet is only
through elastic scattering. The thermodynamics of the jet is described by a relativistic EoS, while it flows through the radiation field of the accretion disc
in Schwarzschild metric. 
The disc assumed has a thick compact corona, which emits through bremsstrahlung and synchrotron processes
like the outer disc, but additionally, through the inverse-Compton process, all of which is implemented via 
a fitting function. 

Generally, most of the studies (as chapter \ref{CH:P1}) on radiatively driven jets are conducted in SR regime
and stronger gravity is mimicked by adding any gravitational potential ad hoc in the momentum balance equation
\citep{ftrt85,vkmc15}.
Even if we overlook the obvious mistake of combining
%\LEt{again, please verify this terminology.}
SR and any gravitational potential from the view point
of the famous Principle of Equivalence, still it produces many unphysical phenomena in the solutions.
For example, the ad hoc gravity in the SR-regime, jets become unrealistically hot, such that sonic points do not form within four Schwarzschild radii. Even in cases where transonic solutions are obtained, the thermal gradient term completely dominates the radiation term. This accelerates the jets to reach their local equilibrium velocity. Hence, further out,
when the jet is cooler, radiation drag becomes more important than radiation driving. In proper GR regime, the radiation drag at moderate distances is much lower.

Since
we are considering curved space-time, consequently the radiative moments
have been computed by implementing the SR and GR transformations on the specific disc intensities and directional derivatives;  as expected, the curvature in space reduces the magnitude of the radiative moments. However, the effect of radiation is more complicated than meets the eye. The radiation
drag term, when computed in GR regime, overwhelms
%\LEt{Strictly, something cannot be "overwhelmed compared to something else": Please check that
%I have retained your intended meaning.    }
near the horizon because of the presence of the $1/g^{rr}$ term, but is lesser than that computed in flat space-time, further out. Crucially, this non-linear difference of drag term between GR and flat space cannot be mimicked by some simple scaling relation.
%\LEt{This sentence is overly convoluted by the term "departure from"; please consider simplifying the sentence so that the meaning is clear.}.}

In the advective disc model, there are two sources of radiation -- the inner compact corona and the outer disc.
The accretion rate not only controls the overall radiative output from the disc, but also determines the
size of the corona. 
Since we are considering Thomson scattering regime, the details of the
spectrum do not matter and frequency integrated moments of the radiation field suffice.
The radiative moments generally have two peaks corresponding to the radiation from the corona and the outer disc
(Fig. \ref{lab:fig2}). A comparison of the moments for an accretion disc
with an inner corona and outer KD \citep{cdc04, c05} with the present disc model shows that the radiative moments computed
from the outer disc of the present model are much stronger.

The expression of relativistic
Bernoulli parameter ($\equiv -hu_t$) for adiabatic and isentropic flow is not conserved along the streamline of a radiatively driven flow or across the shock, but $E$ is a constant of motion. This gives us a great tool to find various classes of solutions. One must not
confuse $E$ with the generalized relativistic Bernoulli parameter obtained for accretion discs \citep{ck16,kc17}.
Since the streamline and various dissipative processes in an accretion disc are different from the jet
(compare $X_f$ of Eq. (\ref{energy3.eq}) and Eq. (18) of Chattopadhyay \& Kumar 2016),
the values of generalized Bernoulli parameters will not be the same for both jet and accretion disc,
even if the jet is launched with the local accretion disc variables on the foot points of the jet. 
 
In this chapter, unlike chapter \ref{CH:P1} \citep{vkmc15}, we consider hotter and therefore geometrically thicker corona.
This has a very interesting radiative flux ($R_1$) distribution. Within the funnel of the corona,
$R_1<0$ and therefore {opposes the out-flowing jet}. Above the height of the corona, $R_1>0$ {and it pushes the jet outward}. That the radiation accelerates can be
understood from the fact that the range of sonic points becomes limited, with the increase of disc luminosity.
If $E$ is high, then the jet is hot at the base
and the effect of radiation is negligible. Thermal driving completely dominates within the funnel
and accelerates the jet such that $v\sim \veq$.
Above the funnel the jet is sufficiently cooled, such that the radiative term starts to become effective, but since the jet has reached up to the local equilibrium speed, radiation deceleration would actually slow the jet down (Figs. \ref{lab:fig9n}b, \ref{lab:fig10n}a). For medium and small values of $E$, thermal and radiation
driving may accelerate jets to relativistic speeds and the speed increases with the disc luminosity.
In fact, for lepton-dominated flow ($\xi=0.01$) jets
do reach $\gamt \gsim 10$. But more than acting simply as an agent of acceleration/deceleration,
radiation does trigger
a shock transition in jets very close to the BH. The shock range is small and the shock strength
is moderate and peaks at certain values of jet energy for a given disc luminosity.
%It may be noted that shocks generated in this paper are triggered by the inwardly directed radiation
%flux within the funnel of the corona, which is different from the shocks generated by "pinching off" the flow geometry in \citet{vc17}.

Radiatively driven fluid jet in relativity has a very rich class of solutions. The `e'-type solutions 
may have one inner-type sonic point, multiple sonic points, and shocks. While the `f'-type
jet is a low-energy solution, such solutions pass through the outer sonic point.
The radiative driving is the most effective for `f'-type jet solutions (Fig. \ref{lab:fig12n}a). This class of solutions can be compared with radiatively driven $\el$ jets in the particle approximation
\citep{cdc04,c05}. Interestingly, discs with sub-Eddington luminosity can power lepton-dominated
jets ($\xi=0.01$) to terminal Lorentz factors $\gamt \sim 3$, but super-Eddington discs can power those
`f'-type jets to $\gamt \sim 10$ (Fig. \ref{lab:fig12n}b). Above, we
argued that the radiation driving of particle jets is more efficient than that of fluid jets due to the presence of
the enthalpy term in the denominator of the radiation term (Eq. \ref{radterm3.eq}). However, the advantage
of considering radiation driving of fluid jets is that wherever the jet has been hot, radiation driving
is not effective,
but the thermal gradient term is. In the region where the temperature falls down, thermal gradient becomes less effective, but radiation takes over, provided the region is relatively close to the disc ($\sim 100 \rg$).
Therefore, the lepton-dominated jets achieve terminal speeds similar to the $\el$ particle jets, and,
in addition, the radiation driving can produce fluid phenomena like shocks in the jet. An unstable
shock can also produce effects like QPOs in the jet, a scenario worth investigating.
Moreover, such internal shocks close to the jet base have been invoked to explain the
high-energy power-law tails in some of the microquasars \citep{l11}.
\citet{ftrt85} also showed the existence of shocks in radiatively driven jets, when the disc is quite
thick and jet geometry deviates from the conical geometry. Although the authors were not considering the
effect of acceleration of radiation on jets,  the $\vt$ quoted by them were all mildly relativistic
($\vt \sim 0.1$). Whereas, we find $\vt$ is a few times higher in general. Because \citet{ftrt85} considered mostly isothermal jets and therefore missed the thermal driving factor for
the jet. 
%Our present work is also different from \citet{mstv04} since the accelerating agent in their
%work was hidden within the equation of state. They also did not find any fluid discontinuities such as shock in the jets.

We conclude by stating that radiation is an important agent in triggering various physical processes
in a jet. The radiation can drive $\ep$ jets to reasonable terminal speeds ($\vt \gsim 0.5$) if the disc is
sub-Eddington. However, for very hot jets under intense radiation field, $\gamt\sim 3$ is achievable.
For lepton-dominated flow and intense radiation field $\gamt \sim 10$ is also possible. The response of jet terminal speed with
disc luminosity is not straight forward; $\vt$ may slightly decrease with increasing luminosity
for high-energy jet, it may decrease and then increase with increasing luminosity for moderate-energy jets,
but will increase with $\ell$ for low-energy jets. It may be worth noting that radiation may accelerate
jets to relativistic terminal speeds, contrary to what is popularly accepted \citep{ggmm02}.
\chapter{Radiatively Driven Jets Under Compton Scattering}
\label{CH:P5}

\section{Overview}
In previous chapters the interaction between radiation and jet matter was under elastic scattering domain where only momentum exchange between radiation and jet takes place. In this chapter we consider Compton scattering which not only accelerates the jet, but also heats it up through energy transfer from radiation onto the jet matter. Hence, along with kinetic acceleration of the jets, we emphasize on radiative heating also. Schwarzschild space-time is considered to take care of curved space-time and relativistic EoS is used \cite{vc19}.
\section{Analysis and results}
\label{sec:res}
\subsection{Nature of radiation field}
\begin {figure}
\begin{center}
 \includegraphics[width=17.cm, trim=20 0 50 300,clip]{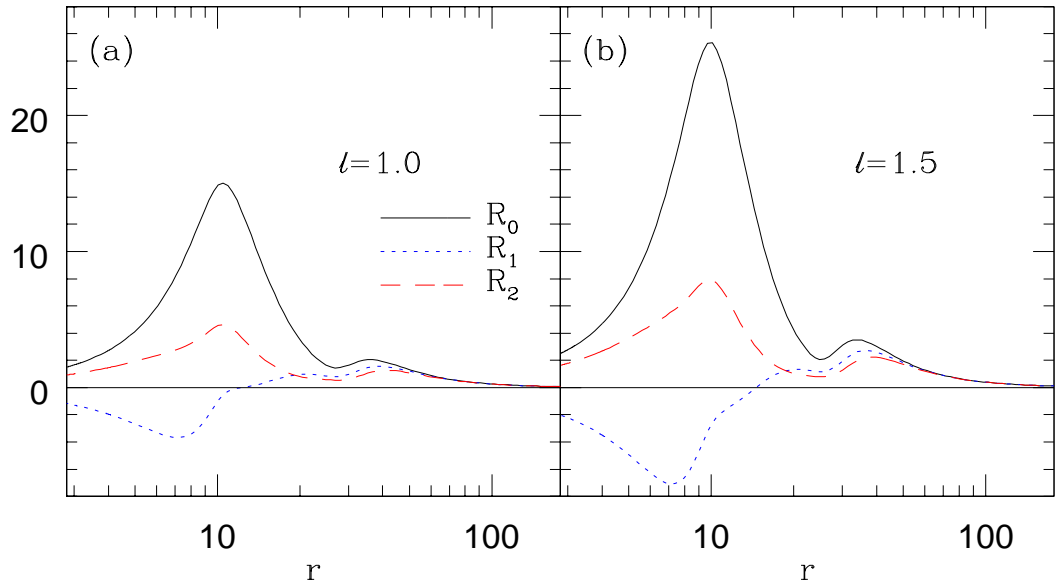}
\vskip -0.5cm
 \caption{Distribution of radiative moments $R_0$ (solid black), $R_1$ (dotted blue) and $R_2$ (red dashed) for (a) $\ell=1.0$ and (b) $\ell=1.5$ along jet length $r$}
\vskip -0.60cm
\label{lab:rad_mom}
 \end{center}
\end{figure}
The accretion disc is assumed to be having thick corona with the height of the corona ($H_{\rm sh}$) is given as \citep{kfm98}
\be
H_{\rm sh}=H^*\left(1-\sqrt{\frac{2}{\xsh}}\right)
%65d0*(1d0-sqrt(2d0/xs))
\ee
Here $H^*$ is upper limit of corona height. Accretion disc parameters are given in table (\ref{tableP5}).
\begin{table}
\caption{Disc parameters}
\label{tableP5}
\centering
 \begin{tabular}{|c c c c c c c|} 
 \hline
 \hline
 $\lambda$ & $x_0$ & $\left[\vartheta_{\rm \small D}\right]_{x_0}$ & $\left[\Theta_{\rm \small D}\right]_{x_0}$ & $\theta_D$ & $H^*$ & $d_0$\\ [0.5ex] 
 \hline%\hline
 $3.6$ & $20000 \rs$ & $0.001$ & $0.03$ & $78.5^0$ & $40$ & $0.4H_{\rm sh}$\\ 
 \hline
 \end{tabular}
\end{table}
In Fig. (\ref{lab:rad_mom}) we show intensity of radiation field along $r$ by plotting radiative energy density $R_0$ (solid black), $r$ component of radiative flux $R_1$ (dotted blue) and $rr$ component of radiative pressure $R_2$ (red dashed) for various disc luminosities of the accretion disc, $\ell=1.0,1.5$ and $0.036$. As expected, the radiation field gets weaker as the luminosity decreases. The radiation flux being negative inside the coronal funnel, enhances the effective radiation drag term resisting the jet flow while it is positive above the corona, resulting into acceleration of the jet. So radiation has twin roles, acceleration and deceleration. We will further explain these effects in next section.
\subsection{Flow variables at sonic points}
\begin {figure}
\begin{center}
 \includegraphics[width=15.cm, trim=0cm 0 3cm 0,clip]{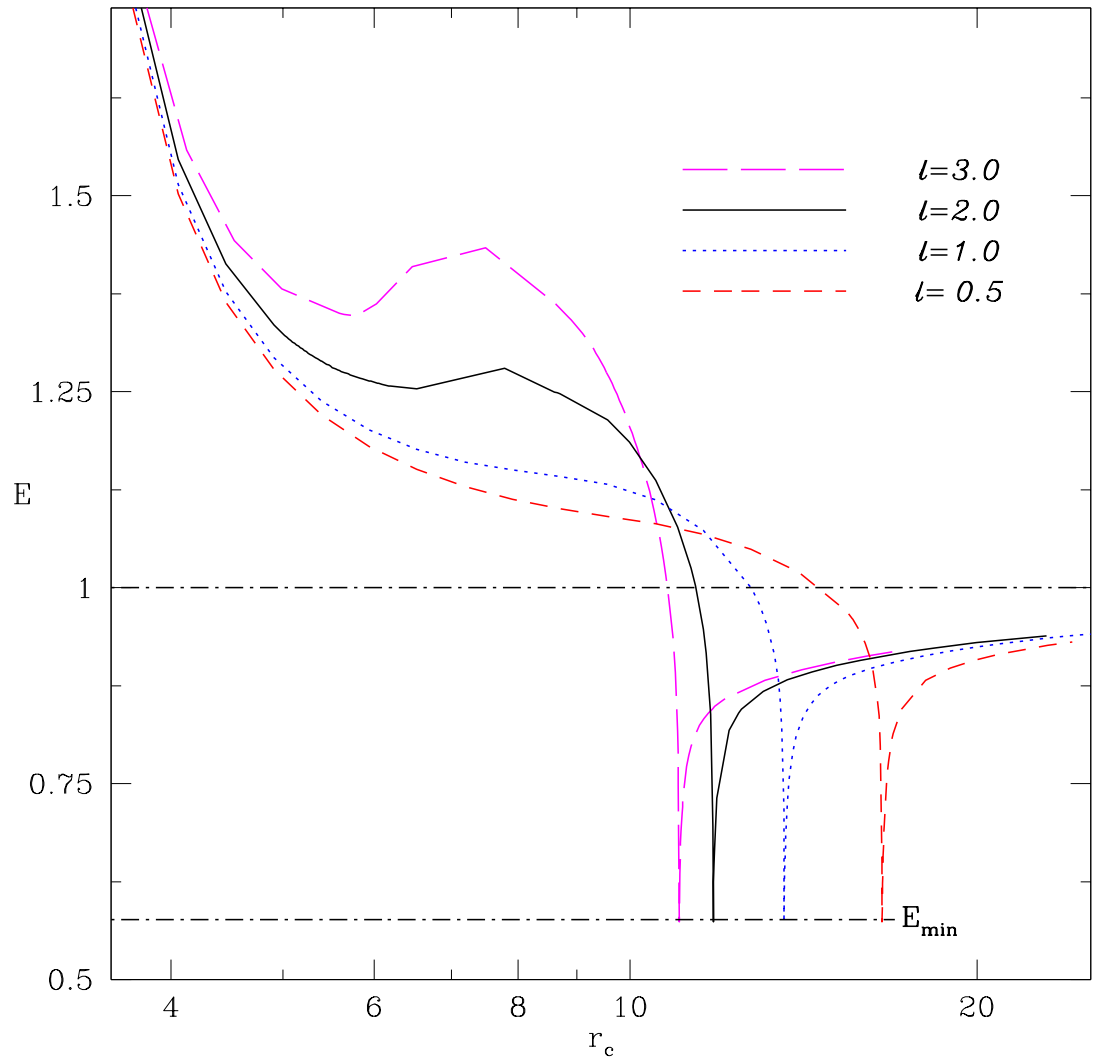}
\vskip -0.5cm
 \caption{Variation of 
 %(a) $\Theta_c$, (b) $a_c$ and (c) 
 $E$ with $r_c$ for $\ell=3.0$ (long dashed magenta) $\ell=2.0$ (solid black), $\ell=1.0$ (dotted blue) and $\ell=0.5$ (dashed red)}
%\vskip -0.75cm
\label{lab:sonic-prop}
 \end{center}
\end{figure}
As shown before, sonic point analysis is an important aspect of obtaining flow solutions because at sonic point flow speed $v$ equals sound speed $a$ providing us a mathematical boundary. Each sonic point corresponds to certain $E$ or defines certain jet base parameters. 
In Fig. (\ref{lab:sonic-prop}) we plot $E$, for $\ep$ ($\xi=1$) flow for different disc luminosities $\ell=3.0$ (long dashed magenta), $2.0$ (solid black), $1.0$ (dotted blue) and $0.5$ (dashed red). The evolution of $E$ indicates that higher $\ell$ makes the flow more energetic and $E$ become non monotonic. This effect is manifestation of radiative terms in equations (\ref{dvdr.eq}) and (\ref{dthdr.eq}) where we observe that radiation not only adds heat to the flow but has a complex term in momentum balance equation with both positive and negative signs which paves way for the jet to posses multiple sonic points for higher luminosities. (Fig. \ref{lab:sonic-prop}). 
\subsection{General pattern of solutions and significance of Compton scattering}
\begin {figure}
\begin{center}
 \includegraphics[width=15cm, trim=0 0 0 7cm,clip]{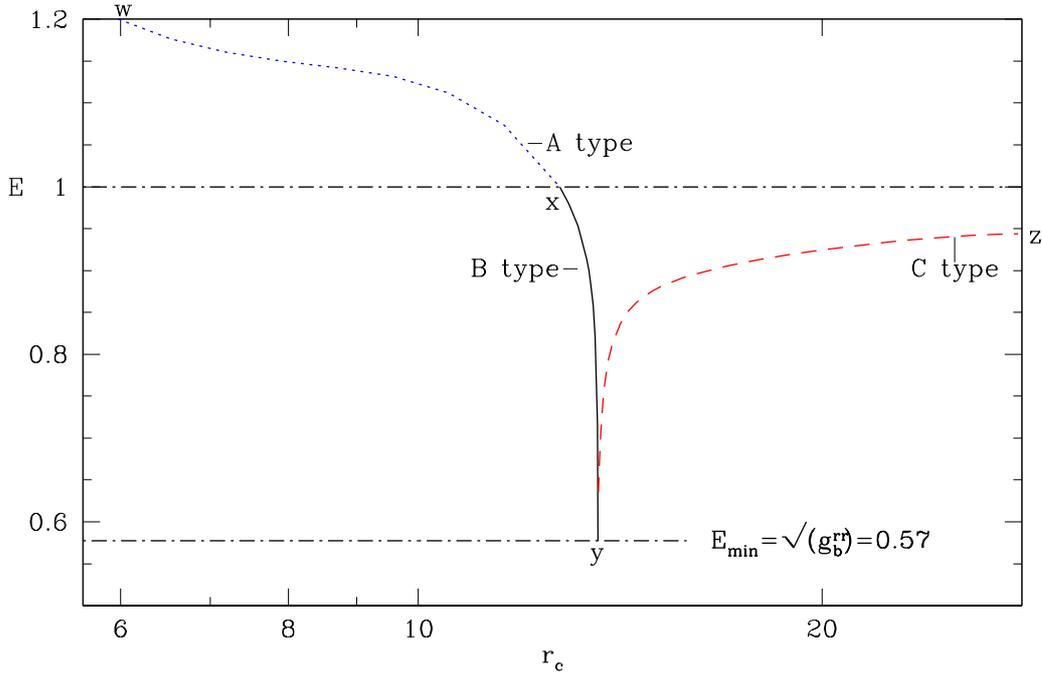}
\vskip -0.5cm
 \caption{Variation of $E$ with $r_c$ for $\ell=1.00$. Depending upon nature of jet base, the solutions are classified in three types: A-type (blue dotted), B-Type (solid black) and C type (red dashed). These are marked in the figure}
 %; Variation of (b) jet three velocity $v$ and (c) Temperature $T$ with $r$ for $E=1.2$ and Variation of (d) jet three velocity $v$ and (e) Temperature $T$ with $r$ for $E=1.07$. Both solutions are of A-type}
%\vskip -0.75cm
\label{lab:gen_sol}
 \end{center}
\end{figure}
 As $E$ is constant of motion, Fig. (\ref{lab:sonic-prop}c) contains information of all types of jet solutions. Each point on the figure corresponds to certain base variables, specifically $v_b$ and $\Theta_b$. Here subscript $b$ denotes quantities at the jet base. Based on the nature of the jet at the base, the solutions can be classified into various categories. To show their classification, we again plot $E-r_c$ curve for $\ell=1$ in Fig (\ref{lab:gen_sol}). Based upon physical state of jet base, we have three types of jet solutions $\rightarrow$ A,B, and C. The collective information of base variables lie in expression of $E$ at base (assuming that no contribution of radiation at $r_b$, $E=E_b=hu_t=h\sqrt{g^{rr}_b}\gamma_b$). Here $h$, $\sqrt{g^{rr}_b}$ and $\gamma_b$ represent thermal, gravitational and kinetic components of $E$ respectively with $h,\gamma_b \geq 1$ and $\sqrt{g^{rr}_b}<1$. Thermal and kinetic components are accelerating factors while gravity decelerates the jet. If thermal and kinetic components dominate over gravitational pull, $E>1$.  
 %We label these solutions as type A.
While if gravity dominates over thermal and kinetic components, $E<1$. It is perceptible that for $E>1$, jets have initial thrust against gravity and radiation just adds to the initial acceleration while for $E<1$, matter at the base has very dominant gravity over acceleration. The very fact that we obtain transonic solutions for $E<1$ shows the active contribution of radiation in jet driving.  
The solutions with $E>1$ are labeled as Type A (blue dotted), for which the jet base is hot ($T_b\sim 10^{12}\rm K$) and the base speeds are relatively greater ($v_b>0.1$).  For such jets, both thermal and radiative driving are effective. $E>1$ signifies that the jets are hot and faster at the time of launching. Solutions that have $E<1$ signifies that gravitational component is dominant at the base over thermal and kinetic components. These have slow jet speeds at the base (typically $v<0.1$). These are further classified into two types, type B (solid black) and type C (red dashed). Type B jets have slow but hot base while type C jets are most interesting as they have very small base speeds ($v<<0.1$) and the base of the jet is very cold ($T_b<<10^{10}\rm K$). All transonic solutions obtained previously in (previous chapters, \cite{vc18a,vc18b,vkmc15}) had $E>1$ because being in Thomson scattering regime, where no energy is transferred by radiation on to the jet, the radiation is unable to push it to the relativistic speeds irrespective of amount of radiation dumped. While in this chapter, we consider effect of Compton scattering and obtaining transonic solutions for $E<1$ means the radiation is able to push the jet even if there is no thermal pressure or initial kinetic energy at the base. It is worth noting that type B and C have no transonic counterparts in studies carried out in elastic scattering domain \citep{vkmc15, vc18b, ftrt85} because no transonic solutions were obtained for $E<1$. In other words, for $E<1$ only multivalued solutions are obtained in elastic scattering domain. Hence one needed hot base in order to drive the jets up to relativistic speeds but in Compton regime, cold and slow jets with dominance of gravity at the base, are heated up to relativistic temperatures and driven up to relativistic speeds.
\subsection*{A-type solutions : Hot and fast jet base}
\begin {figure}%[A Type : Hot and fast jet base]
\begin{center}
 \includegraphics[width=14cm, trim=0.5cm 0 0 3cm,clip]{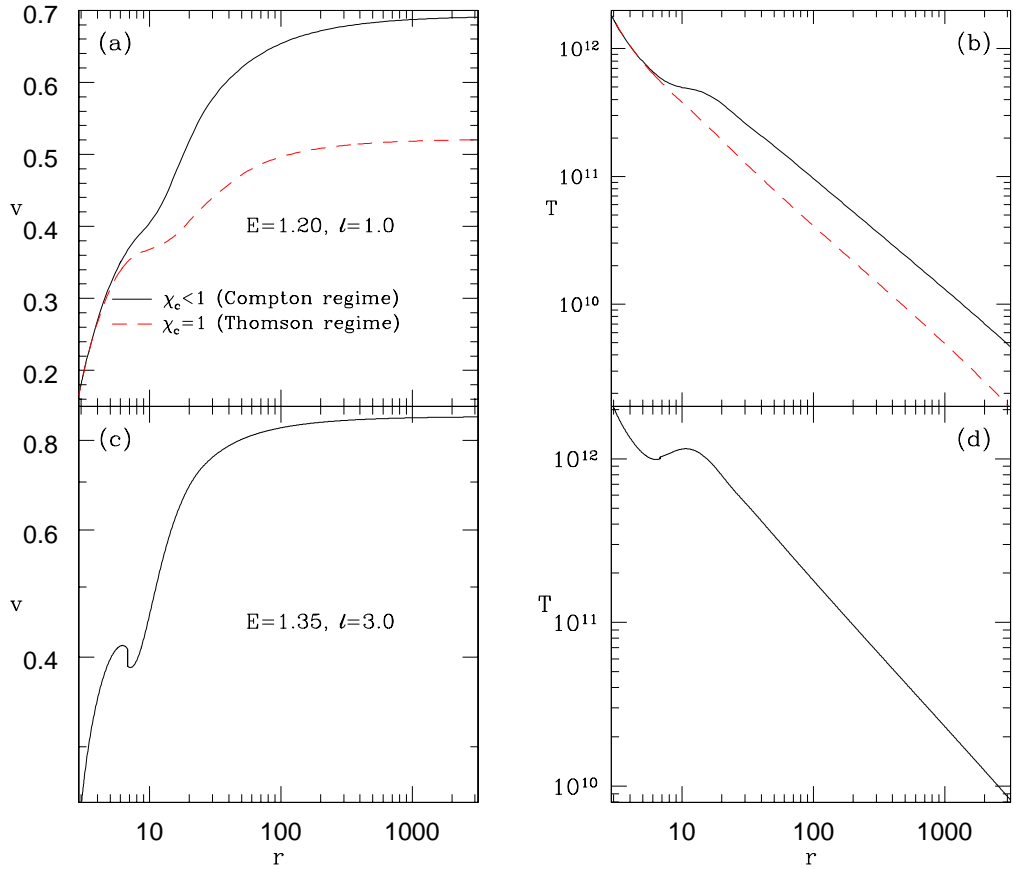}
\vskip -0.5cm
 \caption{Nature of A-type solutions. Variation of (a) jet three velocity $v$ and (b) Temperature $T$ with $r$ for $E=1.2$. Solid curve incorporates Compton scattering while dashed red curve considers Thomson scattering. Variation of jet three velocity $v$ for (c) $E=1.07$ and (d) $E=1.04$. All curves have $\ell=1.0$ and $\xi=1$.}
%\vskip -0.75cm
\label{lab:A_type}
 \end{center}
\end{figure}
In Fig. ({\ref{lab:A_type}a-b), we choose $E=1.2$ (corresponding to type A region in Fig. \ref{lab:gen_sol}) and plot velocity in ({\ref{lab:A_type}a) and temperature in ({\ref{lab:A_type}b) shown by solid black curve. It has base velocity $v_b=0.185$ and base temperature ($T_b>10^{12}$) at chosen base $r_b=3.0$. The jet meets sonic point conditions at $r=5.6$ and passing through it, reaches at terminal speeds of $v_t=0.68$.% Terminal speeds are defined to be at $r=r_\infty=10^6$. 
To show the effective contribution of Compton scattering on jet dynamics, we over-plot similar profiles (red-dashed) for identical energies but considering only Thomson scattering. This is similar to previous chapters. The terminal speeds in elastic scattering regime is only $0.51$. It is clear that Compton driven jets are $25\%$ faster as compared to Thomson scattering. In elastic scattering regime, jet temperature monotonically decreases as it flows outwards while in Compton regime, the jet is heated up at around $r=20$ where radiative moments peak (Fig \ref{lab:rad_mom}b). The rise of the temperature means Compton heating is dominant over jet cooling due to expansion. It again cools down with $r$ as radiation the field gets weaker further away.\\
For lower energies ($E>1.071$), in a certain range of $E$, we obtain multiple sonic points, because of radiation drag and negative flux inside the funnel (Fig. \ref{lab:rad_mom}b, blue dotted) collectively resist the jet and it forms multiple sonic points. Details of shock generation by radiation drag were described in previous chapter. Here we show one shock solution in fig (\ref{lab:A_type}c) for case $E=1.07$. The jet becomes transonic at $r=10.5$ and goes through shock transition under the impact of negative flux inside the funnel at $r=16.01$. Through shock discontinuity, the jet jumps from supersonic branch to subsonic branch and then again accelerates and becomes transonic at $r=20.25$ reaching at terminal speed $v_t=0.632$. 
For even lower energies, the shock disappears. In Fig ({\ref{lab:A_type}d) we plot three velocity $v$ for $E=1.04$. The jet becomes transonic with single sonic point at $r=21.4$ and obtains terminal speed $v_t=0.627$.

\subsection*{B-type solutions : Hot and slow jet base}
\begin {figure}
\begin{center}
 \includegraphics[width=15cm, trim=0 0 0 10cm,clip]{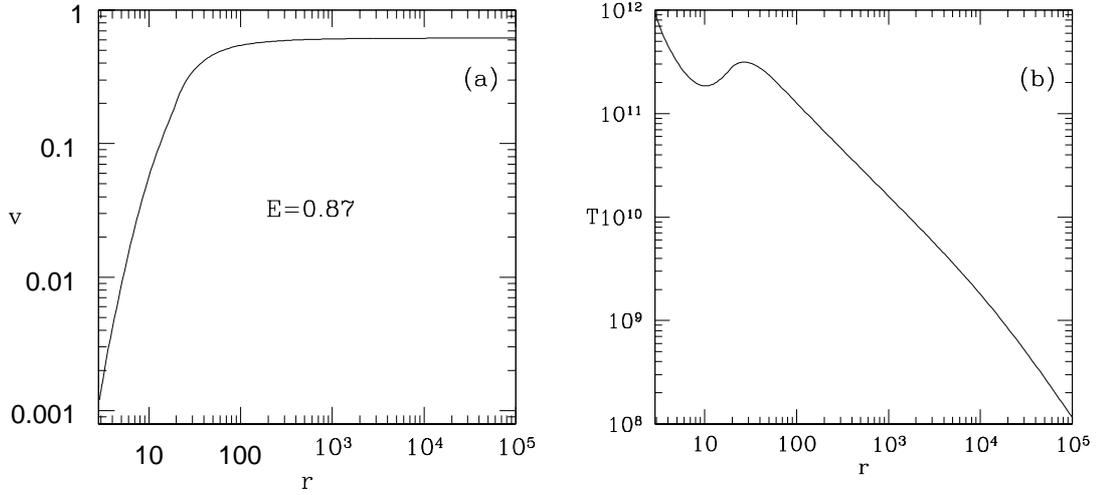}
\vskip -0.5cm
 \caption{Nature of B-type solutions. Variation of (b) $v$ and (c) $T$ with $r$ for $E=0.87$ and $\ell=1.0$}
%\vskip -0.75cm
\label{lab:B_type}
 \end{center}
\end{figure}
For $E<1$ (Fig. \ref{lab:gen_sol}, solid black), the base speeds of the jet speeds are non relativistic at the base but the temperatures are high. In other words, the kinetic component of $E_b$ is ineffective ($\gamma_b\sim1$). The thermal component is effective ($h_b>1$) but dominated by gravity $E\sim h_b \sqrt{g^{rr}_b}<1$. Hence the thermal driving is unable to push the jet outward. These jets are pushed outward by radiation force after launching. These types of solutions were absent in elastic scattering regime (chapter \ref{CH:P3_4}), and we only obtain these in Compton regime.

At $r_b=3$ the minimum energy matter is obtained when $v_b\rightarrow 0$ and $\Theta_b \ll 1$, i. e. 
\begin{equation}
E_{\rm min}=\sqrt{g^{rr}}; \mbox{ i. e., }\gamma_b \rightarrow 1;~~\&~~ h_b\rightarrow 1 
\end{equation}
At $E_{\rm min}$, the B and C class solutions merge. It is precisely for this reason that $E-r_c$
reaches upto $E_{\rm min}$ for any $\ell$ (Fig. \ref{lab:gen_sol}).

On $E_c-r_c$ plane, we choose $E=0.87$ and plot $v$ and $T$ with $r$ in Figs. (\ref{lab:B_type}a) and (\ref{lab:B_type}b) respectively for choice of $\ell=1$. At $r_b=3$, $v_b=10^{-3}$ and $T_b\sim10^{12} \rm K$. Jet passes through a single sonic point at $r=23.05$. The terminal obtained speed is $v_t=0.62$. 

As both type A and B jets have hotter base, such jets in astrophysical scenario depict the cases where jet base is already heated with some process other than Compton heating by radiation% such as Ohmic dissipation, turbulence heating, MHD wave dissipation \citep{be03}, exothermic nucleosynthesis \citep{hp08} etc.

\subsection*{C-type solutions : Cold and very slow jet base}
\begin {figure}
\begin{center}
 \includegraphics[width=15cm, trim=0 0 0 8cm,clip]{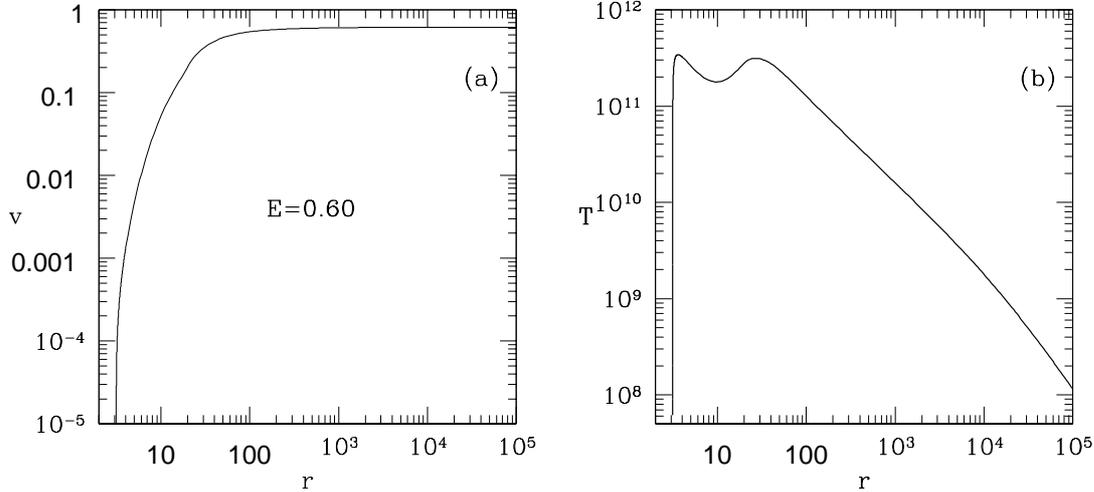}
\vskip -0.5cm
 \caption{Nature of C-type solutions. Variation of (b) $v$ and (c) $T$ with $r$ for $E=0.60$ and $\ell=1.0$}%\vskip -0.75cm
\label{lab:C_type}
 \end{center}
\end{figure}

\begin {figure}
\begin{center}
 \includegraphics[width=15.5cm, trim=0 0 0 1cm,clip]{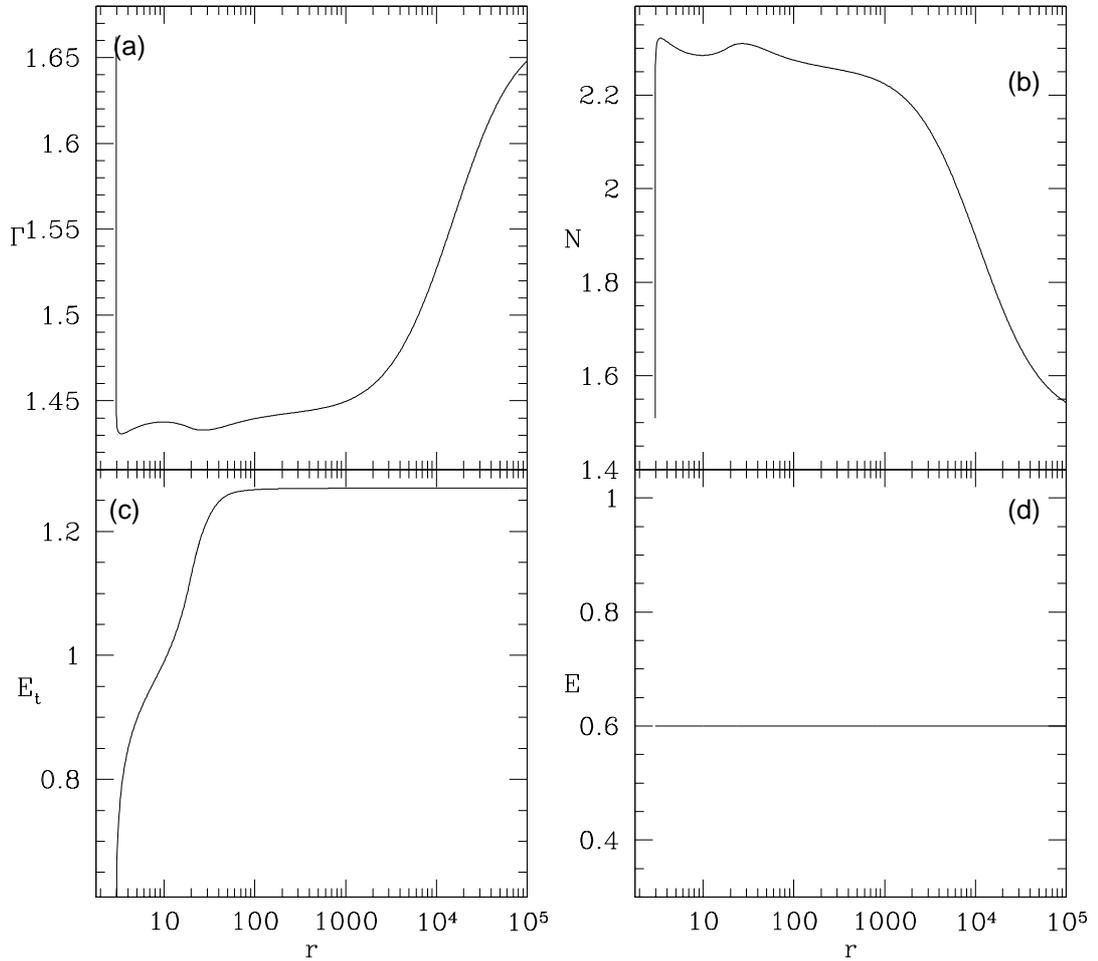}
\vskip -0.5cm
 \caption{Variation of (a) $\Gamma$ and (b) $N$, (c) $E_t$ and (d) $E$ for $\ell=1.0$}%\vskip -0.75cm
\label{lab:gam_poly}
 \end{center}
\end{figure}
Solutions corresponding to red dashed line in $E-r_c$ plot (Fig. \ref{lab:gen_sol}) attract special attention as the jets in these solutions start from $E<1$ and have very cold ($T_b<<10^{10}$) and slow ($v_b\sim10^{-5}$) jet base, that is, both thermal and kinetic components of $E$ are ineffective at the base ($\gamma_b\sim h\sim1$) and gravity is only dominating component of energy ($E_b\rightarrow\sqrt{g^{rr}_b}<1$).

Having very small speeds  they do not need any initial push at the launching. With these base parameters and $E=0.60$, we plotted $v$ and $T$ with $r$ in Fig. (\ref{lab:C_type}a-b) for $\ell=1$. Jet forms a single sonic point at $r=23.10$ and accelerates monotonically obtaining terminal speed $v_t=0.62$ under radiation driving. The temperature (Fig. \ref{lab:C_type}b) profile clearly shows that jets are heated by radiation. The base temperature is non relativistic ($T_b\sim 10^7 \rm K$). The jet is pushed outward only through radiation thrust as thermal driving is negligible at the base. Radiation imparts momentum as well as energy onto the jet and it gets both acceleration and heating. The temperature profile has two peaks. First sharp peak is obtained very close to the base where temperature of the jet becomes relativistic, rising more than three orders of magnitude, upto $3.5\times10^{11}\rm K$. This heating takes place within about half a Schwarzschild radius above the jet base. Then the temperature decreases due to geometric expansion up to $r\sim9$ and above which temperature again rises and obtains second peak with almost similar temperature to the previous peak. The mechanism of the heating and cooling can be understood if we look at equation (\ref{dthdr.eq}). First three terms inside the square bracket are positive and add to cooling of the jet. Mainly this cooling is due to expansion of the jet as it progresses. The last term in the bracket shows radiative heating. For $\ep$ flow, the heating term can be written as :
\be
T_+\propto(\Gamma-1)(1-\chi_c)\gamma\left[\frac{g^{rr} {\cal R}_0}{v}+\frac{v {\cal R}_2}{g^{rr}}-2{\cal R}_1\right]
\label{therm_acc.eq}
\ee
For temperatures above $10^7 \rm K$, $(1-\chi_c)$ is significantly above $0$ that enables radiation heating. As ${\cal R}_1$ is negative inside the funnel, all terms inside the bracket are positive and collectively heat up the jet near base. However, very close to the base,  $v<<1$, hence only first term is dominant and has huge magnitude in the initial heating that is responsible for the first peak and robust amplification in jet temperature. To understand it physically, we see that as the fluid is moving very slowly near the base, it has more time to interact with the radiation field and it is constantly heated up. % until it becomes faster.
 For case $2{\cal R}_1>\frac{g^{rr} {\cal R}_0}{v}+\frac{v {\cal R}_2}{g^{rr}}$, the energy is transferred from matter to the radiation hence Compton cooling becomes effective over heating. Further, 
in our assumption, no emission is considered from the jet giving another reason that heating is dominant. For $3.5<r<10$ cooling dominates over heating. But as the radiation field becomes stronger, the jet is further heated keeping the temperature above $10^{11} \rm K$ up to a distance of $100$ Schwarzschild radii. beyond $100 r_g$ radiation field is very weak and jet continuously cools down as it propagates outward.\\

In Figure (\ref{lab:gam_poly}a-b) we plot variation of $\Gamma$ and $N$ corresponding to parameters of Fig. (\ref{lab:C_type}). Variation of $\Gamma$ delivers similar information that plasma is cold and non-relativistic at the base as well as far away ($r\sim10^5$), but radiation makes it relativistic and hot in between. As expected, the Bernoulli parameter (Fig. \ref{lab:gam_poly}c) evolves and increases with impact of radiation. Starting from $E=0.6<1$ at the base, it reaches at $E\sim1.206>1$ while the generalized relativistic Bernoulli parameter remains conserved and is a constant of motion (Fig. \ref{lab:gam_poly}d).

\subsection*{Effect of luminosity on jet acceleration and heating}
\begin {figure}
\begin{center}
 \includegraphics[width=15cm, trim=0 0 0 0,clip]{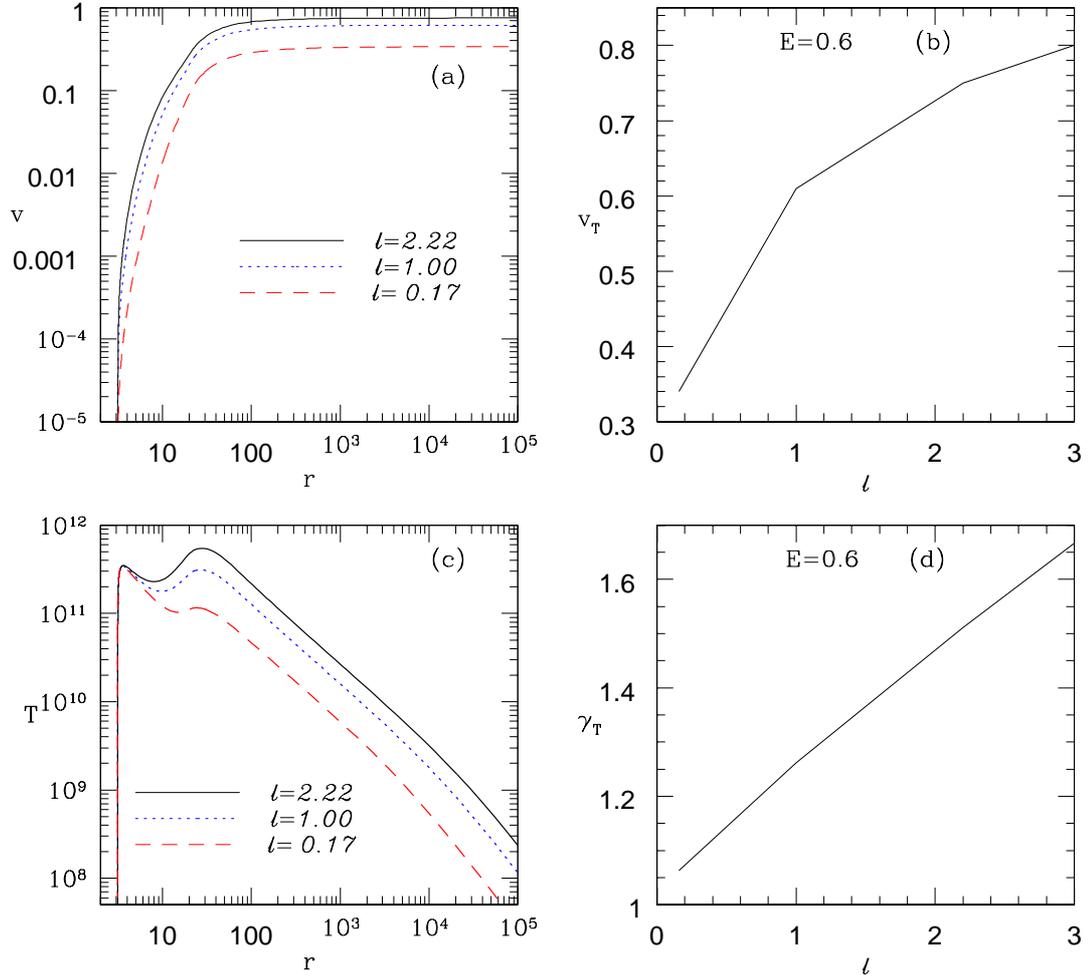}
\vskip -0.5cm
 \caption{(a) velocity profiles for various luminosities for C-type solutions. Corresponding terminal speeds ($v_t$) are plotted in (b). (c) Variation of $T$ with $r$ for various luminosities. (d) Lorentz factors ($\gamma_t$ ) are plotted with $\ell$ (solid black) In panels (a) and (c) different curves are for $\ell=2.22$ (solid black), $\ell=1.00$ (dotted blue) and $\ell=0.17$ (dashed red) and $E=0.6$.}
%\vskip -0.75cm
\label{lab:Temp_profile}
 \end{center}
\end{figure}

We keep same $E=0.6$ and plot velocity profiles for $\ell=2.22$ (solid black), $\ell=1.00$ (dotted blue) and $\ell=0.17$ (dashed red) in Fig. (\ref{lab:Temp_profile}a). As expected, greater acceleration is observed as the radiation field gets more intense. To estimate qualitative magnitude of acceleration and effect of $\ell$, we plot $v_T$ with $\ell$ for $E=0.6$ in Fig (\ref{lab:Temp_profile}b). The terminal speeds range from $v_t=0.34$ to $v_t=0.8$ as $\ell$ goes from $0.17$ to $3.0$.\\
The corresponding temperature profiles for these luminosities are shown in Fig. (\ref{lab:Temp_profile}c). Interestingly the first peak is almost similar for wide range of $\ell$ because near the base, $v(=10^{-5})$ is very small, hence, the jet accumulates greater amount of radiation even for low luminosities (the first term in square bracket in equation \ref{therm_acc.eq}). The radiation field intensity is dominant in deciding second peak and hence the jet under strong radiation field obtain peak temperatures beyond $10^{11} \rm K$.
%We plot the maximum temperatures ($T_m$) with $\ell$ in Fig. (\ref{lab:Temp_profile}d) with solid black curve. To present the effect of heating the base temperatures are shown by blue dotted line and are constant at $T_b\sim 1.5\times 10^7 \rm K$.  This reference limit is considered from example of NGC 1266 which has $\ell\sim10^{-7}$ \citep{all14}, and corresponding brightness temperature is $1.5 \times 10^7 \rm K$ \citep{naw13}\\
%{\bf These solutions are more significant than Type A and B as the accretion models under viscous heating, generate disc temperatures up to $10^8-10^9 \rm K$ but the observed brightness temperatures of radio loud AGNs are generally obtained to be as high as $10^{11}-10^{12} \rm K$. 
%We compare observed brightness temperatures ($T_o$) and calculated temperatures ($T_c$) in some of the radio loud sources in table (\ref{table1}). The temperatures are also over-plotted with calculated temperatures in Fig. (\ref{lab:Temp_profile}d) with filled dots. The calculated temperatures are within an order of magnitude closer to the observed temperatures which shows that radiation heating in jets by Compton scattering can be one of the dominant processes through which jets are heated up after generation.
%However, we observe that even after effective heating by radiation, the observed brightness temperatures of the jet are greater than corresponding calculated upper limits (\ie $T_o>T_c$).}
\subsection*{Effect of composition on jet dynamics}
\begin {figure}
\begin{center}
 \includegraphics[width=15cm, trim=0 0 0 0,clip]{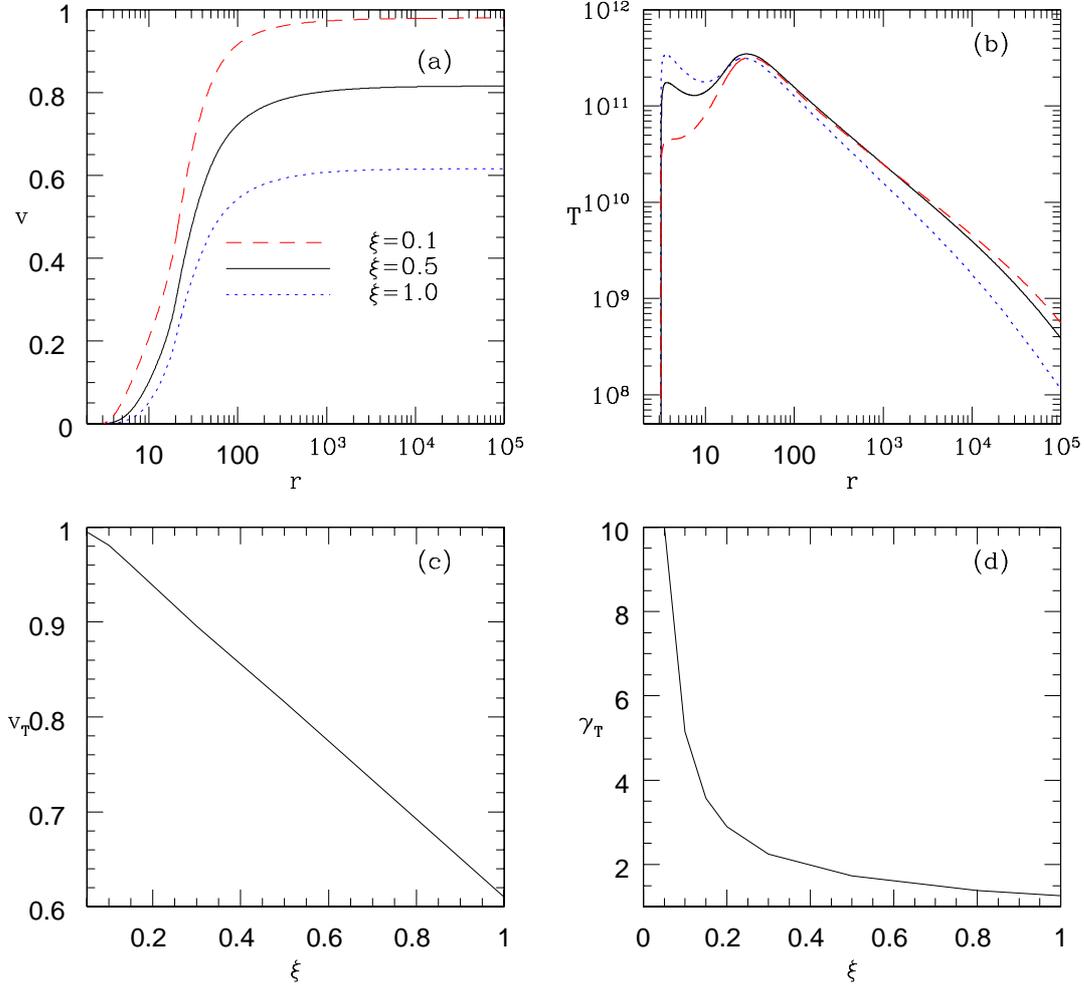}
\vskip -0.5cm
 \caption{(a) $v$ and (b) $T$ profiles for $\xi=1$ (dotted blue), $\xi=0.5$ (solid black), and $\xi=0.1$ (dashed red) for $E=0.6$, $\ell=1.0$, variation of $v_T$ (c) and $\gamma_T$ (d) with $\xi$ for similar parameters}
%\vskip -0.75cm
\label{lab:Temp_profile_xi}
 \end{center}
\end{figure}

The composition of the relativistic jets is a much debated topic. These are either believed to be composed of $\ep$ plasma or pair plasma made of $\el$. As the relativistic EoS considered here, which takes care of composition of the plasma through $\xi$. It permits us to study the jet dynamics with variation of $\xi$. 
To study effect of composition, we vary $\xi$ keeping other parameters such as $E$ and $\ell$ and generate solutions. We plot three velocity $v$ in Fig. (\ref{lab:Temp_profile_xi}a) and temperature $T$ in Fig. (\ref{lab:Temp_profile_xi}b) for $\xi=1.0$ (blue dotted), $\xi=0.5$ (solid black) and $\xi=0.1$ (red dashed). We have $E=0.6$ and $\ell=1.0$. As $\xi$ decreases, lepton fraction in fluid composition increases making the fluid lighter, hence the jet, under radiation acceleration, becomes faster. Corresponding terminal speeds are plotted in Fig. (\ref{lab:Temp_profile_xi}c) which go up to $0.995$ as $\xi$ drops by $0.05$. In terms of terminal Lorentz factors $\gamma_t$ of the jets (Fig.\ref{lab:Temp_profile_xi}d), for very low $\xi(\sim0.05)$, $\gamma_t$ comfortably reaches by $10$. The temperature profiles show that as baryon dominated flows are hotter, heating is also more effective for higher $\xi$.
\begin {figure}
\begin{center}
 \includegraphics[width=14.5cm, trim=0 0 1cm 3cm,clip]{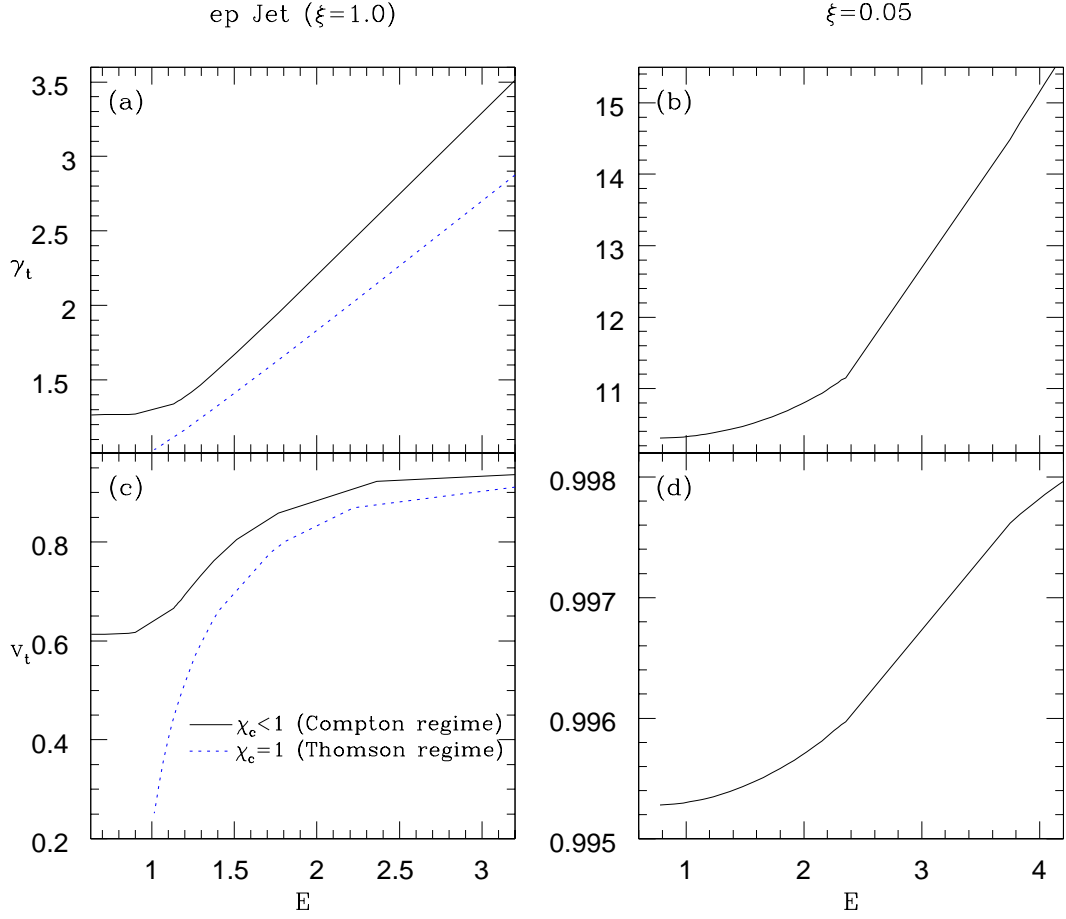}
\vskip -0.5cm
 \caption{Terminal Lorentz factor $\gamma_t$ (a, b) and terminal speed $v_t$ (c, d) as a function of $E$.
The composition chosen for the terminal quantities are $\xi=1$ (a, c) and $\xi=0.05$ (b, d). 
We compare the terminal quantities of jets in the Thomson scattering regime (dotted, blue) with those
in Compton scattering regime (solid, black) in panels (a) and (c).}
%\vskip -0.75cm
\label{lab:term_E_xi}
 \end{center}
\end{figure}

We plot $\gamma_t$ and $v_t$ with $E$ for $\xi=1.0$ (Fig. \ref{lab:term_E_xi} a,c) and $\xi=0.05$ (dashed, blue; Fig. \ref{lab:term_E_xi} b,d). The curves are plotted for $\ell=1.0$. For $\ep$ jets we compare terminal quantities (solid, black) with those obtained in elastic scattering regime %\citep[dotted, blue, similar to][]{vc18b}, 
which reconfirms the fact 
that Compton scattering accelerates the jets more effectively than in the Thomson regime. Further, as there are no 
solutions for $E<1$ in the Thomson scattering regime, the terminal speeds approach very low values as
$E\rightarrow 1$. However, Compton driven jets maintain $v_t>0.6$ even for $E<1$. This lower limit off $v_t$ or
$\gamma_t$, is highly relativistic ($v_t>0.99$) for lepton dominated jets (dashed, blue).
It may be noted that Fig. (\ref{lab:Temp_profile_xi}) is for C-type jets, while Fig. (\ref{lab:term_E_xi})
represents all types of jets for given $E$ and $\ell$.

\section{Concluding remarks}
\label{sec:con}
We have studied role of accretion disc radiation on dynamics of non-rotating relativistic jets under relativistic regime. Impact of radiation is studied considering Compton scattering. This work is in continuation of our previous efforts where we studied interaction between radiation and matter under elastic scattering regime. We could show that there are significant differences as one considers Compton scattering in radiative driving. As there is energy transfer between radiation and jet matter compared to mere momentum transfer in elastic scattering regime, the terminal Lorentz factors obtained are relatively higher for jets. A more important consequence is evident from the fact that we could obtain transonic solutions with relativistic terminal speeds even for $E<1$, where gravity is dominant over thermal driving at the base. In Thomson scattering regime, at such $E$, only multivalued solutions could be obtained. Hence the role of radiation is more than jet acceleration as it not only accelerates the jet at larger distances but provides initial push inside the accretion funnel. For $E<1$, the terminal Lorentz factor for super Eddington luminosities is below 2 for $\ep$ jets but with ease it reaches beyond 10 for $\el$ composition. It is reiterated that these results have no counterpart in the previously studies jets in Thomson scattering regime \citep{vkmc15,vc18a,vc18b} as there $E$ is always required to be greater than 1 for transonic solutions.

Radiatively driven jets are found to be possessing multiple sonic points and internal shocks in certain range of the parameters. These internal shocks may produce by various reasons. We showed in \cite{vc17} that non radial cross section may harbour internal shocks, In \cite{vc18a}, we showed that in presence of radiation, even radial jets may go under shock transitions. The shocks have similar features as obtained in \cite{vc18a}. A number of processes that give rise to internal shocks give theoretical support and strengthen the attempts that assumed internal shocks to explain various observed features of the jets \citep{bk79, l11}.

Along with these, another consequence is seen in heating of the jet by radiation. The theoretical upper limit of brightness temperature in extragalactic sources has been set to be at $T_{\rm max}(=10^{12} \rm K)$. This limit is an outcome of inverse Compton catastrophe \citep{r94}. Beyond this limit, it was established that inverse Compton scattering of radiation from the matter dominates and the matter looses energy through synchrotron emission. Another theoretical limit was put by considering equipartition between radiating plasma and magnetic field. The assumption of equipartition constraints the limit at $T_{\rm eq}=5 \times 10^{10} \rm K$ \citep{r94, n17} . Interestingly, the observed brightness temperature of powerful extragalactic sources in low frequency synchrotron radiation is generally obtained to be of the order $~$100 MeV or $10^{12}\rm K$ \citep{mfp96,tml98}. Hence the observations comfortably exceed the set theoretical limit.

If we consider radio quit sources (that either do not have jet or have very weak jets) we see that the inferred temperatures are commonly in the range $10^6 \rm K$-$10^9 \rm K$ \citep{uab05,wh06,uwt05,naw13}, which comfortably sets the allowed temperature range for the accretion discs. If jets are to be accelerated up to ultra relativistic temperatures, plasma heating is evident. 
Numerous processes have been quoted in the literature that heat up the plasma at the jet base or in accretion funnel, like Ohmic dissipation, turbulence heating, MHD wave dissipation \citep{be03}, exothermic nucleosynthesis \citep{hp08} etc. The heating of gas inside radio sources to reach at such temperatures is attempted through Compton scattering of radiation but only moderated impact was obtained \citep{ss01}. 
In this analysis, taking dynamic nature of jet into consideration, we have obtained that through Compton scattering the jets can be heated comfortably up to temperatures $>10^{11}{\rm K}$, for jets launched with base temperatures and velocities keeping in agreement with the inner boundary conditions dictated by accretion discs.

%\chapter{Radiation hydrodynamic simulations of astrophysical jets}
%\label{CH:P6}
\chapter{Conclusions}
\section{Major outcomes}
In this dissertation, detailed study of jets has been carried out considering radiation and thermal driving. Across the chapters, the investigation is made in various regimes, flat space-time to curved space-time, fixed $\Gamma$ to variable $\Gamma$ relativistic EoS, radial to non radial cross section of the jet, Thomson scattering to Compton scattering regime etc. Study in various regimes not only enables us to understand the sequential understanding of jet behaviour as the approximations improve but also we could distinguish the impact of these regimes and relative differences in the outcomes. Detailed conclusions of various chapters are already described. Here we summarize and outline major outcomes showing impact of the work.
\subsection{Terminal speeds of the jets}
The terminal speeds of the jet depend upon driving agent as well as base parameters. In thermally driven jets, the terminal speeds are relatively less, while in radiatively driven jets, it is relatively greater but the terminal speeds of the jets further increase significantly as we consider Compton scattering in stead of Thomson scattering. As radiation from accretion disc imparts momentum as well as energy onto the jet, it not only accelerates it but also heats it up giving rise to additional thermal driving. Hence in the Compton regime, thermal and radiative driving are coupled. The baryon dominated jets ($\ep$) comfortably achieve terminal Lorentz factors up to a few while lepton dominated jets easily go beyond Lorentz factors $10$. We quoted earlier in section (\ref{sec_motivation}) via \cite{ggmm02}, that radiation acceleration hasn't been considered to be dominant factor for jet acceleration. Now following above results, we can state that a fluid jet, generated in acceleration funnel, can be accelerated up to relativistic speeds, even if there is no other significant accelerating agent present except radiation. This is more evident from the results of Compton driving (chapter \ref{CH:P5}).
\subsection{Significance of considering curvature of space-time for outflows}
To simplify analysis, approximating gravity by pNp with special relativity (SR+PW) has been generally used to treat astrophysical flows around BHs. We find that though such consideration makes the analysis relatively easier but leads to some other bigger problems. Such flows are hotter by few orders of magnitude compared to study carried out in general relativistic analysis. Further, the solutions having sonic point close to the jet base are missing in absence of GR. Including GR, we also show that the radiation field depends upon gravitational field in non linear manner. Hence we conclude that when one deals with the region closer to the BH, general relativistic analysis is inevitable.
\subsection{Internal shocks and their implications}
In various circumstances, jets harbour multiple sonic points as well as internal shocks. In chapter \ref{CH:P2}}, we obtained internal shocks induced by non-radial geometry of the jets. Further, in chapters \ref{CH:P3_4} and \ref{CH:P5}, we obtained shocks in radial jets that are induced by specific nature of the radiation field. To the best of our knowledge, radiation driven shocks in jets are shown for the first time. The significance of such shocks is evident as they are repeatedly shown to be important to explain various observed features of radio loud sources such as very high energy emissions (GeV to TeV) in their spectra. 

We showed that these shocks are able to account for high energy tail in microquasars.
\subsection{Significance of relativistic EoS}
Except one part of chapter \ref{CH:P3_4}, we have considered relativistic EoS throughout the work. The approximated relativistic EoS takes into account the variable nature of adiabatic index $\Gamma$. $\Gamma$ depends upon temperature and thus the thermodynamic nature of the matter changes with the temperature. Further, the used relativistic EoS also considers information of composition in the flow through defined $\xi$ parameter. It allowed us to study the impact of flow composition on jet dynamics. The differences on the outcomes of the consideration are apparent in chapter \ref{CH:P3_4}
%$\bullet$ {\bf Rich class of solutions under radiation hydrodynamics and their implications :}\\
\subsection{Compton scattering and relativistic jets for $E<1$}
An important consequence of Compton scattering in jets (chapter \ref{CH:P5}) is evident from the fact that we obtained transonic solutions leading to relativistic terminal speeds for $E<1$, where gravity is dominant over thermal driving at the base. In elastic scattering regime, at such $E$, only multivalued solutions could be obtained. This shows greater role of radiation as it not only accelerates the jet by several magnitudes but also provides initial push to kick the jets out of the accretion funnel when the matter is essentially bound at the base.
\subsection{Radiation heating and observed brightness temperatures of radio loud AGNs}
In chapter \ref{CH:P5}, we studied impact of radiation on jets considering Compton scattering. The brightness temperature of powerful extragalactic sources in low frequency synchrotron radiation is generally obtained to be of the order $~$100 MeV ($10^{12}\rm K$). The heating of gas inside radio sources to reach at such high temperatures is attempted through Compton scattering of radiation but only moderated impact was obtained \citep{ss01}. Taking dynamic nature of jet into consideration, we have obtained that the jets which are launched with non relativistic temperatures at the base, can be heated comfortably up to temperatures $>10^{11}{\rm K}$, which is within one order of magnitude of obtained temperatures. The study provides a hint that radiation is one of the major factors that may heat up the jets.
\section{Future prospects}
The results are encouraging as they show important and few unexplored aspects of radiation driving in jets and pave a way forward for us. The further plans of the work include investigation of following aspects :
\begin{itemize}
\item Following this work, which is a steady state analysis, corresponding simulations can be carried out which will give time dependent nature of the results.
\item Inclusion of magnetic field in radiation driven relativistic jets (GRMHD) will be interesting. 
\item Having discussed the observed brightness temperatures and their connection with obtained high temperature by Compton heating, we need to check the results carefully in further work for establishing a firm connection between the results and spectral features of the radio sources.
\item The role of internal shocks in explaining observed features are to be investigated deeper.
\item As the whole analysis is carried out considering generated jets in the accretion funnel and in chapter \ref{CH:P5} we saw that the base parameters are consistent with the typically obtained inner boundary parameters of the accretion solutions. This gives us confidence that accretion ejection study can be carried out that will {connect accretion solutions to the jet solutions}.
%\item 

\end{itemize}
%\appendix
\appendix % Cue to tell LaTeX that the following 'chapters' are Appendices

\input{AppendixA}
%\chapter{Appendix}
%\section{Accretion Disc and associated radiation parameters}

% Include the chapters of the thesis, as separate files
% Just uncomment the lines as you write the chapters

%\input{Chapters/Chapter1} % Introduction

%\input{Chapters/Chapter2} % Background Theory 

%\input{Chapters/Chapter3} % Experimental Setup

%\input{Chapters/Chapter4} % Experiment 1

%\input{Chapters/Chapter5} % Experiment 2

%\input{Chapters/Chapter6} % Results and Discussion

%\input{Chapters/Chapter7} % Conclusion

%% ----------------------------------------------------------------
% Now begin the Appendices, including them as separate files

\addtocontents{toc}{\vspace{2em}} % Add a gap in the Contents, for aesthetics

\appendix % Cue to tell LaTeX that the following 'chapters' are Appendices

\addtocontents{toc}{\vspace{2em}}  % Add a gap in the Contents, for aesthetics
\backmatter

%% ----------------------------------------------------------------
%\label{Bibliography}
%\lhead{\emph{Bibliography}}  % Change the left side page header to "Bibliography"
%\bibliographystyle{unsrtnat}  % Use the "unsrtnat" BibTeX style for formatting the Bibliography
%\bibliography{Bibliography}  % The references (bibliography) information are stored in the file named "Bibliography.bib"

\end{document}

%% file: abs.tex
%\frontmatter
%\lhead{{ABSTRACT}}
\section*{ABSTRACT}
%\vspace{2em}

%\begin{center}
%	\Large \bf ABSTRACT
%\end{center}
Detailed study of relativistic jets around black hole sources, such as microquasars and active galactic nuclei (AGNs) is carried out and the impact of radiation and thermal driving is investigated. Across the chapters, the analysis is done in various
regimes, flat space-time to curved space-time, fixed to variable adiabatic index
relativistic equation of state, radial to non radial cross section of the jet, Thomson (elastic) scattering to Compton scattering regime etc.\\
Accretion disc
plays auxiliary role for generating radiation field, as well as, geometrically shaping the jet near its base. The disc radiates through thermal, bremsstrahlung, synchrotron and inverse {Compton} processes. Jets are taken to be non rotating
and axis-symmetric. Further, relativistic equation of state is used that takes care
of variable adiabatic index at relativistic {temperatures, as well as, composition} of
the flow. The radiative moments are calculated using their full relativistic transformations. Bending of photon path is
also considered.\\
We show that special relativistic study of jets using pseudo Newtonian potential is inadequate. Such flows are hotter by one or two orders of {magnitude, }compared to those considered in curved space-time. The family of solutions having sonic point close to the jet base are missing in absence of general relativistic
analysis. Including general relativity, we also show that the radiation field depends upon gravitational field in non linear {manner, } making general relativistic
analysis more significant.
%\frontmatter
We obtained internal shocks in jets close to the {base,} as a result of non-conical
cross section and nature of radiation field on jet dynamics. Theoretical evidence
of internal shocks is {significant,} as these are required to explain high energy tail
of the spectra of radio sources. Under thermal and radiative driving, jets with
electron-positron composition are obtained to be achieving relativistic speeds up
to Lorentz factors $\gamma \sim 10$ while for electron-proton composition it is $\gamma \sim 2$ for
luminous discs. We also showed that extragalactic jets around AGNs are faster
than those around microquasars. We obtain that through Compton {scattering, relativistic jets} generated with non relativistic temperatures, are efficiently heated by temperatures up to $T \sim 10^{11-12} \rm K$.
Another important consequence of Compton scattering {is,} that we obtained transonic solutions with relativistic terminal {speeds,} for bound state of the jet at the
base (that is, generalized Bernoulli parameter $E < 1$){, where} gravity is dominant over thermal
driving. In Thomson scattering regime, at such energies, no transonic solutions could
be obtained. This shows that radiation, in fact, has greater role as it not only accelerates the unbound matter of the {jets,} but the acceleration is effective {even} in absence of any other accelerating agent such as thermal acceleration. % or when the matter is bound.\\
Apart from these {features,} a detailed analysis of dependence of jet
variables upon various parameters like plasma composition, magnetic pressure
in the disc, luminosity, accretion rate, jet geometry and jet launching energy {etc,}
is carried out.

%% file: AppendixA.tex
\chapter{Accretion Disc and Associated Radiation Parameters}
%\section{Accretion Disc and associated radiation parameters}
\label{app:rad_mom}
\section{Estimating approximate accretion disc variables}
%\subsubsection{Estimating approximate accretion disc variables}
$U^\mu$ are the components of accretion
four-velocity, and the corresponding three-velocity components are
${v}\equiv(\vartheta_x,0,\vartheta_\phi)$, where $x,~\theta,~\phi$ are usual spatial coordinates. We define
$\vartheta=\vartheta_x/\sqrt{(1-\vartheta_\phi^2)}$ as the radial three-velocity measured
by a local rotating observer. Following this, one can present
the velocity distribution of the outer disc and the corona in a compact form
\citep[see Appendix A of][]{vkmc15}
\be
\vartheta_{\rm i}=\left[1-\frac{(x-2)x^2}{\{x^3-[(x-2)\lambda^2]\}U_t^2|_{x_{0\rm i}}}
\right]^{1/2}.
\label{accvel.eq}
\ee
Here, the suffix ${\rm i}$ denotes variables of the corona (i. e.,
i$=${\small C}) or the outer disc or SKD (i. e., i$=${\small D}) and
$U_t|_{x_{0\rm i}}$ is the covariant time component of the $U^\mu$s at the outer edge.
For the corona, $x_{0\rm i}=\xsh$ and for the outer disc $x_{0\rm i}=x_0$.
At $x_0$, $ [\vartheta_{\rm \small D}]_{x_0} \approx 0$ but increases as it falls towards the BH till it reaches
$\xsh$, where the flow speed reduces by one-third. In shocked accretion disc, this reduction is automatic, but
even in shock free discs centrifugal barrier, radiation pressure all can impede the inflow, making it hot and thereby forming the corona. 
Assuming a slow variation of the adiabatic index the temperature distribution can also be assumed as \citep{vkmc15}
\be
\Theta_{\rm i}=\Theta_0\left(\frac{U^x_0x_0H_0}{U^x_{\rm i} xH_{\rm i}}\right)^{\Gamma -1}.
\label{acctemp.eq}
\ee
Moreover, \citet{vkmc15} proposed an approximate relation between $\xsh$ and the
accretion rate, given by
%\bea
%\xsh=125.31326199423765-24.602485347397550{\dot m} \nonumber \\
%+1.7646624611542214{\dot m}^2-0.043447536123535316{\dot m}^3
%\label{xsdotm.eq}
%\eea
\be
\xsh=125.313-24.603{\dot m_\sk}+1.765{\dot m_\sk}^2-0.043{\dot m_\sk}^3
\label{xsdotm.eq}
\ee
Here $\xsh$ is in geometric units and $\dot m_\sk$ is the accretion rate in units of Eddington
rate (Eddington rate $\equiv {\dot M}_{\rm Edd}=1.4\times10^{17}\mbh/\msol$gs$^{-1}$).
In order to completely specify $\vartheta_{\rm i}$ and $\Theta_{\rm i}$ at all $x$, one also needs to know the local height $H_{\rm i}$.
Numerical simulations show that the outer disc has a flatter structure than that predicted by
assumptions of vertical equilibrium and the inner torus like corona is basically a thick
disc (with advection terms) and the height to radius ratio can vary anything between 1.5 to 10  \citep{dcnm14,lckhr16}. Therefore, we define $H_{\rm 0}=0.4H_{\rm sh}+\tan \theta_{\rm D}x_0$.
If we supply $[\vartheta_{\rm \small D}]_{x_0},~\rho_0,~H_0$ and ${\dot m_\sk}$ at $x_0$, then the distribution
of velocity, temperature, density at all $x_{\rm i}$ and the location of $\xsh$ can be estimated. This
allows us to compute the density profile for a given ${\dot m}_\sk$.% Typical accretion disc parameters
%are given in table \ref{11}.
\section{Radiative intensity and luminosity from the accretion flow}
We assume that there is stochastic magnetic field in the outer disc which is in partial equipartition with the gas pressure.
The ratio of magnetic pressure ($p_{\rm mag}$)
and the gas pressure ($p_{\rm gas}$) is also assumed to be constant $\beta$
{\ie} $p_{\rm mag}=B^2/8\pi=\beta p_{\rm gas}=\beta n_\sk k T_\sk$, where, $n\sk$ and $T_\sk$ are the SKD
local number density and temperature, respectively.
The outer disc emits mainly via synchrotron and bremsstrahlung processes and the corona
additionally via inverse-Compton process. 
The functional form of the frequency integrated, local intensity of the outer disc is given by \citep{kc14,vkmc15}, 
$$
{\tilde I}_{\od}={\tilde I}_{\rm syn}+{\tilde I}_{\rm brem}
$$
\bea
=\left[\frac{16}{3}\frac{e^2}{c}\left( \frac{eB_{\od}}{m_e c} \right)^2
 \Theta^2_{\od} n_{\od} x+ 1.4\times 10^{-27}n_{\od}^2g_bc \sqrt{\frac{\Theta_{\od} 
m_e}{k}}\right] \nonumber \\
\times \frac{\left(d_0 \sin \theta_{\od}+x\cos \theta_{\od}
\right)}{3} \ \  {\rm erg} \
{\rm cm}^{-2} {\rm s}^{-1}
\label{skint.eq}
\eea
Here, $\Theta_{\od}, n_{\od}, x, \theta_{\od}$, $B_{\od}$ and $g_b(=1+1.78\Theta_{\od}^{1.34})$ are the 
local dimensionless temperature, electron number density, radial distance of the disc, the semi-vertical
angle of the outer disc surface, 
the magnetic field and relativistic Gaunt factor, respectively. Intensity is measured in the disc local rest
frame.
The factor outside square brackets
converts emissivity (${\rm erg}~{\rm cm}^{-3}{\rm s}^{-1}$) into intensity  (${\rm erg}~ {\rm cm}^{-2}{\rm s}^{-1}$).
The luminosity of the outer disc is obtained by integrating $I_0$ over the disc surface, {\ie}
\be
L_{\od}=2\int^{x_{0}}_{\xsh} \int^{2\pi}_{0} I_{\od}r~{\rm cosec}^2\theta_{\od}~d\phi dx
\label{lum.eq}
\ee
which, can be presented in units of $L_{\rm Edd}(\equiv 1.38\times10^{38}\mbh/\msol~{\rm erg}~{\rm s}^{-1})$ as $\ell_{\od}=L_{\od}/L_{\rm Edd}$.

Now in chapter \ref{CH:P1}, since the accretion disc solution has been approximated, we do not calculate the radiation from corona directly, but instead estimate it from the enhancement
factor computed from self-consistent two temperature solutions \citep{mc08}. 
The ratio of corona and outer disc luminosities, is computed following \citet{mc08} and was presented
in \citet{vkmc15},
%\begin{eqnarray}
%\chi=-5.974+1.996 \xsh -0.166 \xsh^2 + 6.653 \times 10^{-3} \xsh^3 \nonumber \\
%   -1.280 \times 10^{-4} \xsh^4 + 9.455\times 10^{-7} \xsh^5~(\xsh < 35); \nonumber \\
%\chi=2.693+0.096 \xsh-3.465 \times 10^{-3} \xsh^2 + 3.898\times 10^{-5} \xsh^3  \nonumber \\
%-1.439\times 10^{-7} \xsh^4 ~(\xsh \geq 35)% \nonumber
%\label{chi1.eq}
%\end{eqnarray}
%We assume that these functions are generic.
So the luminosity from the corona can be estimated as $L_{\rm \small C}= \chi L_{\od}$,
and the dimensionless total luminosity is given by
\begin{eqnarray}
\ell=\ell_{\rm \small C}+\ell_{\rm \small D}=(1+\chi)\ell_{\rm \small D}
%;~~\mbox{for}\mbh=\Ma \nonumber \\
%~~~  =\chi_8 L_{\od};~~\mbox{for}\mbh=\Mb
\label{ell.eq}
%\label{coronalum.eq}
\end{eqnarray}
$\chi$ is depends upon mass of the BH. \citet{vkmc15} showed that the magnitude is more for suparmassive BH then that for stellar mass BH. This method of obtaining corona luminosity is used in chapters \ref{CH:P1} and \ref{CH:P3_4}. However, in chapter \ref{CH:P5}, the corona luminosity is calculated using corona counterpart of equation (\ref{lum.eq}) with calculating specific intensity of corona similar to outer disc (Eq. \ref{skint.eq}) \citep{vc18a}. %The total disc luminosity $\ell$ (in units of $L_{\rm Edd}$) is given by,
%\be
%%\ell=\ell_{\rm \small C}+\ell_{\rm \small D}=(1+\chi)\ell_{\rm \small D}
%
%\ee

The specific intensity measured in the local rest frame of the corona is given by 
%${\tilde I}_{\rm \small C}= L_C/{\pi}{A_C}$. 
\be
{\tilde I}_{\rm \small C}=L_{\rm \small C}/{\pi}{A_{\rm \small C}}={\ell_{\rm \small C}}L_{\rm Edd}/{\pi}{A_{\rm \small C}}~({\rm erg}~{\rm cm}^{-2}~{\rm s}^{-1}),
\label{coronaint.eq}
\ee
Here $A_{\rm \small C}$
is the surface area of the corona. 
To obtain the specific radiation intensities (equations \ref{skint.eq} and \ref{coronaint.eq}) from the accretion disc, we need the number density and temperature distribution of the disc. Here, ${\tilde I}_{\rm \small C}$ is obtained from ${\tilde I}_{\od}$, in which $n_{\rm \small D}$ 
is obtained by supplying ${\dot m} $ and equation(\ref{accvel.eq}), and $\Theta_{\rm \small D}$
from equation(\ref{acctemp.eq}). 
%In this paper, we have only concentrated on accretion discs around $\mbh=10\msol$.
Similarly the intrinsic Keplerian disc (KD) with accretion rate to be ${\dot M}_{\rm KD}={\dot m}_{\rm KD}{\dot M}_{\rm Edd}$, the specific intensity is given by \citep{ss73} 
\begin{equation}
I_{\kd 0}=\frac{3GM_{B}{\dot M}_{\rm K}}{8{\pi}^2r^3}\left(1-
{\sqrt{\frac{3r_g}{r}}} \right) \ \  {\rm erg} \
{\rm cm}^{-2} {\rm s}^{-1}
\label{kdint.eq}
\end{equation}
%\end{center} 
%\section*{APPENDIX B : Paczy\'nski-Wiita potential and general relativity}
\section{Paczy\'nski-Wiita potential (PW) and relativistic flows}
\label{app:pw_gr}
More often, radiatively driven relativistic jets are studied in the SR plus PW regime and not in GR regime.
Although Equivalence principle strictly precludes this possibility, but in astrophysics this trend has been 
followed by a number of researchers because it is assumed that GR affects only in the region outside the
BH and not at moderate to large distances. Here we show that, the differences in equations in the two approach
affects the solutions close to the BH, as well as at moderate distances (few$\times~10\rg$). Moreover, the temperature produced in SR+PW regime is unphysically high. Furthermore, radiation effects in curved space time
cannot be properly taken into account by any scaling relations, if the moments are computed in the flat space.
The effect of curvature in the radiation term has been addressed in section \ref{sec:rad_curved}.
So we list the first two points below. 
\subsection{Equations of motion}
%\textbf{(i) Paczynski-Wiita (PW) potential vs GR calculations}\\
The equations of motion in the two approaches, can be written down as,
\begin{equation}
\left[\frac{dv}{dr}\right]_{GR}=\frac{2a^2r-{ \frac{\left(1-a^2\right)r}{r-2}}+\frac{(2-\xi)\gamma r^2}{(f+2\Theta){ \sqrt{g^{rr}}}}\left[(1+v^2){{\cal R}_1}-v
{\left(g^{rr} {\cal R}_0+\frac{{\cal R}_2}{g^{rr}}\right)}\right]}{r^2\gamma^2v\left(1-\frac{a^2}{v^2}\right)}
\end{equation}
\begin{equation}
\begin{split}
\gamma^4v&\left(1-\frac{a^2}{v^2}\right)\left[\frac{dv}{dr}\right]_{GR}=\frac{2a^2 \gamma^2}{r}-{\frac{\left(1-a^2\right)\gamma^2}{r(r-2)}} \\
& +\frac{(2-\xi)\gamma^3}{(f+2\Theta){ \sqrt{g^{rr}}}}\left[(1+v^2){{\cal R}_1}-v
{\left(g^{rr} {\cal R}_0+\frac{{\cal R}_2}{g^{rr}}\right)}\right]
\end{split}
\label{appb1.eq}
\end{equation}
and
%\begin{equation}
%\left[\frac{dv}{dr}\right]_{PW}=\frac{2a^2r-{ \frac{(1-v^2)r^2}{{(r-2)^2}}}+\frac{(2-\xi)\gamma r^2}{(f+2\Theta)}\left[(1+v^2){{\cal R}_{1F}}-v
%{ \left({\cal R}_{0F}+{\cal R}_{2F}\right)}\right]}{r^2\gamma^2v\left(1-\frac{a^2}{v^2}\right)}
%\end{equation}
\begin{equation}
\begin{split}
\gamma^4v&\left(1-\frac{a^2}{v^2}\right)\left[\frac{dv}{dr}\right]_{PW}=\frac{2a^2 \gamma^2}{r}-{\frac{1}{{(r-2)^2}}} \\
&+\frac{(2-\xi)\gamma^3}{(f+2\Theta)}\left[(1+v^2){{\cal R}_{1F}}-v
{ \left({\cal R}_{0F}+{\cal R}_{2F}\right)}\right]
\end{split}
\label{appb2.eq}
\end{equation}
The subscript PW signifies equations of motion in SR+PW regime, while GR represent the equations in
Schwarzschild metric.
Values with subscript $F$ are calculated in flat space.
The r. h. s of the two equations (\ref{appb1.eq} and \ref{appb2.eq}) algebraically differ in the second term of the numerator,
and the presence of curvature $g^{rr}$ in the radiation terms. 
Since the form of the first terms are same, let us take the ratio of the 2$^{nd}$
terms on R.H.S of the above two equations,
$$
\frac{(1-a^2)g^{rr}}{{(1-v^2)}}=\delta
$$
The calculations with SR+PW potential will be comparable to GR if $\delta\sim1$. 
%This condition reveals that PW potential has been successful in solving accretion problems but will give large error in study of relativistic outflows. In spherical accretion, at larger distances, both $v$ and $a$ are much smaller than 1 and as $g^{rr}\rightarrow1$, this condition is satisfied by remarkable precision. As fluid comes closer to the BH, $(1-v^2)$ decreases faster than $(1-a^2)$ and it is compensated by decrease in $g^{rr}$. On the contrary
For relativistic winds or jets, at large distances, $g^{rr}\approx 1$, $a$ is very low but $v\rightarrow 1$, then
$$
\delta \gg 1
$$
Similarly close to the BH i.e., $r\rightarrow 2$, $g^{rr} \rightarrow 0$,
$v\approx 0$, but $a$ is large, therefore
$$
\delta \ll 1
$$
This analysis points to the possibility
of a large deviation from GR solutions even at larger distances. \\
This difference in the second term of the EoMs arise because in SR+PW, regime the gravity enters as an additive
term (equation \ref{appb2.eq}), while in GR it is a space time phenomena (equation \ref{appb1.eq})
so it affects any source of energy. Therefore, the curvature term ($g^{\mu \nu}$s) should couple with the thermal and the kinetic
terms as is seen above, i. e., the curvature term is coupled with the thermal term in the form of sound speed $a$
and also the Lorentz factor $\gamma$. Consequently, if there is a discontinuity like shock in the flow,
then the gravity term in SR+PW will not change across the shock, but in GR it will change, since
both $a$ and $v$ jump across a shock.
%If one may add further,
%\citet{abci96} also showed that even in accretion problems, SR and PW potential are not compatible.
\subsection{Overestimated thermal content in PW analysis}
\begin{figure}[h]
\begin{center}
 \includegraphics[width=15.cm, angle=0, trim=0 0 7cm 5cm,clip]{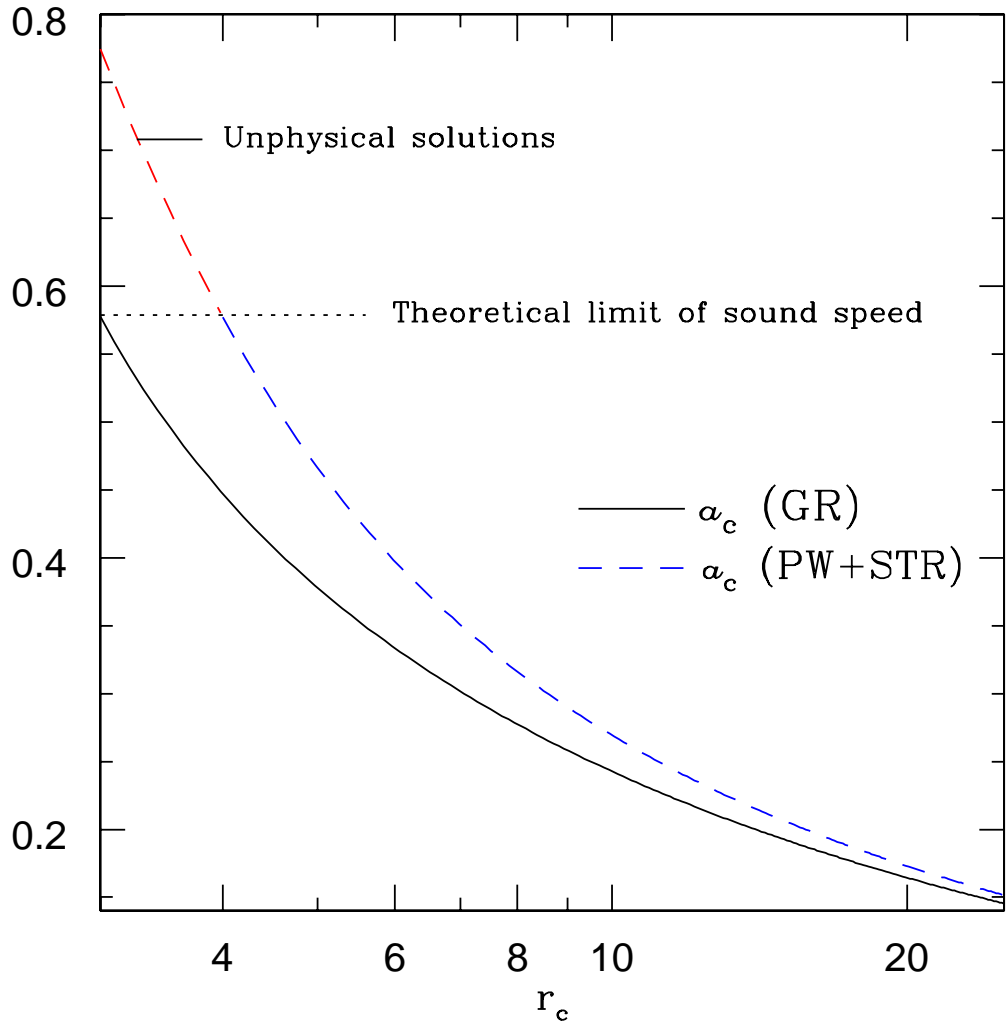}
\vskip -0.5cm
 \caption{$a_c$ as a function of $r_c$ for $\ell=0$. % Right : $\Theta_c$ as a function of $r_c$ for $\ell=2.85$. 
 Solid curve is for GR solutions while dotted shows solutions with PW potential
}
\label{lab:figB1}
 \end{center}
\end{figure}
Causality imposes an upper limit of sound speed which is $a<1/\sqrt{3}$. In GR this translates to a lower bound
in the location of sonic points ($r_c>3$). This is clear in Figs. (\ref{lab:fig5n}). Since pseudo potentials makes the flow unphysically hot, so the lower limit
of sonic point in pNp+SR regime extends to a larger distance ($r_c>4$).  
For thermally driven flow this can be very easily shown. From equations (\ref{appb1.eq}) and (\ref{appb2.eq})
and ignoring radiation, we obtain $a_c$ as a function of $r_c$. 
$$
[a_c]^2	_{GR}=\frac{1}{2r_c-3}
$$
$$
[a_c]^2	_{PW}=\frac{r_c}{2(r_c-2)^2+r_c}
$$
These are algebraic relations, and $[a_c]_{PW}$
is higher than $[a_c]_{GR}$ (Fig. \ref{lab:figB1}). In other words, the jet in SR+PW description
is much hotter than the GR one. This also means the SR+PW jet is subjected to a much stronger
thermal gradient push than it happens in reality. Moreover, all jet solutions with $3<r_c\leq 4$ are absent in
SR+PW solutions.